\documentclass{PoS}
\usepackage{amsmath,amssymb,amsfonts}
\usepackage{graphicx, graphics, subfigure} 

\title{Introduction to universality and renormalization group techniques}

\ShortTitle{Introduction to universality and renormalization group techniques}

\author{\speaker{Alessandro Sfondrini}%
         \\
	Institute for Theoretical Physics and Spinoza Institute,
        Utrecht University,\\
	Leuvenlaan 4, 3584 CE, Utrecht, The Netherlands.\\
        E-mail: \email{A.Sfondrini@uu.nl}}

\abstract{These lecture notes have been written for a short introductory course on universality and renormalization group techniques given at the VIII Modave School in Mathematical Physics by the author, intended for PhD students and researchers new to these topics.\\
First the basic ideas of dynamical systems (fixed points, stability, etc.) are recalled, and an example of universality is discussed in this context: this is Feigenbaum's universality of the period doubling cascade for iterated maps on the interval. It is shown how renormalization ideas can be applied to explain universality and compute Feigenbaum's constants.\\
Then, universality is presented in the scenario of quantum field theories, and studied by means of functional renormalization group equations, which allow for a close comparison with the case of dynamical systems. In particular, Wetterich equation for a scalar field is derived and discussed, and then applied to the computation of the Wilson-Fisher fixed point and critical exponent for the Ising universality class.\\
References to more advanced topics and applications are provided. }

\FullConference{Eighth Modave Summer School in Mathematical Physics\\
		 26th august -  1st september 2012\\
		 Modave, Belgium}

\newtheorem{theorem}{Theorem}[section]

\newtheorem{exercise}[theorem]{Exercise}

\def\M{\mathcal{M}}
\def\U{\mathcal{U}}
\def\V{\mathcal{V}}
\def\Ws{\mathcal{W}^s}
\def\Wu{\mathcal{W}^u}
\def\R{\mathbb{R}}
\def\N{\mathbb{N}}
\def\Z{\mathbb{Z}}
\def\Ren{\mathfrak{R}}
\def\ren{\mathsf{R}}

\def\eps{\varepsilon}

\def\a{\alpha}
\def\b{\beta}
\def\g{\gamma}
\def\d{\delta}

\def\be{\begin{equation}}
\def\ee{\end{equation}}

\newcommand{\bea}{\begin{eqnarray}}
\newcommand{\eea}{\end{eqnarray}}

\begin{document}

\section*{Introduction and plan of the notes}

These notes have been  written for a short course at the VIII Modave school of Mathematical Physics, and aim to illustrate some aspects of universality and Renormalization Group (RG) techniques in Physics.

The topic is broad and complex. Even focusing on a particular aspect would take a whole monograph, and indeed there are several dedicated to this purpose. Here the focus is on the relatively new approach of Functional Renormalization Group Equations (FRGE, sometimes called, misleadingly in our opinion, ``exact''), and in particular on Wetterich's equation, which is being widely applied in statistical Physics, Standard Model Physics, and to study the asymptotic safety conjecture of Weinberg for Quantum Gravity.

The course from which these notes originated was intended for young researchers that are not necessarily familiar with renormalization group ideas, a topic that often finds little space in master courses in theoretical Physics. Therefore, rather that focusing on Quantum Field Theory (QFT) alone and going deep into the many technical tools needed to employ the  FRGE in the current topics of research, we preferred to keep the discussion as elementary as possible. In this spirit, the ideas of universality and renormalization group are first presented in the simple context of dynamical systems.

The first part of the notes introduces the basic notions related to dynamical systems, including fixed points, attractors and asymptotic manifolds, which are crucial for understanding FRGE in quantum field theory. Then Feigenbaum's universality for iterated maps on the interval is presented, and explained in terms of renormalization group, following the exposition of Collet and Eckmann. This has the advantage of introducing RG in a relatively simple context, where the mechanism has been completely understood and mathematically formalized.

In the second part we focus on QFT, restricting to the case of scalar fields. First we briefly recall some textbook notions about QFT: divergences, renormalization, anomalous scaling, etc., fixing the notation for the following sections. Then we present the formal derivation of Wetterich's FRGE and apply it to obtain the perturbative $\b$-functions of $\lambda\phi^4$ theory, commenting on the differences with perturbative calculations as obtained \textsl{e.g} in dimensional regularization. Finally, we use the FRGE to investigate the Wilson-Fisher fixed point for three-dimensional Ising-like magnets, a calculation quite reminiscent of the one for the Feigenbaum fixed point.

Many things are left out of this presentation. The general spirit is to present all ideas in the simplest possible context, and extend them to more realistic scenarios by analogy. The mathematical formalization is minimal by necessity of elucidating the more conceptual points. This can be a bit unsatisfactory in particular in the first part, where the such formalization could be used to provide rigorous proof of universality. In the second part, many important topics are omitted, such as the application of FRGE to theories with (gauge) symmetries and on curved backgrounds, as well as a discussion of the asymptotic safety conjecture for gravity; however, such topics may be found elsewhere in the literature. Good companions of these note are the ones of two parallel Modave courses that  introduce conformal field theories \cite{rovai} and elucidate the relation between RG flows and scale invariance in QFT \cite{moskovic}, topics that unfortunately do not find space here.

Finally, here the FRGE is presented as an useful tool but no statements are made about its reliability for rigorously investigating non-perturbative scenarios and establishing renormalizability in a constructive sense, which we believe is yet to be completely understood. We hope to return to these issues in the future.

To make the presentation more pedagogical, several figures are included, as well as a selection of short  exercises.
\bigskip

\subsubsection*{Acknowledgments}
It is a pleasure to thank the organizers of the VIII Modave School on Mathematical Physics for their invitation, as well as the other lecturers and the participants for stimulating discussions in Modave. I especially thank Antonin Rovai, thanks to whose effort these proceedings exist. I acknowledge support by the Netherlands Organization for Scientific Research (NWO) under the VICI grant 680-47-602. My work is  also part of the 
ERC Advanced grant research programme No. 246974, ``Supersymmetry: a window to non-perturbative physics''.\\
I am also grateful for the warm hospitality of CPHT of \'Ecole Polytechnique in Paris, during which part of these notes has been written, and for the interesting discussions on this subject that I had there and at the ITF in Utrecht.

\newpage

\section*{Part I: Dynamical systems}
Dynamical systems constitute a branch of mathematical Physics concerned, roughly speaking, with study of the time evolution of some space. They can be used to study many types of problems, from chemical reactions to fluctuations of the stock market, and including all of Hamiltonian mechanics.

The type of time evolution (deterministic or stochastic) and the nature of the space under consideration (a smooth manifold, a measure space, etc.) dictate the kind of tools to be employed in this study. Here we will consider only the simplest type of dynamical systems, where evolution is deterministic and  the space is some (open subset of) $\R^n$. General references are e.g. \cite{strogatz, devaney, abraham, rouelle}.

These simple cases can be split in two categories depending on which notion of time we decide to adopt that is, whether the system will have \emph{continuous time} $t\in\R$ or \emph{discrete time} $t\in\mathbb{Z}$. Let us start from the former.

\section{Dynamical systems with continuous time}
\subsection{Generalities}
In this case we are with dealing autonomous, first order ordinary differential equations (ODEs) on an open set $\M\in\R^n$, of the form
\be
\label{eq:ODE}
\dot{x}=f(x)\,,\quad\quad x\in\M \,.
\ee
To keep our discussion as simple as possible, we assume everything to be suitably regular, so that Cauchy's theorem guarantees the existence of the unique solution with initial condition $x\in\M$ at time $t=0$. We will call this solution $\Phi^t(x)$. Furthermore we will assume that this solution exists for all $t\in\R$ (or at least for all $t>0$). In fact, in what follows, we will not be interested in solving one specific Cauchy problem, but on understanding the generic motion of a generic point $x\in\M$, and in particular in what happens asymptotically, i.e. when $t\to\infty$. To this end, we can study the map
\be
\Phi^t:\quad \M \to \M\,,
\ee
which will be smooth under our assumptions. In the case where the solution exists for all $t\in\R$, one immediately notices the following properties
\be
\Phi^0=\mathrm{Id}\,,\quad
(\Phi^t)^{-1}=\Phi^{-t}\,,\quad
\Phi^t\circ\Phi^s=\Phi^{t+s}\,,
\ee
that imply that we can define a one-parameter abelian group of diffeomorphisms
\be
\Phi=\{\Phi^t,\quad t\in\R\}\,.
\ee
One could also distinguish the case where $\Phi(x)$ is not invertible, and consequently $\Phi$ is only a \emph{semigroup}, but this is not important now.
We will say that $\Phi$ is the \emph{flow} of the differential equation (\ref{eq:ODE}).

We are now in a position to give a more formal definition. A \emph{continuous time regular dynamical system} is a couple $(\M,\Phi)$ where $\M$ is\footnote{More generally, $\M$ can be a $n$-dimensional smooth manifold.} a regular open subset of $\R^n$ and $\Phi$ is a one parameter group of diffeomorphisms.

Before proceeding further, it is worth pausing for some comments. We said that we will be mostly interested in the asymptotic behavior of these systems. Why? Let us be concrete, and pick the study of the solar system as a specific case. In principle, we could solve Cauchy's problem for given initial conditions on the positions and velocities of the planets. In practice this is impossible analytically, but for the sake of the argument let us assume that we can do it somehow (e.g. numerically) to any desired accuracy. Then, since Newton's equations are deterministic, we should know exactly the orbits of any planet at any instant.\footnote{ This reasoning could be called Laplacian determinism, as the XIX century mathematician Laplace was among the ones that emphasized it most.}

In modern times, after the works of Poincar\'e, this idealization has been considered less and less convincing, for reasons that are probably clear to the reader. First, we cannot solve any ODE to any accuracy, and besides, when we decide to treat the solar system as isolated we overlook the possibility of a small perturbation occurring at some point (e.g. a small asteroid passing through it) that would alter the equations. 
Even if we could eliminate these problems, we still have to specify some initial conditions, that come from a physical measurement, and carry some uncertainty, i.e. $x(t=0)=x_0\pm\delta x$. Using regularity of the ODEs, one can find a bound on the propagation of this uncertainty: schematically
\be 
\label{eq:expbehav}
\|\Phi^t(x_0+\delta x)-\Phi^t(x_0)\|\lesssim C\, e^{\lambda\,t}\,\delta x\,,\quad C>0,\ \lambda>0\,.
\ee
In practice, due to the wild exponential behavior, even when $\delta x$ is very small, the separation will become large after a relatively short interval of time.

For these reasons it is more interesting to study the properties of $\Phi(x)$ for a generic $x\in\M$, and ask asymptotic and often qualitative questions, such as whether the solar system is stable, or some planet will be ejected from it at some point. This will be our general approach, not only in the context of dynamical systems, but of Quantum Field Theories (QFTs) as well.

\begin{exercise}[First order ODEs] 
We are restricting here to (systems of) first order, autonomous ODEs, which could seem to be a strong limitation. On the contrary, show that an (autonomous or not) ODE of any order can be recast as system of  first order autonomous ODEs.
\end{exercise}

\begin{exercise}[ODEs, solutions, and existence for all times] 
We are restricting to ODEs such that $\Phi^t$ exists and is unique for all $t\in\R$. One should realize that this is truly a restriction.
\begin{enumerate}
\item Recalling Cauchy's theorem, write down a Cauchy problem that has infinitely many solutions.
\item Consider the ODE with $\M=\R$ and given by $\dot{x}=1+x^2$. Does the solution exist for all $t\in\R$?
\end{enumerate}
\end{exercise} 

\begin{exercise}[Dependance on initial conditions]
\label{ex:pendulumtip}
We want to illustrate the magnitude of the effects in (\ref{eq:expbehav}). A simple, perhaps a bit trivial example, is a pencil balanced on its tip (inverse pendulum). The pencil has length $\rm L$, and it is balanced on its tip with no speed and up to a great accuracy in the position of its top, $\rm \delta$. After how long a time $\rm \Delta t$ we won't be able to predict the position of the pencil with a precision of at least $1\ {\rm rad}\approx57^\circ$?\\
You can take ${\rm L=10\,cm}$, and try ${\rm \delta=10^{-4}m}$, ${\rm \delta=10^{-15}m}$.\\
{\rm (Answer:  $\rm \Delta t \approx 0.7\,s$ and $\rm \Delta t \approx 3.2\,s$.)}
\end{exercise}

\subsection{Asymptotic behavior: fixed points}
Here we will discuss the asymptotic behavior for autonomous ODEs. The simplest case is when there exists one point $x^*\in\M$ such that
\be
\Phi^t(x^*)=x^*\,,\quad \forall\ t>T\,,
\ee
for some $T$. It is clear that this is equivalent to requiring the above condition to hold for all $t$ and, in terms of (\ref{eq:ODE}), to requiring that
\be
f(x^*)=0\,.
\ee
We will call such an $x^*$ a \emph{fixed point} or \emph{critical point} for the dynamical system.

Of course it is not typical that the initial condition for a Cauchy problem is precisely $x(0)=x^*$. These points are interesting because they influence the flow for any initial condition close to them. Consider, for instance, the case where $\M=\R$ and there exists an unique critical point $x^*$ such that\footnote{We regard the case $f'(x^*)=0$ as non-generic.} $f'(x^*)\neq0$. Then, two things may happen:
\begin{itemize}
\item $f'(x^*)>0$: then $f(x)<0$ to the left of $x^*$, and the flow pushes these points to smaller values, away from $x^*$; similarly $f(x)>0$ to the right of $x^*$ and the flow pushes them to the right, again away from $x^*$. 
\item $f'(x^*)<0$: then $f(x)>0$ to the left of $x^*$, and the flow pushes these points to larger values, towards $x^*$; similarly $f(x)<0$ to the right of $x^*$ and the flow pushes them to the left, again towards $x^*$. Intuitively,\footnote{See Exercise \ref{ex:flow}.} the flow cannot cross $x^*$, so that the motion tends to $x^*$.  
\end{itemize}
We leave it to the reader that is new to these ideas to familiarize himself with this idea, e.g. by drawing more general scenarios for $f(x)$ when  $\M=\R$.  

This simple example motivates the need to classify fixed points depending on their property to attract or repel the points in their neighborhood under the flow $\Phi$. This classification, together with many useful criteria, was first put forward by Lyapunov.
Let $x^*$ be a critical point for $\Phi$, and let $\Phi$ exists for all $t\in\R$. Then
\begin{enumerate}
\item $x^*$ is \emph{attractive} (or \emph{asymptotically stable}) if there exists a neighborhood $\V$ of $x^*$ such that
\[x\in \V\quad\quad\Rightarrow\quad\quad \lim_{t\to+\infty}\Phi^t(x)=x^*\,.\]
\item $x^*$ is \emph{stable} for all times\footnote{Stability only in the future or past amounts to restricting to $t>t_0$ or $t<t_0$ respectively.} if for any neighborhood $\U$ of $x^*$ there exists a neighborhood $\V_0$ of $x^*$ such that
\[x\in \V_0\quad\quad\Rightarrow\quad\quad\Phi^t(x)\in\U\quad\quad\forall\ t\in\R\,.\]
\item $x^*$ is \emph{unstable} if it is not stable. 
\item $x^*$ is \emph{repulsive} if there exists a neighborhood $\V$ of $x^*$ such that
\[x\in \V\quad\quad\Rightarrow\quad\quad \lim_{t\to-\infty}\Phi^t(x)=x^*\,.\] 
\end{enumerate}

With this terminology, it is easy to classify the dynamical systems given by a \emph{linear} ODEs. As it is well known, we have in that case
\be
\dot{x}=A\,x\,,\quad\quad \Phi^t(x)=e^{t\,A}\,x\,,
\ee
where $e^{t\,A}$ is the exponential of a matrix, defined by the convergent series $e^{t\,A}=\sum_{k=0}^\infty t^k\,A^k/k!\,$.
Furthermore, if $A$ is symmetric\footnote{More generally, similar considerations can be made using the Jordan form of $A$, but they will not be important for us.} then it can be diagonalized, and it is enough to consider its eigenvalues $\lambda_1,\dots,\lambda_n$.
Clearly, there is just one fixed point $x^*=0$, and it is not hard to verify the following statements:
\begin{itemize}
\item If Re$(\lambda_i)<0$ for all eigenvalues, then $x^*$ is attractive.
\item If Re$(\lambda_i)>0$ for all eigenvalues, then $x^*$ is repulsive.
\item If Re$(\lambda_i)=0$ for all eigenvalues, then $x^*$ is stable.
\item If Re$(\lambda_i)<0$ for some eigenvalues and Re$(\lambda_j)>0$ for others, then $x^*$ is unstable.
\end{itemize}
We will also say that $x^*$ is \emph{hyperbolic} if Re$(\lambda_i)\neq0$ for all $i$.

Of course linear equations are not very interesting \textit{per se}. However, consider any ODE with (at least) one fixed point $x^*$. Then we can write
\be
\dot{x}=f(x)=A\,(x-x^*)+O\big(\|x-x^*\|^2\big)\,,\quad\quad{\rm with}\quad A={\rm Jac}\left.f\right|_{x=x^*}\,,
\ee
simply by expanding around $x^*$. Our intuition suggests then that, at least if $x^*$ is hyperbolic, then the linearized analysis should be enough to ``understand'' the flow, at least close to $x^*$. This is actually the case, as it follows from an important theorem that we will state without proof (see e.g. \cite{rouelle}).

\emph{Grobman-Hartman theorem:} if $\dot{x}=f(x)$ has a hyperbolic fixed point $x^*\in\M=\R^n$, then there is a neighborhood of $x^*$ such that the flow there is homeomorphic to the flow of the linear system $\dot{x}=A\,x$. In other words, locally the nonlinear flow is conjugated by a continuous invertible map to the linear one. There are extensions of this theorem that guarantee (under additional hypotheses)  that the two flows are diffeomorphic, but they are more subtle and we will not discuss them here.

As a corollary of this theorem, it follows that the stability property of an hyperbolic fixed point can be found from the corresponding linearized ODE, i.e. from the eigenvalues of the Jacobian at the fixed point, which is also known as Lyapunov's spectral method.

Before proceeding further, it is worth illustrating all this on a simple example. Consider the following dynamical system, which we spell out in coordinates:
\be
\dot{x}_1=x_1-x_1\,x_2\,,\quad\quad\dot{x}_2=-x_2+x_1^2\,.
\ee 
One easily finds three fixed points
\be
x^*_{(1)}=(0,0)\,,\quad\quad
x^*_{(2)}=(1,1)\,,\quad\quad
x^*_{(3)}=(-1,1)\,.
\ee
Consider for instance $x^*_{(1)}$. It is clearly hyperbolic and unstable, because the Jacobian there is the matrix $A={\rm diag}(1,-1)$. Therefore, by Grobman-Hartman theorem, it follows that the flow in the vicinity of $x^*_{(1)}$ is conjugated to the one of the associated linear system. The latter is very simple: in the linear system, the axis $x_1$ supports an exponentially repulsive motion, whereas the axis $x_2$ supports an exponentially attractive one. Generic initial conditions yield hyperbolae that asymptote to the coordinate axes.

\begin{figure}[!t]
  \begin{center}
    \subfigure[Full nonlinear flow.]{\label{fig:nonlinearflow}\includegraphics[width=0.4\textwidth]{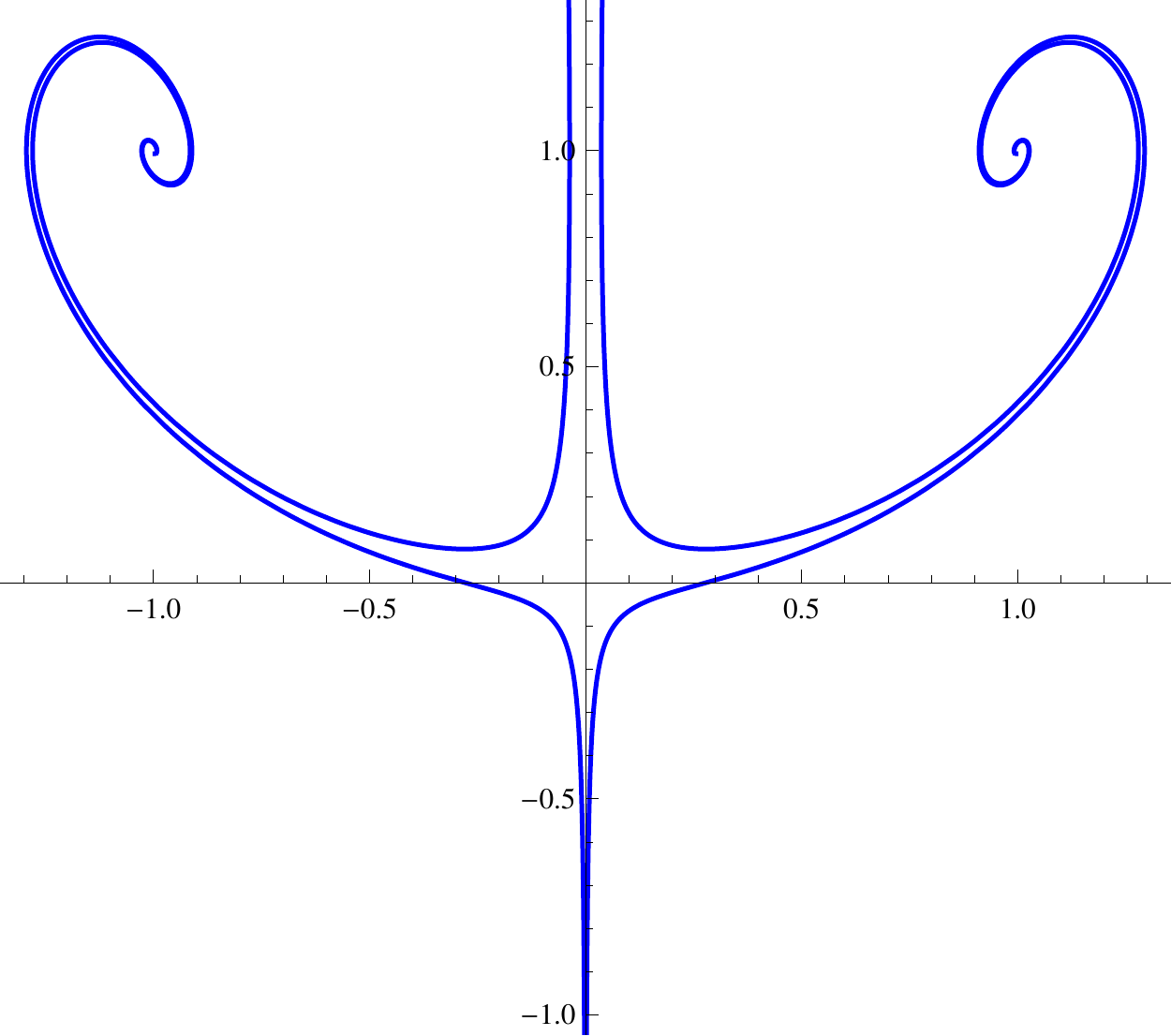}}
    $\quad\quad$
    \subfigure[Linearization around $(0,0)$.]{\label{fig:linearflow}\includegraphics[width=0.4\textwidth]{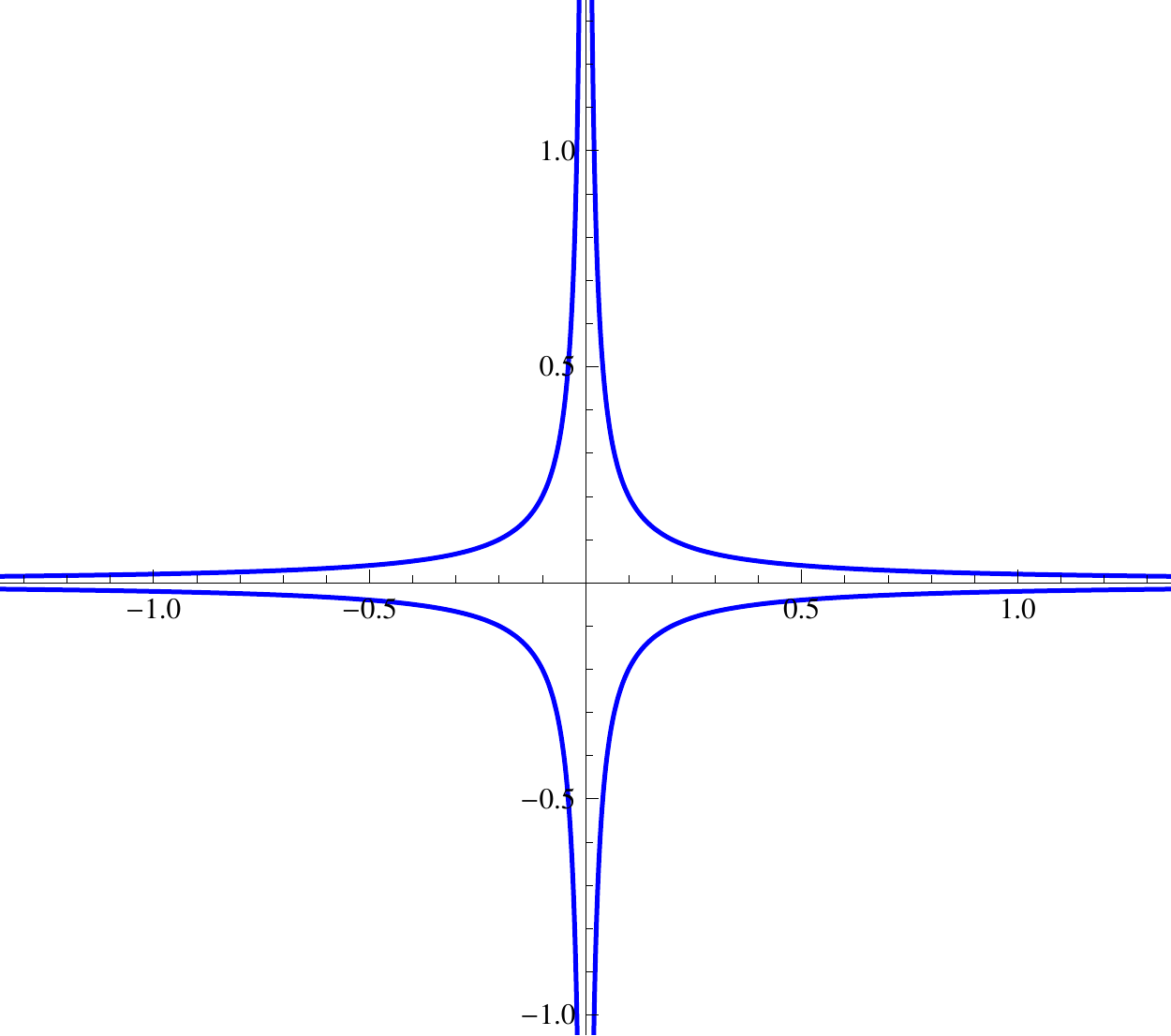}}
  \end{center}
  \caption{Flow for a nonlinear and linearized system.}
  \label{fig:flow}
\end{figure}

Looking at Figure \ref{fig:flow} it is clear that the nonlinear flow around  $x_{(1)}^*$ is similar to the linear one. It is interesting to look for some curves that play a role similar to the coordinate axes in the linear system, which helped us to understand the motion of a generic initial condition. A very important and useful result guarantees their existence.

\emph{Stable manifold theorem (Perron-Hadamard):} if $x^*$ is a hyperbolic fixed point, then the two sets
\bea
\Ws&=&\left\{x\in\M:\ \lim_{t\to+\infty}\Phi^t(x)=x^*\right\}\,,\\
\Wu&=&\left\{x\in\M:\ \lim_{t\to-\infty}\Phi^t(x)=x^*\right\}\,,
\nonumber
\eea 
are regular manifolds called stable and unstable manifold, and are tangent to the hyperplanes generated by the eigenvectors of the linearized system corresponding resp. to negative and positive eigenvalues (see e.g. \cite{abraham,HPS}).

We shall not prove this theorem, but it is worth noticing that the difficulty is in proving that a regular stable (unstable) manifold exists locally. Once one has constructed such $\Ws_{\rm loc.}$ ($\Wu_{\rm loc.}$) it is easy to obtain the whole manifolds by defining
\be
\Ws=\bigcup_{t\leq0} \Phi^t(\Ws_{\rm loc.})\,,\quad\quad\Wu=\bigcup_{t\geq0} \Phi^t(\Wu_{\rm loc.})\,.
\ee

We conclude this section with a word of warning: we discussed only the dynamics around fixed points, but these are not the only objects that influence the asymptotic behavior. From dimension $n\geq2$, dynamical systems may present \emph{limit cycles}, and for $n\geq3$ they may have \emph{chaotic behavior}. Even giving an appropriate definition of these concepts will take us too far, and we invite the interested reader to consult, e.g. \cite{strogatz}. 

\begin{exercise}[Qualitative analysis of ODEs]
In case the reader is not familiar with the qualitative analysis of ODEs, we suggest some exercises here.
\begin{enumerate}
\item Consider the dynamical system $\dot{x}=A\,x$ on $\M=\R^2$, where $A$ is a symmetric, constant matrix, and draw the trajectories (the phase portrait) of the system.
\item Consider a mechanical system of the form $\ddot{x}=-V'(x)$. What are the stable and unstable fixed points in terms of $V(x)$? Can the system have attractive fixed points?
\item Perform the qualitative study of the Lotka-Volterra system
\be
\dot{x}_1=\a\, x_1-\b\, x_1 x_2\,,\quad\dot{x}_2=\g\, x_2+\d\, x_1 x_2\,,\quad \a,\,\b\,,\g\,,\d\,>0\,.
\ee
You will find a detailed and commented analysis of this system e.g. in \cite{strogatz}.
\end{enumerate}
\end{exercise} 

\begin{exercise}[Some short proofs]
\label{ex:flow}
In stating our propositions we accepted some facts that the interested reader could prove, for $(\M,\phi)$ a regular continuous time dynamical system.
\begin{enumerate}
\item Prove that if $x,y\in\M$, $x\neq y$, then $\phi^t(x)\neq\phi^t(y)$. Conclude that a trajectory can never go past a fixed point. 
\item Prove that hyperbolic fixed points are always isolated.
\item List reasonable properties that $\Ws_{\rm loc.}$  and $\Wu_{\rm loc.}$ should have, and assume them to prove the stable manifold theorem in two dimensions.
\end{enumerate}
\end{exercise} 

\subsection{Bifurcations}
We introduce here the notion of \emph{family of dynamical systems}, that we will analyze in more detail for discrete-time systems. Here we simply allow for a (regular) dependence of (\ref{eq:ODE}) on one or more real parameters $\mu_1,\dots,\mu_n$, so that we have
\be
\dot{x}=f(x;\,\mu_1,\dots,\mu_n)\,,\quad\quad x\in\M \,.
\ee
This generalization is quite natural, as it allows to study the evolution of a system depending on some external condition, such as the demographics as a function of resources, etc.

\begin{figure}[!t]
  \begin{center}
    \subfigure[Tangent bifurcation.]{\label{fig:tangentbif}\includegraphics[width=0.48\textwidth]{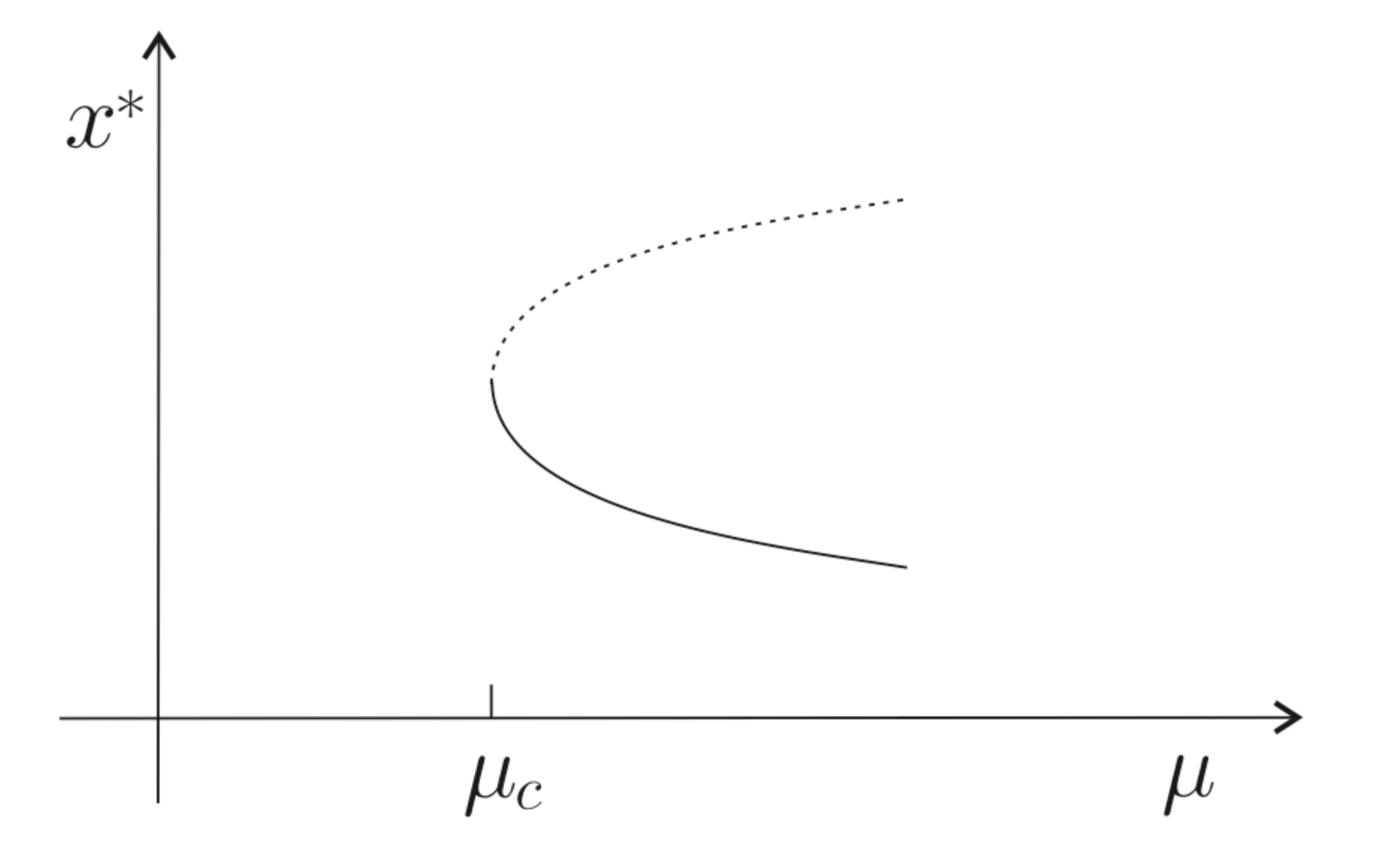}}
    \subfigure[Pitchfork bifurcation.]{\label{fig:pitchforkbif}\includegraphics[width=0.48\textwidth]{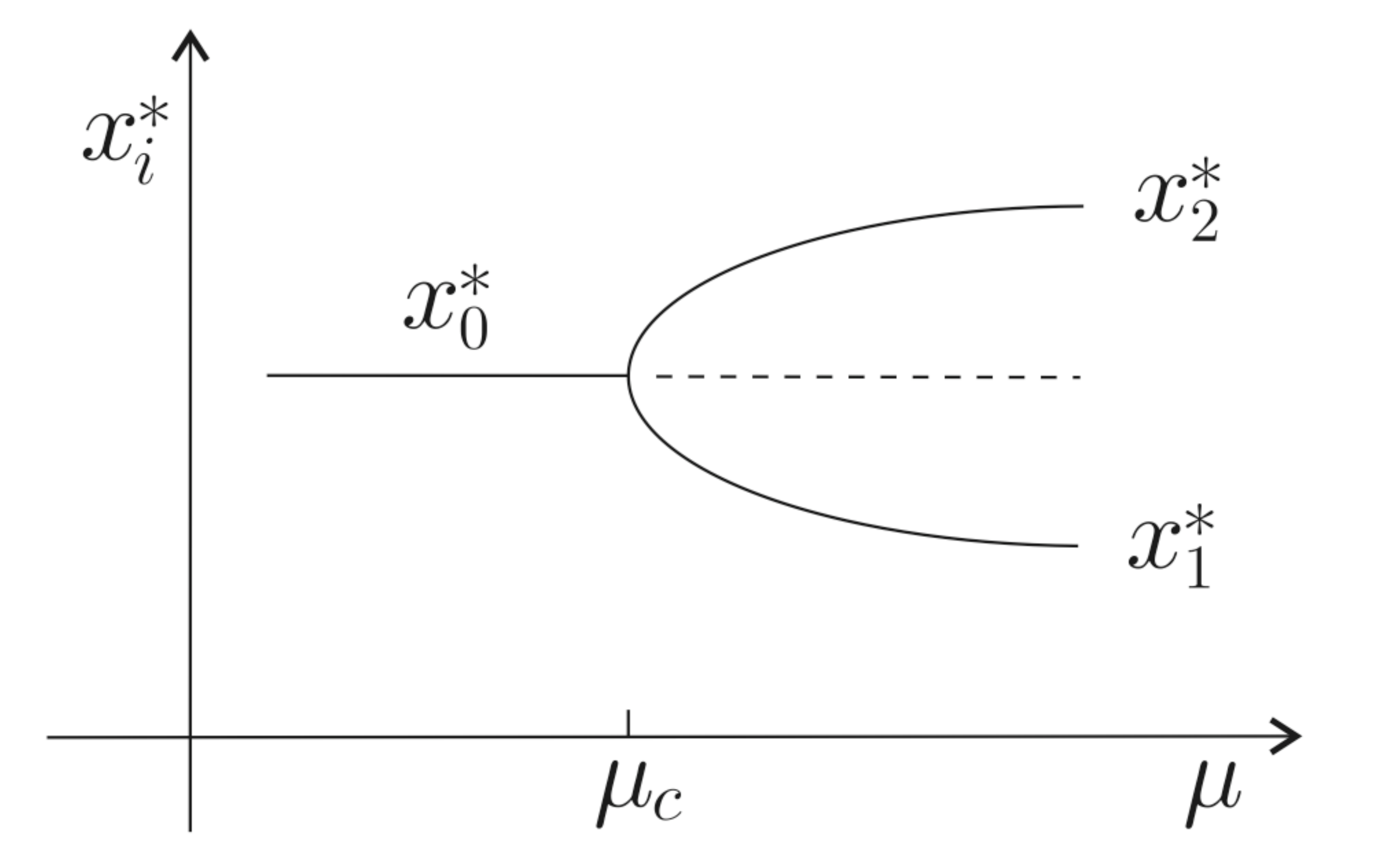}}
  \end{center}
  \caption{Bifurcations diagrams showing the fixed points as functions of $\mu$.}
  \label{fig:bifurcs}
\end{figure}

Probably the simplest example is the linear equation
\be
\dot{x}=\mu\,x\,,\quad\quad x\in\M=\R\,,\quad \mu\in\R\,,
\ee
and it is clear that the asymptotic properties of the system depend on the sign of $\mu$.

A similar, but somewhat less trivial example is e.g.
\be
\dot{x}=(x-\mu_1)^2-\mu_2\,,\quad\quad x\in\M=\R\,,\quad \mu_i\in\R\,.
\ee
Here the parameter $\mu_1$ is largely inessential, as it can be reabsorbed by a translation. However, again the sign of $\mu_2$ is important: when it is negative there are no critical points, whereas when it turns positive a couple of critical points (one stable, one unstable) is created. This is called the \emph{tangent bifurcation}. It is useful to plot the \emph{bifurcation diagram} (Figure \ref{fig:tangentbif}), where one draws the position of the critical points as a function of the relevant parameter (in this case, $\mu_2$). This gives an overview of the asymptotic properties of the system as the function of the external parameters. 

Another example, on which we will return later, is the so-called \emph{pitchfork bifurcation}, given for instance by
\be
\dot{x}=-x^3+\mu\,x\,,\quad\quad x\in\M=\R\,,\quad \mu\in\R\,.
\ee
Here we have a stable critical point for $\mu<0$ which splits up into two stable and one unstable critical points when $\mu$ becomes positive. The bifurcation diagram of Figure \ref{fig:pitchforkbif} suggests the name.

\begin{exercise}[A bifurcation diagram]
Draw the bifurcation diagram for 
\be
\dot{x}=\mu+x-\frac{1}{3}x^3,\quad\quad x\in\M=\R\,,\quad \mu\in\R\,.
\ee
Argue that, by changing $\mu$, this system can be used as a model of hysteresis in a magnet. 
\end{exercise}

\begin{exercise}[Redundant parameters]
\label{ex:logistic}
Consider the \emph{logistic equation}\footnote{This describes the growth of a population with limited amount of resources, as opposed to Malthusian, or exponential, growth (see next section).}
\be
\dot{x}=\mu_1\, x-\mu_2\,x^2,\quad\quad x\in\M=\R\,,\quad \mu_i>0\,.
\ee
Argue that by appropriate rescaling, both parameters can be eliminated to yield
\be
\dot{x}= x-x^2,\quad\quad x\in\M=\R\,.
\ee
\end{exercise}

\section{Dynamical systems with discrete time}
Some phenomena where observables can be measured only at some moment in time are more naturally described by using a discrete time variable. Examples are the abundance of a certain species after each reproductive cycle, the amount of crops collected every year, etc.

To discuss dynamical systems with discrete time we just have to rephrase what we said in the previous section.

\subsection{Generalities, fixed points, Lyapunov exponents}
Rather than being defined by an ODE such as (\ref{eq:ODE}), the typical definition of a discrete-time dynamical system is a recursion law
\be
x_{n+1}=f(x_n)\,,\quad\quad f:\ \M\to\M\,.
\ee
We immediately obtain that the (discrete-time) flow $\Phi$ satisfies, for $n\in\N$,
\be
\label{eq:discreteflow}
\Phi^0={\rm Id}\,,\quad\quad \Phi^1(x)=f(x)\,,\quad\quad \Phi^n=\left(\Phi^1\right)^n=f\circ\dots\circ f\,,
\ee
where the last equation indicates the $n$-fold composition of functions. Depending on whether $f(x)$ is invertible, one can add the additional property
\be
 \Phi^{-1}=\left(\Phi^1\right)^{-1}=f^{-1}\,,
\ee
and extend (\ref{eq:discreteflow}) to $n\in\Z$, in which case the flow will be a group, rather than just a semigroup.

As seen in the previous section, it is interesting to look at \emph{fixed points} $x^*\in\M$ that satisfy
\be
\Phi^n(x^*)=x^*\quad\quad\Longleftrightarrow\quad\quad x^*=f(x^*)\,.
\ee
Again, it will be important to understand whether $x^*$ attracts or repels the nearby points. Let us consider the case where $\M=I\subset\R$ is a (possibly unbounded) interval on the real line, and study the flow of $x_0=x^*+\eps$. Then
\be
\label{eq:slope}
x_1=f(x_0)\approx x^*+\eps\,f'(x^*)\,,\quad\quad x_n=f^n(x_0)\approx x^*+\eps\,\left(f'(x^*)\right)^n\,.
\ee
Clearly, the asymptotic behavior around $x^*$ depends on whether the modulus of the slope $|f'(x^*)|$ is larger than one (repulsive fixed point) or smaller than one (attractive fixed point). Similar notions can be formulated also when $\M$ is more general, i.e. in higher dimensions, but we will not need them in what follows; in fact, from now on, we will restrict to the simple case where $\M$ is an interval.

Let us introduce a notion that will be useful later on, the one of \emph{Lyapunov exponent}. This is a useful tool to understand the behavior of two neighboring ($\eps$-close) generic points in $\M$: will they remain close together, and eventually get squeezed to the same attractive fixed point (as most of the cases that we have encountered studied so far), or will they become more and more separated\footnote{ The latter is a typical feature of \emph{chaotic systems}, which we will not define in detail. The idea is that in these systems  the unpredictable behavior seen in Exercise \ref{ex:pendulumtip} for a particular initial condition (a pencil balanced on its tip) will be true \emph{for any initial condition}.}? The natural quantity to consider is the the limit
\be
\d(x,n)=\lim_{\eps\to0}\frac{\left|\Phi^n(x+\eps)-\Phi^n(x)\right|}{|\eps|}=\left|f'\left(\Phi^{n-1}(x)\right)\right|\,.
\ee
To remove the dependence on $n$ one can take the average of $\d(x,n)$ along the orbit. Finally, (\ref{eq:slope}) suggests that the separation grows geometrically, so that we write
\be
\label{eq:lyapunovexp}
\g(x)=\lim_{N\to\infty} \frac{1}{N}\sum_{n=0}^{N-1}\log\left|f'\left(\Phi^{n}(x)\right)\right|\,.
\ee
We will say (if the above limit exists) that $\g(x)$ is \emph{the Lyapunov exponent of $x$}. A theorem by Oseledec \cite{oseledec} guarantees that indeed the limit exists for almost every $x\in\M$, and it is immediate to see that $\g(x)$ will be the same for any $x\in\M$ with the same asymptotic behavior. In the cases of our interest, in fact, there will be only one Lyapunov exponent, so that we will from now on drop the dependence on~$x$.

We conclude this section with two simple examples of one-dimensional discrete time-systems, that are also called \emph{iterated maps of the interval}. As we already seen it is interesting to allow for dependence on one or more parameters $\mu_i$.  The simplest example is the \emph{Malthusian growth}, a simple model for population expansion with unlimited resources. The law is linear
\be
x_{n+1}=f_\mu(x_n)\,,\quad\quad f_\mu(x)=\mu\,x\,,\quad\quad x\in[0,+\infty)\,,\quad\mu>0\,,
\ee
and the recursion can be solved immediately to give
\be
\Phi^n(x)=\mu^n\,x\,,\quad\quad n\in\Z\,.
\ee
As indicated, the flow can be extended to negative $n$, and dictates a simple geometric behavior, similar to the one of the continuous-time system $\dot{x}=\log\mu\ x$.

One could have the impression that discrete time systems are just a trivial modification of ODEs, but this is not the case. In fact, discrete-time systems given by a simple law can yield an extremely complex behavior. The model immediately more complicated than Malthusian growth is the \emph{logistic growth}, which is a modification of the former to include finite resources. It can be written as
\be
\label{eq:logisitc}
x_{n+1}=f_\mu(x_n)\,,\quad\quad f_\mu(x)=\mu\,x\,(1-x)\,,\quad\quad x\in[0,1]\,,\quad0<\mu\leq4\,,
\ee
and a partial study of it, in its simplest regime, will occupy the rest of this chapter.

This surprising complexity comes from the fact that discrete-time systems can be seen as arising from continuos-time ones in higher dimensions, by defining $\Phi^n(x)$ as the intersection of $\Phi^t(x)$ with some submanifold embedded in $\M$, a procedure called Poincar\'e section. It is then clear that discrete-time systems do not suffer the same topological limitation of continuous time ones, and therefore can exhibit a richer behavior also in low dimension\footnote{To better understand what ``topological limitations'' we are referring to, the interested reader is invited to do Exercise \ref{ex:fixedpoints}, or read the proof of Poincar\'e-Bendixon theorem in e.g. \cite{strogatz}.}.

\begin{exercise}[Lyapunov exponent]
Consider the iterated map on the interval $\M=[0,1]$ defined by
\be
x_{n+1}=f(x_n)\,,\quad\quad f(x)=6 (x - x^2)\, (1 - 2x +2 x^2)\,.
\ee
What is the Lyapunov exponent? Check it numerically too.\\
{\rm (Answer: $\log (2^{4/3}-2)$.)}
\end{exercise}

\begin{exercise}[Redundant parameters for the logistic map]
The most general form of the discrete-time logistic map is 
\be
f_{\mu_1,\mu_2}(x)=\mu_1\,x-\mu_2\,x^2\,,\quad\quad x\in\left[0,{\mu_1^2}\big/{4\mu_2}\right]\,,\quad\mu_i>0\,.
\ee
Show that this can be reduced, by a rescaling, to (\ref{eq:logisitc}), but the remaining parameter $\mu$ cannot be eliminated. Explain what is the difference between this case and the one of Exercise \ref{ex:logistic}.
\end{exercise}

\begin{exercise}[Limit sets in $D=1$]
\label{ex:fixedpoints}
The set $L_\omega\subset\M$ is called the limit set (more properly the \emph{$\omega$-limit set}) for $\Phi$ if 
\be
L_\omega=\left\{y:\,y=\lim_{n\to\infty} \Phi^{t_n}(x)\right\},
\ee
for some sequences $\{t_n\}_{n\in\N}$ and some points $x\in\M$. Prove that is $\Phi$ is a \emph{continuos-time, one-dimensional} dynamical system, then if $L_\omega\neq\emptyset$ it always contains a fixed point. Show with an example that this  is not the case for discrete-time systems.
\end{exercise}

\subsection{The logistic map}
Let us start the study of the logistic map. As we said this is defined by the one-parameter family of functions on $\M=[0,1]$
\be
f_\mu\,:\ [0,1] \to [0,1]\,,\quad\quad f_\mu\,:\ x\mapsto \mu\,x\,(1-x)\,.
\ee 
In order to have that $f_\mu([0,1])\subset[0,1]$, it must be $0\leq\mu\leq4$. Let us start our analysis from small $\mu$.

\begin{figure}[!t]
  \begin{center}
    \subfigure[$\mu=0.9$, equilibrium at $x^*=0$.]{\label{fig:09}\includegraphics[width=0.4\textwidth]{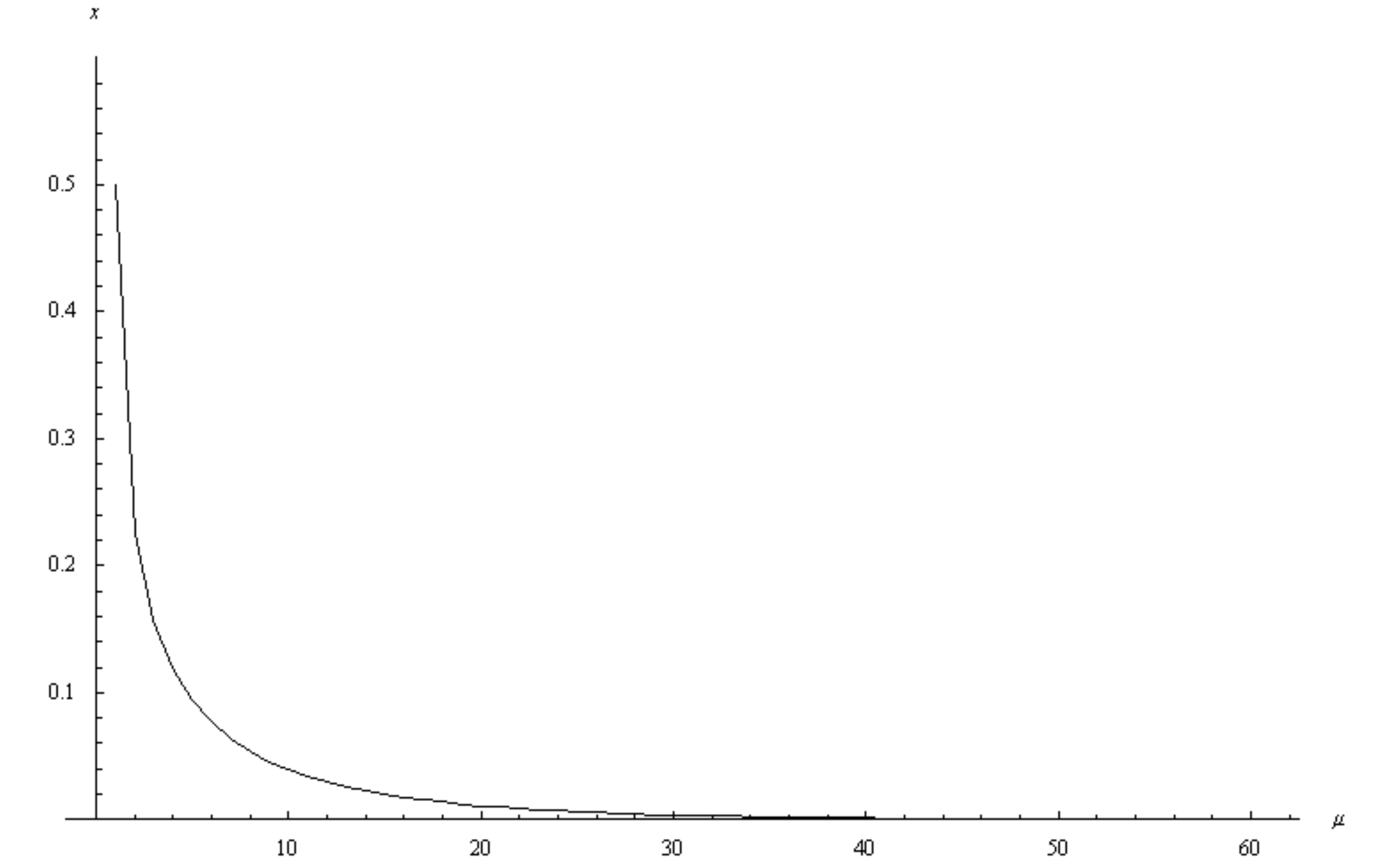}}
    \subfigure[$\mu=2.85$, equilibrium at $x^*\neq0$.]{\label{fig:285}\includegraphics[width=0.4\textwidth]{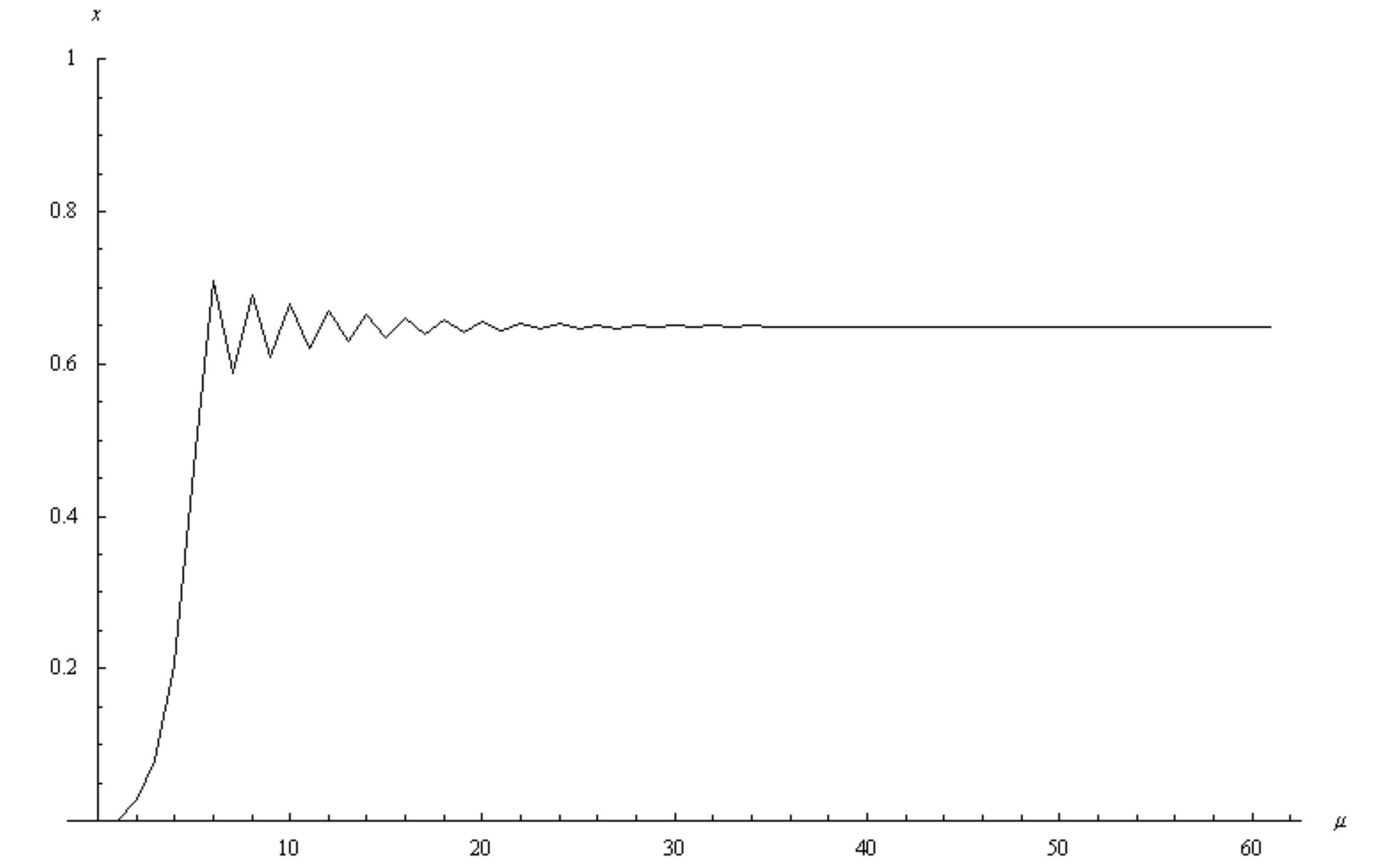}}\\
    \subfigure[$\mu=3.2$, two-cycle.]{\label{fig:32}\includegraphics[width=0.4\textwidth]{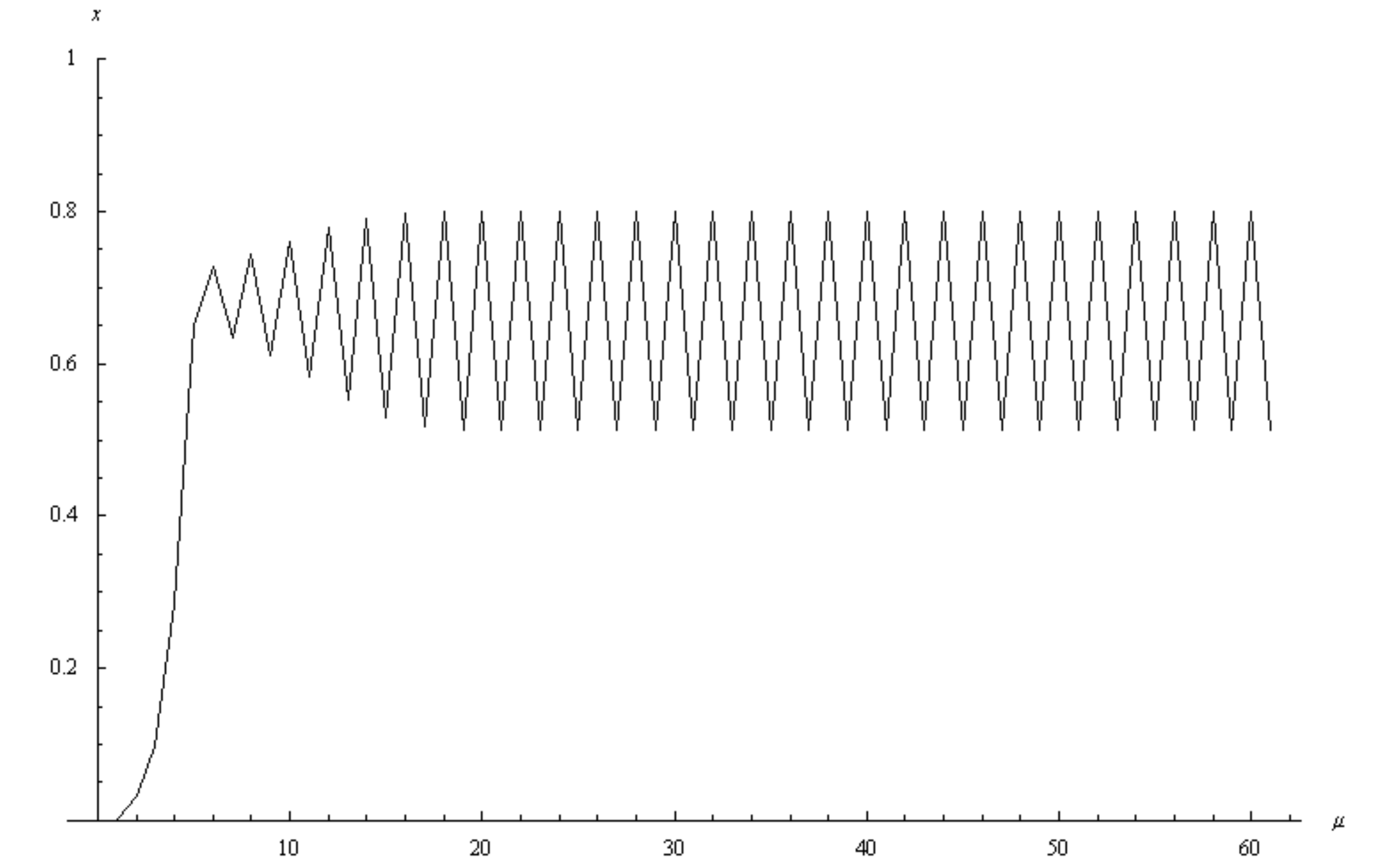}}
    \subfigure[$\mu=3.5$, four-cycle.]{\label{fig:35}\includegraphics[width=0.4\textwidth]{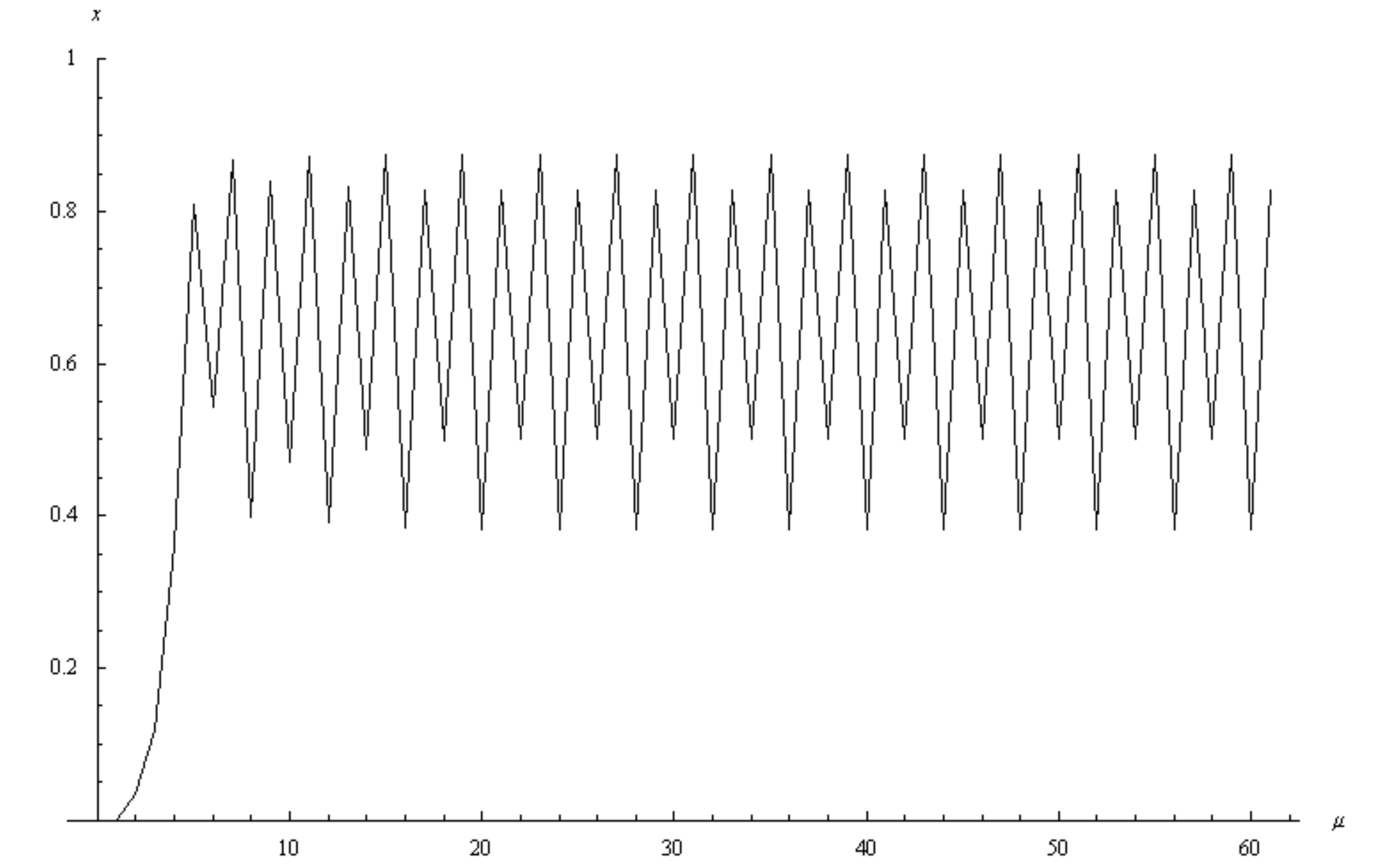}}\\
    \subfigure[$\mu=3.9$, chaotic behavior.]{\label{fig:39}\includegraphics[width=0.7\textwidth]{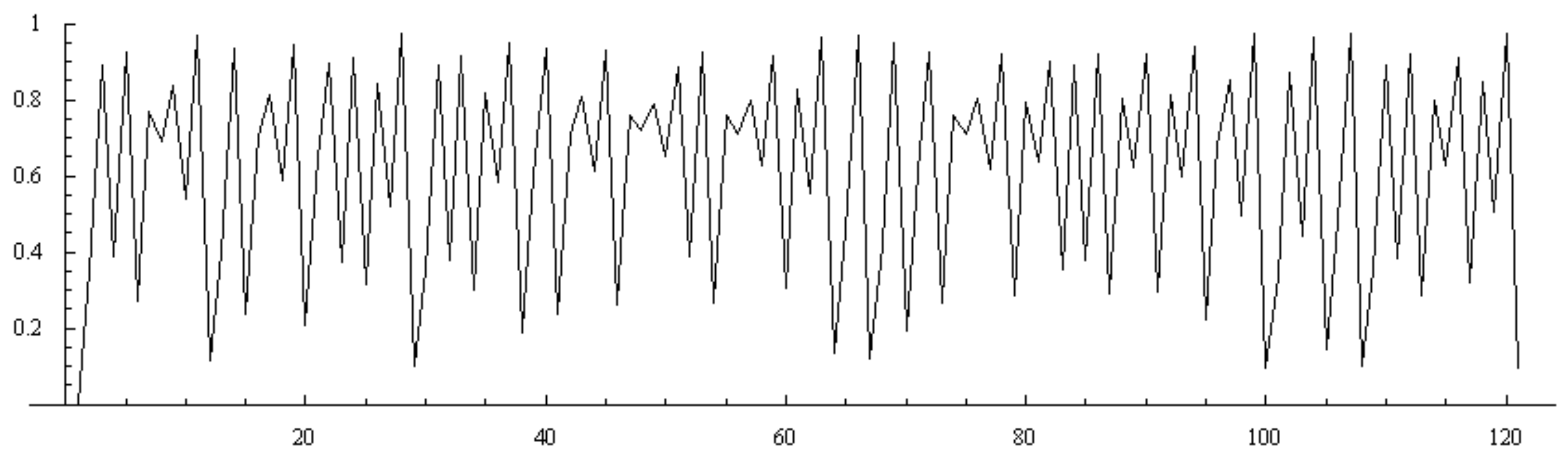}}
  \end{center}
  \caption{Asymptotic behavior of the logistic map for some $\mu$'s.}
  \label{fig:bifurcations}
\end{figure}

When $\mu=0$, clearly nothing happens. When $0<\mu<1$ the fixed point equation $f_\mu(x)=x$ has only one solution in $x^*=0$, which is a stable fixed point. For larger values of $\mu$, $x^*=0$ becomes unstable and a new stable fixed point is generated at $x^*=\frac{\mu-1}{\mu}$. Its stability can be checked by looking at
\be
|f'(x^*)|=|2-\mu|\,,
\ee
so that this fixed point is stable for $1< \mu<3$. It is not easy to intuitively understand what happens when $\mu$ gets larger than $3$; a way to proceed is to simulate the behavior of this system with a computer and plot the resulting orbits.

In Figure \ref{fig:bifurcations} some orbits are plotted. As expected, for $\mu=0.9$ one has that $x_n\to x^*=0$, whereas for $\mu=2.85$ the attractive fixed point is at $x^*\approx0.65$. It is interesting to notice that at $\mu=3.2$ there are no fixed points, but $x_n$ oscillates between two points. We will say in this case that there is an \emph{attractive 2-cycle}. When $\mu$ is further increased to $\mu=3.5$, the motion oscillates between four points--an attractive 4-cycle. Finally, for very large values of $\mu$ such as $\mu=3.9$, there is no apparent pattern for $x_n$; indeed it will turn out that there the motion is \emph{chaotic}.

\begin{figure}[t!]
  \begin{center}
  \includegraphics[width=0.85\textwidth]{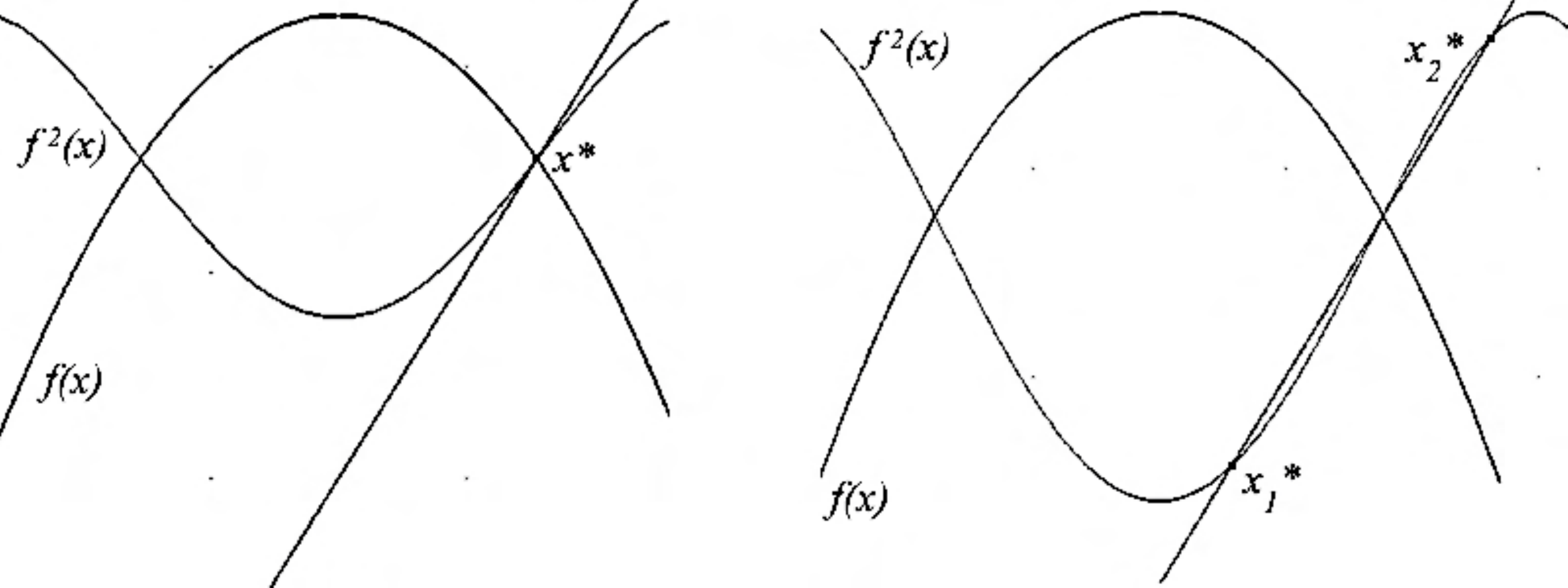}
  \end{center}
  \caption{A period-doubling bifurcation.}
  \label{fig:perioddoubling}
\end{figure}

It is worth investigating in more detail what happens when passing $\mu=3$. First, let us remark that when $f_\mu$ has a two-cycle, $(f_\mu)^2=f_\mu\circ f_\mu$ must have two fixed points:
\be
f_\mu(x_1)=x_2\,,\quad f_\mu(x_2)=x_1 \quad\quad
\Longleftrightarrow\quad\quad (f_\mu)^2(x_1)=x_1\,,\quad(f_\mu)^2(x_2)=x_2\,.
\ee
Let us now set $\mu=3-\eps$. Then $x^*=\frac{\mu-1}{\mu}$ is a stable fixed point with slope $f_\mu'(x^*)=\eps-1$. Correspondingly, the composition $f_\mu\circ f_\mu$ has also a fixed point there, with slope $(f_\mu\circ f_\mu)'(x^*)=(\eps-1)^2<1$, again stable.\\
Let now $\mu=3+\eps$. Then $x^*=\frac{\mu-1}{\mu}$ is unstable, with slope $f_\mu'(x^*)=-\eps-1<-1$. The same fixed point for the composition has then slope $(f_\mu\circ f_\mu)'(x^*)=(\eps+1)^2>1$, and it is also unstable. Furthermore, as depicted in Figure \ref{fig:perioddoubling}, by continuity a couple of fixed points are created to the left and to the right of $x^*$, and it is easy to see that they are stable.

\begin{figure}[t!]
  \begin{center}
  \includegraphics[width=0.9\textwidth]{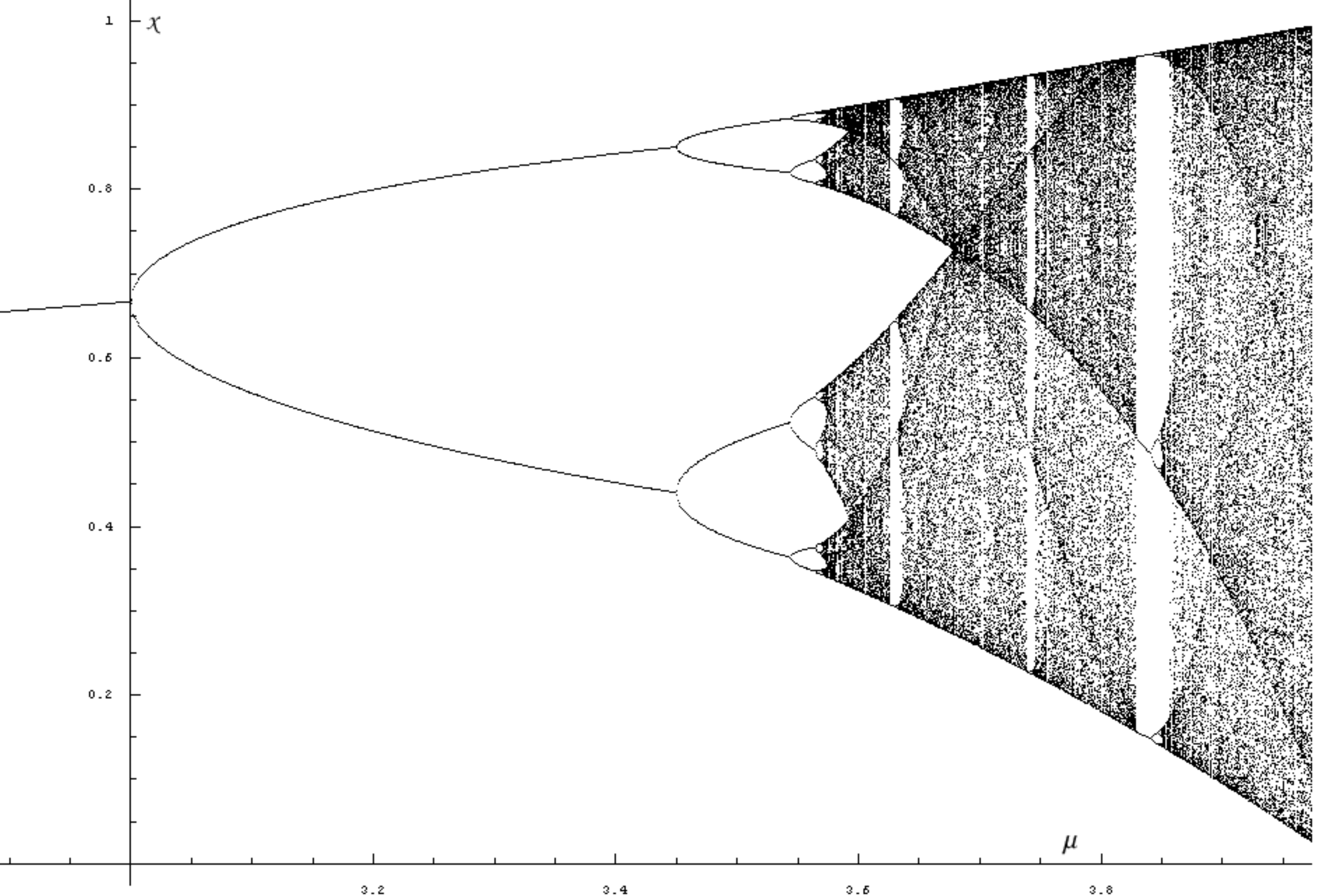}
  \end{center}
  \caption{Bifurcation diagram for the logistic map for $\mu\geq3$, where the first bifurcation occurs.}
  \label{fig:bifurcationdiag}
\end{figure}

The above reasoning seems to be applicable not only when going from period-one to period-two, but every time we double the period of the attractive cycle. We have already seen that for larger $\mu$ there exists an attractive four-cycle. It is worth plotting the \emph{bifurcation diagram} for the logistic map, that indicates for any $\mu$ the set to which the motion is attracted. Looking at Figure \ref{fig:bifurcationdiag} we se that at several points $\mu_0,\mu_1,\dots,\mu_n,\dots$ a bifurcation occurs, where the numebr of attractive points doubles (i.e. one goes from a $2^n$-cycle to a $2^{n+1}$ one). As we discussed, $\mu_0=3$, and one can see from the plot that $\mu_1\approx3.45$, $\mu_2\approx3.55$, etc.

\begin{figure}[ht!]
  \begin{center}
  \includegraphics[width=0.75\textwidth]{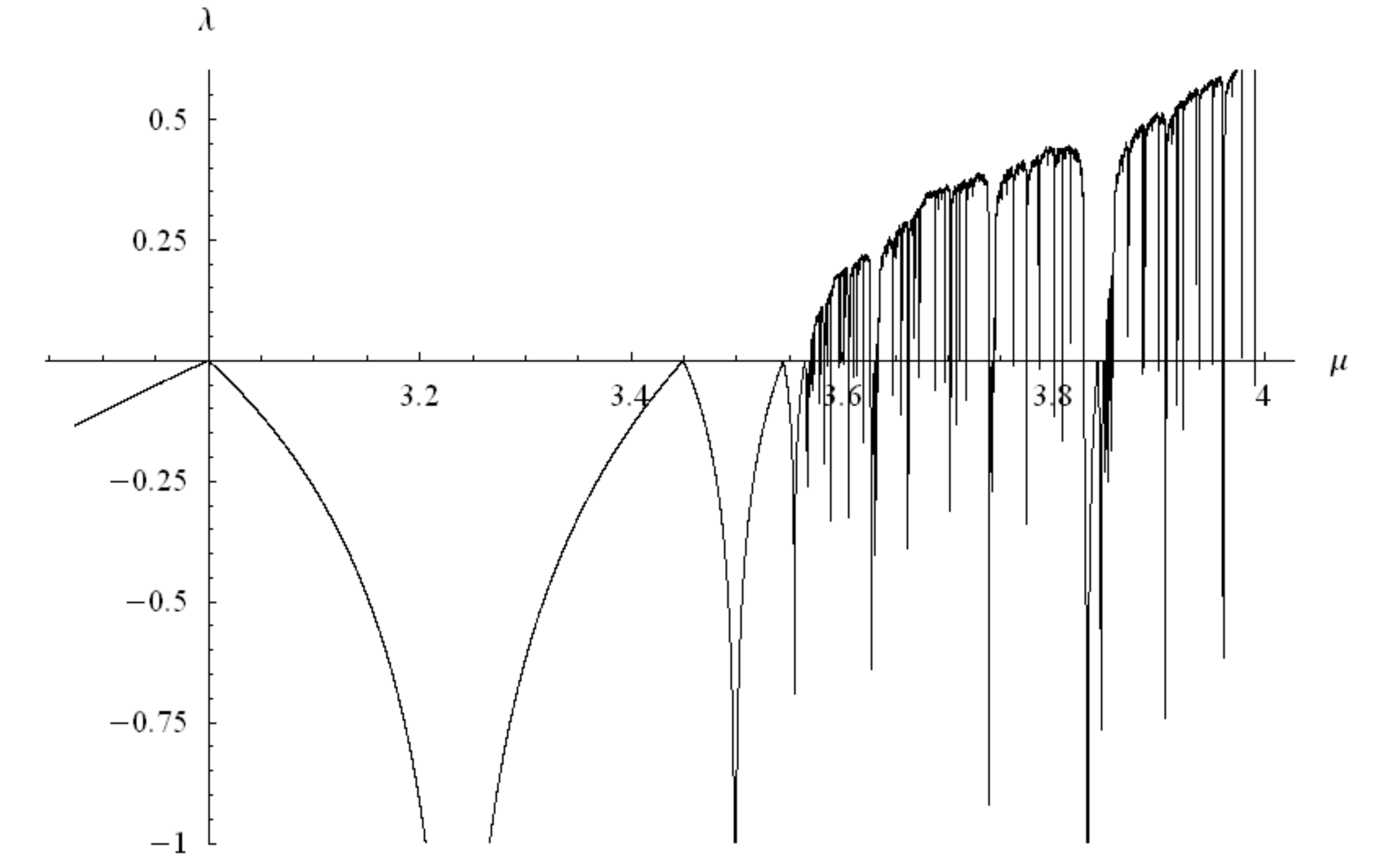}
  \end{center}
  \caption{Lyapunov exponent for the logistic map for $\mu\geq3$.}
  \label{fig:lyapunov}
\end{figure}

Furthermore, these points seem to accumulate to some $\mu_\infty\approx3.6$ after which the trajectory jumps wildly between many points. To better understand this, let us look at Figure \ref{fig:lyapunov}. One sees that the Lyapunov exponent $\g(\mu)$ is smaller than zero in presence of a $2^n$-cycle, and vanishing at the bifurcation point as it should. However, for $\mu>\mu_\infty$ one sees that $\g(\mu)>0$ which signals the beginning of chaotic behavior\footnote{For $\mu$ even larger there exists some interval where the behavior is back to periodic, with $\g(\mu)<0$. These are the so-called periodic-windows, which we will not discuss here.}. It is also interesting to notice that between any couple of bifurcation points there is a point where $\g(\tilde{\mu}_n)=-\infty$, with $\mu_{n-1}<\tilde{\mu}_n<\mu_{n}$. We will say that there the system has $2^n$-\emph{superstable cycle}.

We are not interested in what happens when $\mu>\mu_\infty$. What is important to us, and we have established numerically, is that there is a sequence of period-doubling bifurcations at $\mu_n$ and, correspondingly, a sequence of superstable $2^n$-cycles at $\tilde{\mu}_n$, that converge to some $\mu_\infty$.

\begin{table}
\begin{minipage}[b]{0.4\linewidth}
\centering
\begin{tabular}{ l | l | l }
$j$&$\tilde{\mu}_j$&$\delta_j$\\ \hline
0 & 3.00000000$\dots$   &                 \\
1 & 3.44948974$\dots$   &  4.751        \\
2 & 3.54409035$\dots$   &  4.656        \\
3 & 3.56440726$\dots$   &  4.668        \\
4 & 3.56875941$\dots$   &  4.668        \\
5 & 3.56969160$\dots$   &  4.669        \\
6 & 3.56989125$\dots$   & 4.669 \\
7 & 3.56993401$\dots$   & 4.668  \\
8 & 3.56994317$\dots$   & 4.667  \\
9 & 3.56994514$\dots$   & 4.671  \\
10& 3.5699455573883578  & 4.673  \\
11& 3.5699456473525193  &  4.66955 \\
12& 3.5699456666186404  &  4.66966\\
13& 3.5699456707444445  &  4.66935\\
14& 3.5699456716280371  &  4.66917\\
15& 3.5699456718175778  &                 \\
\end{tabular}
\caption{Values of ${\mu}_j$ and $\delta_j$.}
\label{tab:deltas}
\end{minipage}
\hspace{0.5cm}
\begin{minipage}[b]{0.5\linewidth}
\centering
\begin{tabular}{ l | r | r | r }
$n$&$\tilde{\mu}_n$&$d_n$&$\alpha_n$ \\ \hline
1 & 3.236067977 & -0.190983 & -2.68555\\
2 & 3.498561698 & 0.0711151 &  -2.52528\\
3 & 3.554640862 & -0.028161 & -2.50880\\
4 & 3.566667379 & 0.0112250 &  -2.50400\\
5 & 3.569243531 & -0.004482 & -2.50316\\
6 & 3.569795293 & 0.0017908 &  -2.50296\\
7 & 3.569913465 & -0.000715 & -2.50295\\
8 & 3.569938774 & 0.0017908 &  -2.50293\\
9 & 3.569944194 & -0.0007155 & -2.50293\\
\end{tabular}
\vspace{3.37cm}
\caption{Values of $\tilde{\mu}_n$, $d_n$ and $\alpha_n$.}
\label{tab:alphas}
\end{minipage}
\end{table}

It is not hard to compute numerically (e.g. by Newton's method) the values of the first few $\mu_n$'s to a good precision. These are written in Table \ref{tab:deltas}, and it is not hard to see that the sequence $\mu_n\to\mu_\infty\approx3.5699$ converges, at least approximately, geometrically. We will call the number
\be
\delta=\lim_{n\to\infty}\frac{\mu_{n}-\mu_{n-1}}{\mu_{n+1}-\mu_{n}}\approx4.669201609\,,
\ee
\emph{Feigenbaum's $\delta$}.\footnote{It is interesting how Mitchell Feigenbaum found this rate of convergence; he was studying the sequence $\mu_n$ on a pocket calculator, and needed to guess the next bifurcation point as well as he could in order not to waste computer time. In doing so, he realized that the convergence was geometric.} Notice that $\frac{\mu_{n}-\mu_{n-1}}{\mu_{n+1}-\mu_{n}}$ is not exactly equal to $\delta$ when $n$ is finite. Clearly the same rate dictates the convergence of $\tilde{\mu}_n$ as well.

\begin{figure}[t]
\centering
\includegraphics[width=9cm]{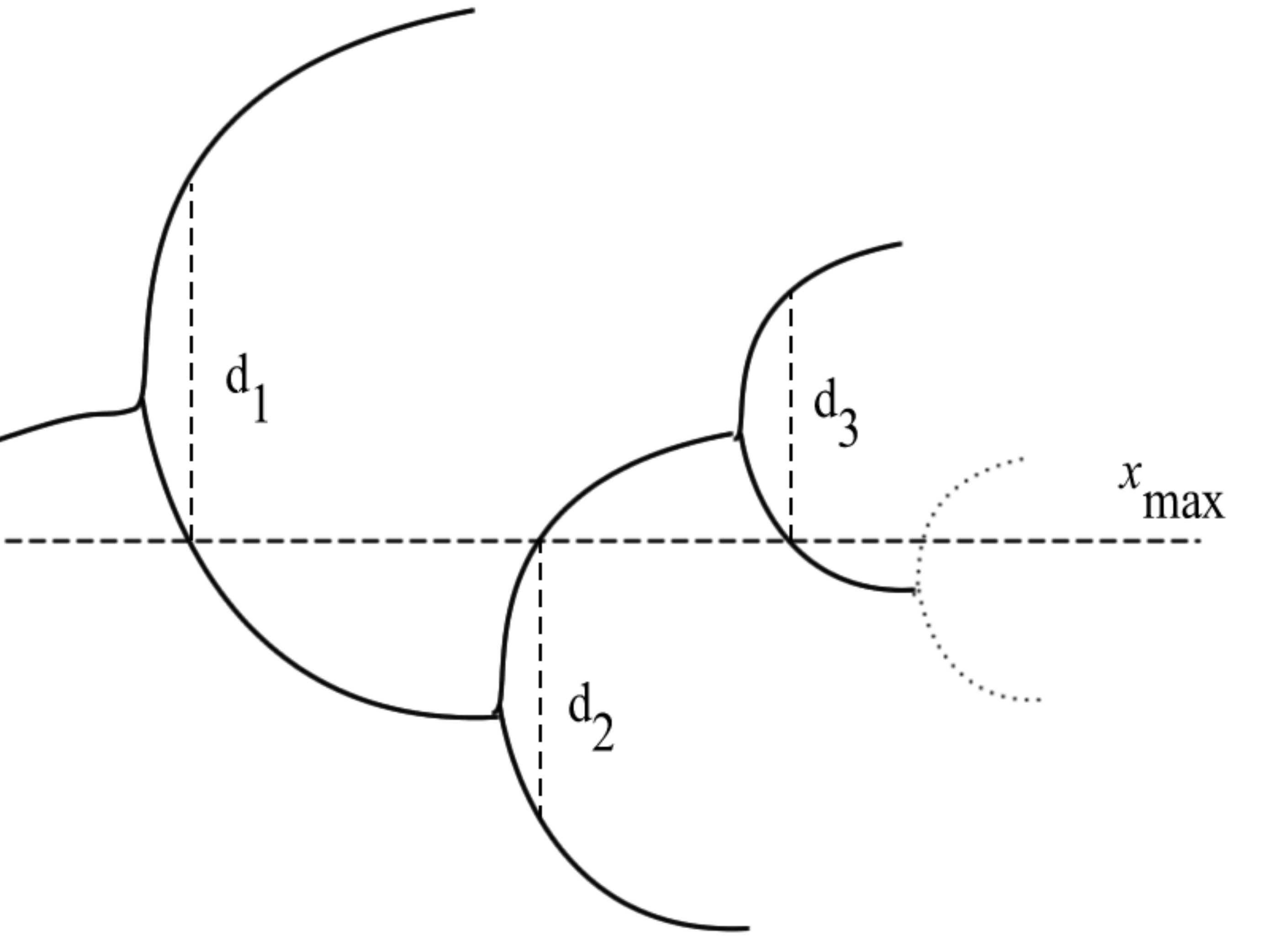}
\caption{The $d_n$ sequence used to define $\alpha$.}
\label{fig:Usequence}
\end{figure}

Looking back at Figure \ref{fig:bifurcationdiag} we see that the constant  $\delta$ dictates the horizontal scale in the sequence of bifurcations. Clearly enough, there is also a vertical scale: in fact, after each bifurcation, the couple of new attractive points generated that are generated spread out in a $\sf C$-shaped figure\footnote{Or $\sf U$-shaped: in fact historically this goes under the name of $\sf U$-sequence.} as $\mu$ increases. The size of this $\sf C$ shrinks bifurcation after bifurcation, and it makes a lot of sense to suspect that it does so geometrically. The ``size'' $d_n$ can be defined as the distance between two neighboring points in a superstable cycle. This can be found by  looking at the intersection of the bifurcation sequence with the line $x=x_{max}=1/2$, as we will see in Exercise \ref{ex:superstable} and as depicted in Figure \ref{fig:Usequence}.

In Table \ref{tab:alphas} the first few values of $d_n$, as found numerically, are written. Indeed they converge geometrically, and one can define \emph{Feigenbaum's $\alpha$} as
\be
\alpha=\lim_{n\to\infty}\frac{d_{n}}{d_{n+1}}\approx-2.502907875\,,
\ee
which is negative because we keep into account the sign of $d_n$, see Figure  \ref{fig:Usequence}.

\begin{exercise}[The period-doubling mechanism]
Looking at Figure \ref{fig:perioddoubling} we argued that, for the logistic map, when an equilibrium for $f_\mu$ becomes unstable, then two new fixed points are generated for $f_\mu^2$. Argue that these are stable, and find the properties of $f_\mu$ that are important for this mechanism to happen. 
\end{exercise}

\begin{exercise}[Superstable cycles]
\label{ex:superstable}
Prove that a $2^n$-cycle for the logistic map
\be
f_{\tilde{\mu}_{n}}(x_1)=x_2\,,\quad\dots\quad f_{\tilde{\mu}_{n}}(x_j)=x_{j+1}\,,\quad\dots\quad f_{\tilde{\mu}_{n}}(x_{2^n})=x_1\,,
\ee
is superstable if and only if $x_j=1/2$ for some $j$. Argue that indeed this happens in between of any couple of bifurcations.
\end{exercise}

\subsection{Universality}
So far we have established that a certain family of maps of the interval into itself exhibits a sequence of period-doubling bifurcations, with geometric rate tending to $\delta\approx4.669$, leading to a chaotic behavior. What makes this story more interesting is that this behavior is \emph{common to many maps as well as real-world systems}, i.e. it is, at least to some extent, \emph{universal}.

In fact, one can check numerically that the same sequence of period-doubling bifurcations occurs for dynamical systems defined by the maps on $I=[0,1]$
\bea
\nonumber
f_\mu(x)&=&\mu\,(1-x^2)(2x-x^2)\,,\quad\quad 0\leq\mu\leq\frac{9}{16}\,,\\
f_\mu(x)&=&\mu\,\sin(\pi x)\,,\quad\quad\quad\quad\quad\quad\quad 0\leq\mu\leq1\,,
\label{eq:othermaps}
\eea
as well as many others. What is more remarkable, the bifurcations points $\mu_n$ converge geometrically to some $\mu_\infty$ with a rate that tends to $\delta\approx4.669$. This also happens also for dynamical systems of different kind, such as Mandelbrot's set, or R\"ossler's system of ODEs, see e.g. \cite{strogatz}.

Furthermore, and what is more important from a physicist's perspective, the period-doubling cascade towards chaos occurs also in real-world systems. For instance, let us consider a fluid-dynamics experiment of Rayleigh-B\'enard convection, following Libchaber and Maurer \cite{libchaber}.

Consider a box containing a fluid. The bottom of the box is kept at temperature $T$, whereas the top is kept at $T+\Delta T$. The temperature difference (or rather a related dimensionless quantity called Rayleigh's number) is the external parameter that the experimenter may vary. When the temperature difference is small, heat is conduced to the colder upper surface. However, increasing the gradient, the familiar convective motions are generated. These consists of several counter-rotating cylinders that drive steadily the hotter fluid upwards.\footnote{In practice, in order to obtain a stable enough convective motion, great care has to be taken in setting the experiment, such as picking appropriate shape and dimension of the box, and of course an appropriate fluid.} 
Further increasing $\Delta T$ leads to a more complicated dynamics of the fluid:  the heat flow is not steady anymore, but fluctuates, as it can be seen by measuring the time-evolution of the local temperature at a given point in the upper surface.

This is a discrete-time dynamical system\footnote{In fact, it makes sense to make measurements with time scales that are large with respect to the ones of the microscopic degrees of freedom.}: $\Delta T$ plays the role of $\mu$, and the oscillations of the local temperature the role of $\Phi^n(x)$. What was found then is that, as one increases $\Delta T\sim \mu$, one goes from the steady temperature (fixed point) to a two-cycle, then to a four-cycle, and so on. Even more strikingly, the period-doubling bifurcations occur geometrically with rate $4.4\pm0.1$ compatible with $\delta$. This has been shown to occur in a number of experiments in hydrodynamics \cite{libchaber,libchaber83,berge,giglio}, electronics \cite{linsay,testa,arecchi,yeh}, charged gases \cite{braun} and chemistry \cite{simoyi}.

 It is worth pointing out that obtaining these experimental results is quite hard. On top of difficulties such as suppressing the noise and avoid generating chaotic behavior due to other kind of turbulence, a key obstacle is that  (due to the geometric progression) it is in practice possible to measure only the first few bifurcations. On the other hand, $\frac{\mu_{n}-\mu_{n-1}}{\mu_{n+1}-\mu_{n}}$ will approach $\d$ only asymptotically. This makes these result even more remarkable.
 
 Of course, the word ``universal'' should be taken with a pinch of salt. The maps (\ref{eq:othermaps}) are not terribly general, and indeed they share some features with the logistic map:
 \begin{enumerate}
\item $f_\mu(x)$ is regular\footnote{All our examples are analytic functions, which is a very strong requirement. We will not go into the details of \emph{how regular} we need $f_\mu(x)$ to be.}.
\item $f_\mu(x)$ is unimodal, i.e. it has one maximum $x_{max}$ and satisfies
\be
f'_\mu(x)>0\,,\quad x<x_{max}\,\quad\quad\&\quad\quad 
f'_\mu(x)<0\,,\quad x>x_{max}\,.
\ee
\item $f_\mu(x)$ has a quadratic maximum
\be
f''_\mu(x_{max})<0\,.
\ee
\end{enumerate}

It turns out that these requisites are indeed necessary, as we will also see in the exercises by considering some examples.  More abstractly, it is reasonable to require some kind of regularity, since we used it in the previous section to explain period-doubling. As for unimodality, it is also quite clear that a very general $f_\mu(x)$ with many maxima and minima may have a richer dynamics than the simpler examples we considered. As for the third requirement, it is hard to justify it \textsl{a priori}, and we will take it as an ``experimental'' evidence. What turns out is that maps that satisfy all requisites but have a maximum of higher order show the same period-doubling cascade, but with \emph{different universal constants} in place of $\delta$ and $\alpha$.\footnote{Accurate values of such constants can be found e.g. in \cite{briggs}.}

In the next section we will see how all these features can be explained in a ``renormalization group'' framework, which also yields quantitative predictions for $\delta$ and $\alpha$.

\begin{figure}[t]
\begin{minipage}[t]{0.54\linewidth}
\centering
\includegraphics[width=\linewidth]{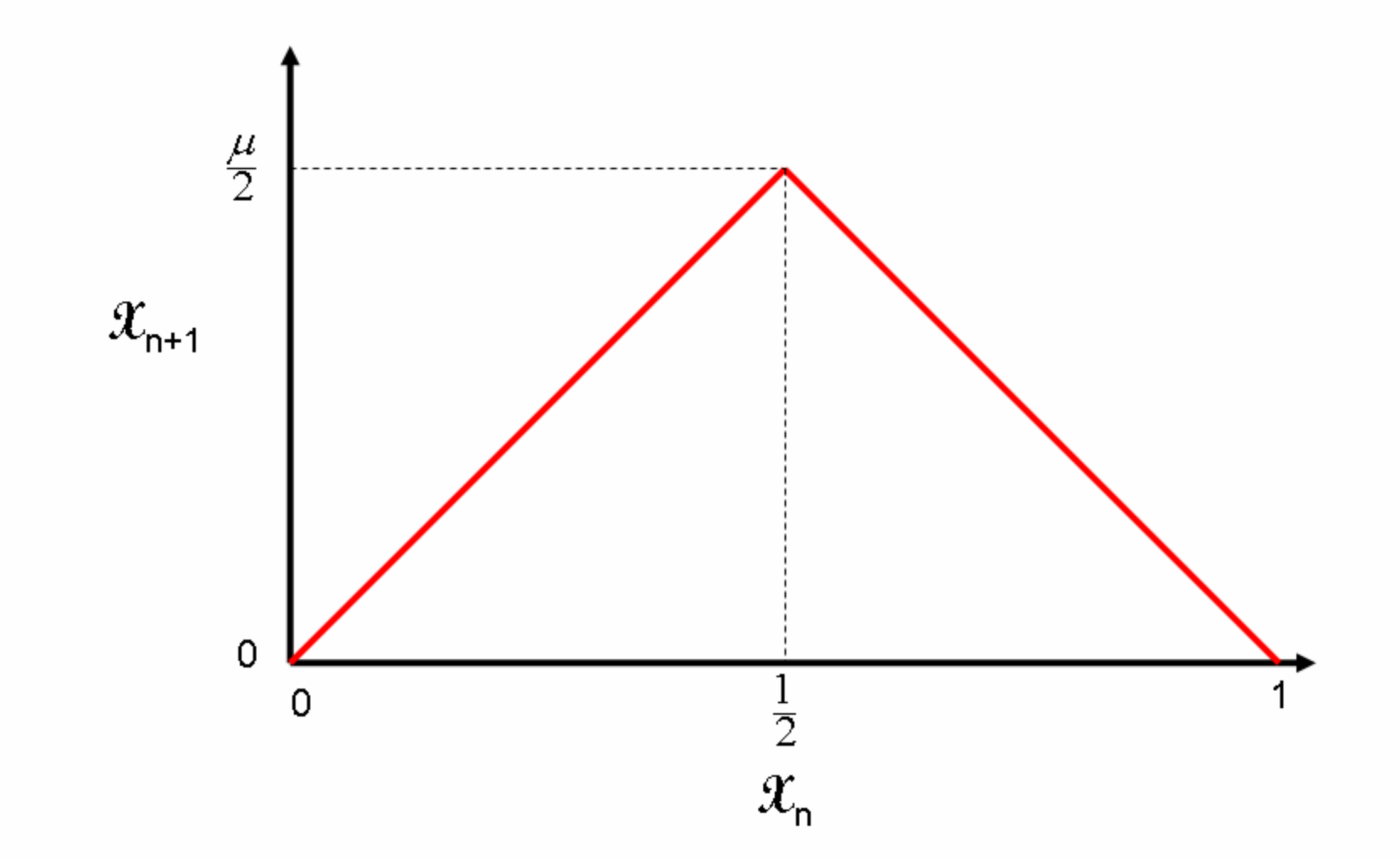}
\caption{The tent map.}
\label{fig:tent}
\end{minipage}
\begin{minipage}[t]{0.44\linewidth}
\centering
\includegraphics[width=\linewidth]{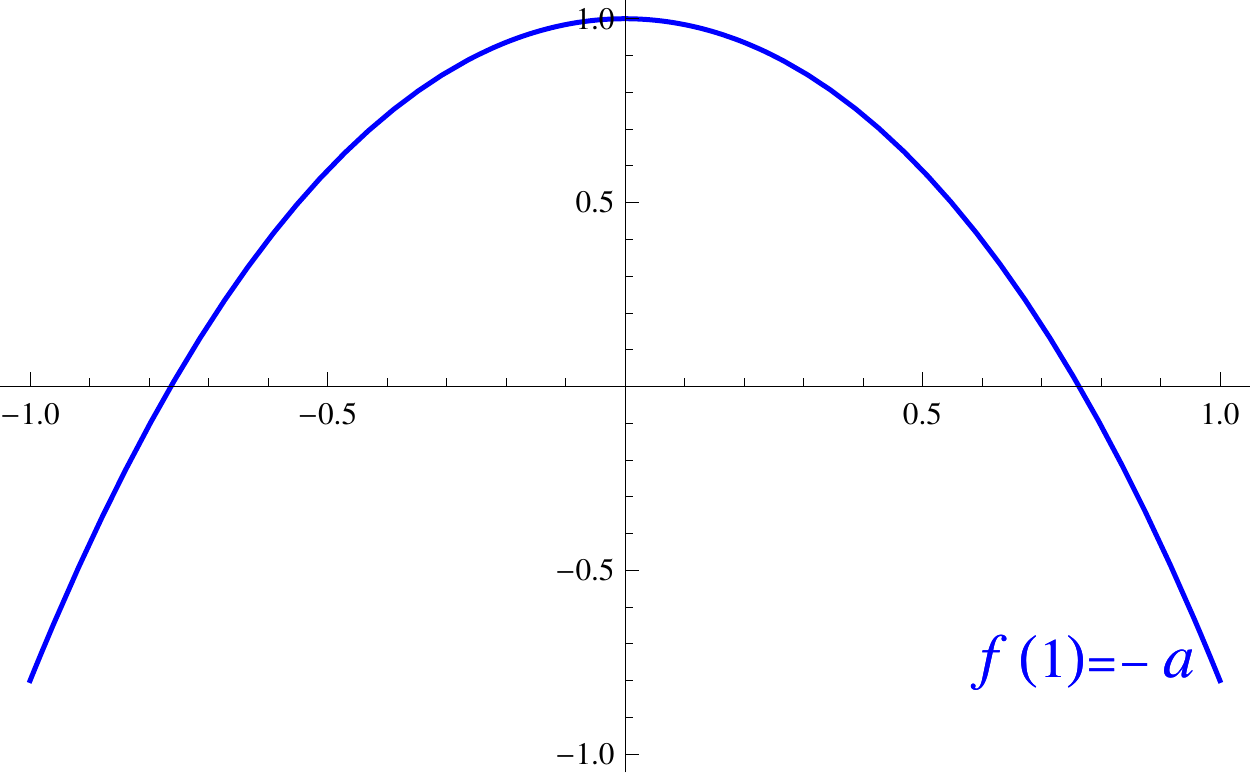}
\caption{Logistic map rescaled.}
\label{fig:rescaledmap}
\end{minipage}
\end{figure}

\begin{exercise}[The tent map]
Consider the ``tent map'' of Figure \ref{fig:tent}
\be
f_\mu(x)=\begin{cases} \mu\,x & x \leq 1/2 \\
\mu(1-x) & x>1/2 \end{cases}\,,\quad\quad 0\leq\mu\leq2\,.
\ee
Prove that no period-doubling bifurcation occurs, and that in fact there is a sharp transition to chaotic behavior.
\end{exercise}

\begin{exercise}[Universality for quartic tips]
Consider the map on the interval $I=[0,1]$
\be
 f_\mu(x)=\mu\, (x - x^2)\, (1 - 2x +2 x^2)\,,\quad\quad 0\leq\mu\leq8\,,
\ee
 which, as you can verify, has a quartic maximum. Show numerically that there is a period-doubling cascade, and get a (rough!) estimate for the analogs of $\d$ and $\a$.
Note that this exercise can take some time.\\
{\rm(Answer: $\delta\approx7.3$ and $\alpha\approx-1.7$)}
\end{exercise}

\section{Renormalization group approach}
\label{sec:feigenbaum}
In this section we will discuss how the universal behavior of period-doubling bifurcations can be explained in terms of renormalization group (RG) techniques. In this context, both words ``renormalization'' and ``group'' make little sense (to be fair, this name is a bit misleading in practically any context). They just indicate a rather general set of ideas of broad application, from Quantum Field Theory to the Physics of phase transitions.

The reader that has some familiarity with  RG in statistical Physics will find many similarities with what we are doing now. We will come back to this at the end of the section.

\subsection{Heuristic}
We are concerned with several one-parameter families of maps $f_\mu$ of the interval $I$ into itself that exhibit similar period-doubling cascade of bifurcations. It makes sense to consider the space $\U$ of all maps with ``good properties'',
\be
\U=\{f:\ I\to I\,,\quad f\ {\rm regular,\ unimodal\ and\ with\ quadratic\ tip}\}\,.
\ee
Observe that the families $\{f_\mu\}\subset\U$ are curves in $\U$.
Clearly $\U$ is a subset of some space of functions, and to proceed rigorously further formalization (e.g. on the metric of this space) would be needed; here we will be qualitative. 

Let us chose explicitly $I$. Earlier we picked $I=[0,1]$, but to better keep track of the maximum of $f_\mu(x)$ we will set $I=[-1,1]$ in such a way that the maximum is in $x=0$ and takes value $f_\mu(0)=1$; we further restrict to even maps to simplify figures and discussion. A typical map is shown in Figure \ref{fig:rescaledmap}.

To single out the fundamental characteristics of the universal behavior, it is easier to think in terms of superstable maps: we have that, for any $n=1,2,\dots$ at $\mu=\tilde{\mu}_n$ there exists a superstable map of period $2^n$, and of characteristic size $d_n$. The convergence of $\tilde{\mu}_n\to\mu_\infty$ has universal rate $\delta$, whereas the one of $d_n\to0$ has rate $\alpha$. 

Therefore, universality should emerge as a property of $\U$ under the action of some ``renormalization'' operator $\Ren:\U\to\U$. This operator
\begin{itemize}
\item Should relate maps with a $2^n$-cycle to maps with a $2^{n-1}$ cycle,
\item Should relate superstable maps to superstable maps, up to a rescaling of $\alpha$.
\end{itemize}
The first property suggests that it must be $\Ren(f)\sim f\circ f$, which however is not an operator on $\U$. In fact, it is easy to see that if $f(x)$ is unimodal, $f\circ f(x)$ is not. It is clear that some kind of rescaling is needed.

\begin{figure}[t!]
  \begin{center}
  \includegraphics[width=\textwidth]{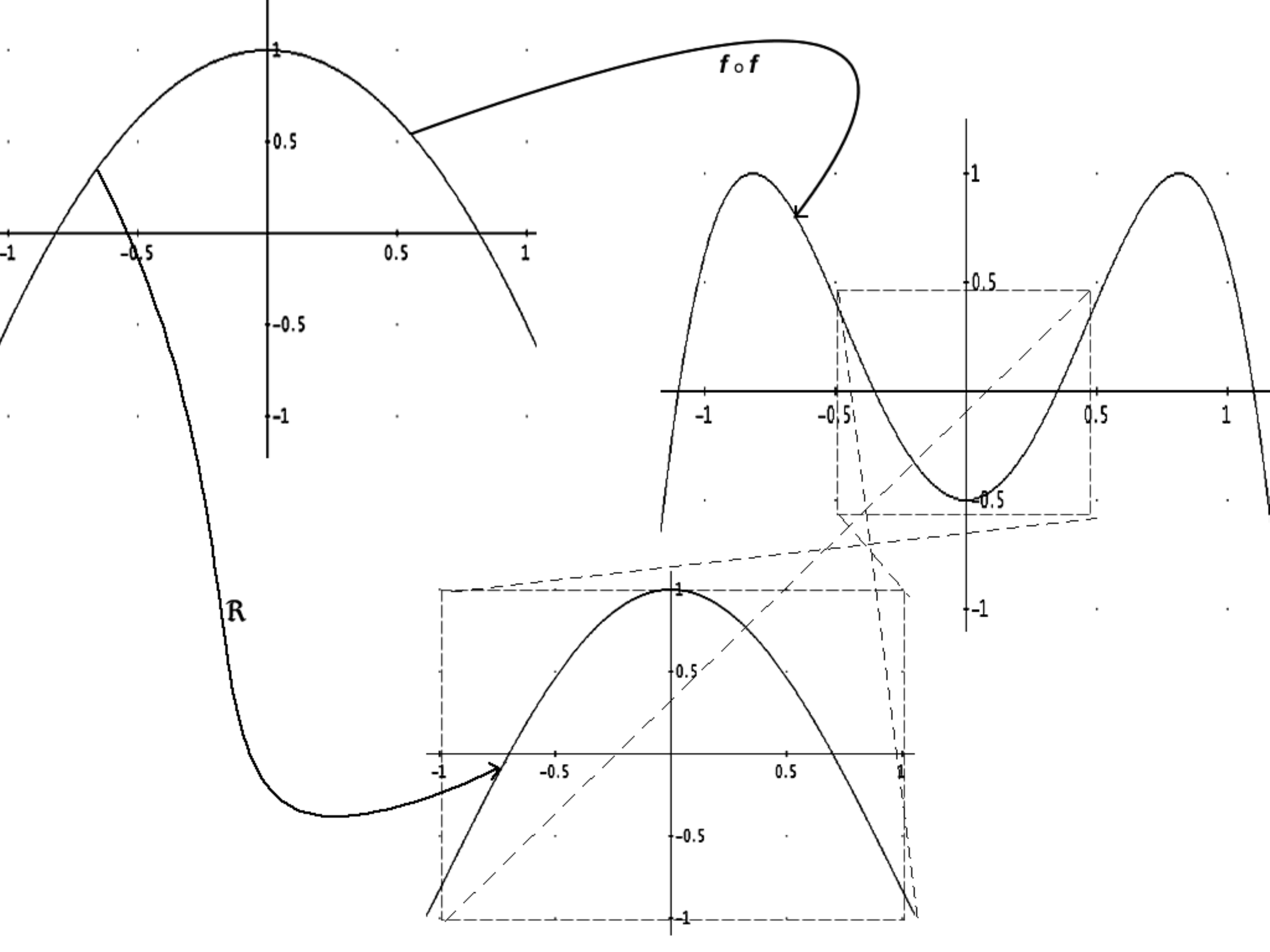}
  \end{center}
  \caption{The renormalization operator $\mathfrak{R}$.}
  \label{fig:R}
\end{figure}

A generic map $f\in\U$ has its minimum on $I$ exactly at the boundaries of the interval. Looking at Figure \ref{fig:rescaledmap} define 
\be
a=-f(1)\,,\quad\quad b=f(a)\,.
\ee
With a drawing, one can convince himself that the following inclusions hold
\be
f([-1,1])=[-a,1]\,,\quad f([-a,a])=[b,1]\,,\quad f([b,1])=[-a,f(b)]\subset[-a,a]\,,
\ee
 provided that it is
 \be
 \label{eq:renormalizability}
 0<a<b<1\,,\quad\quad f(b)<a\,.
 \ee
In this case therefore one has that
\be
f\circ f:\ [-a,a] \to [-a,a]\,.
\ee
Then, there is no problem to act with $\Ren$ on $f$, if we rescale all variables in an appropriate way (see Figure \ref{fig:R}):
\be
\label{eq:ren}
\Ren (f)\,(x)=-\frac{1}{a}f\circ f(-a\,x)\,,
\ee
provided that (\ref{eq:renormalizability}) holds. In other words,  (\ref{eq:renormalizability}) are conditions of $f\in\U$ to be in the domain of $\Ren$ or, as sometimes it is said, for $f$ to be renormalizable.

\vspace*{1cm}

Even if in general it is not immediate to see whether also $\Ren(f)$ is in the domain of $\Ren$, for superstable maps things are easier. In fact, superstable maps are renormalizable, and if $f$ is a period-$2^n$ superstable map, then $\Ren(f)$ is a period-$2^{n-1}$ superstable map, see Exercise \ref{ex:superstable}.
The values $\mu=\mu_\infty$, where each family of maps gets to the onset of chaos, identify maps that have $2^\infty$-period, and are therefore infinitely renormalizable.

It is now the time to change point of view, and start studying a new dynamical system, where $\M=\U$ and $\Phi=\Ren$; here we want to understand the orbits of the ``points'' $f\in\U$ (that are, in fact, functions in an infinite-dimensional space)  under action of $\Ren$. This analysis may be more mathematically complicated and subtle than the previous ones, but we will reason by analogy with what we have seen until now.

Unfortunately, it is very hard to establish the properties of this infinite dimensional dynamical system, but a lot of progress can be made if we accept some conjectures, originally put forward by Feigenbaum \cite{feigenbaum78,feigenbaum79a,feigenbaum79b}, see also \cite{feigenbaum83,cvtanovic}.
\begin{enumerate}
\item There is a fixed point $\phi^*\in\U$, i.e. $\Ren(\phi^*)=\phi^*$.
\item The fixed point is hyperbolic, meaning that the derivative of the renormalization operator $\ren=d\,\Ren$ at $\phi^*$ has no eigenvalue of modulus one.
\item Only one of its eigenvalues has modulus larger than one; we will call it $\d$.
\end{enumerate}

Under these assumptions, it is reasonable to assume that there exists an \emph{unstable manifold} $\Wu$ of dimension one which generalizes the eigenspace relative to $\d$, and a \emph{stable manifold} $\Ws$ of codimension one, that generalizes the eigenspace of stable eigenvectors. Let us make an additional assumption.
\begin{enumerate}
\item[4.] Let $\Sigma_n$ be the manifolds of period-$2^n$ superstable maps in $\U$. Then $\Wu$ intersects $\Sigma_1$ transversally at $\phi^*_0\in\Wu\cap\Sigma_1$.
\end{enumerate}

We have already remarked that $\Ren$ sends superstable maps into superstable maps. Therefore we have the inclusions
\be
\Ren^n (\Sigma_{n+1})\subset\Sigma_1\,,
\ee
and it is not hard to imagine that all the $\Sigma_n$'s will intersect transversally $\Wu$, at points $\phi^*_{n-1}$. It is also clear that the sequence $\phi^*_n$ converges geometrically to the fixed point $\phi^*$, with rate $\d$. One can imagine that not only the points $\phi^*_n\to \phi^*$, but also the manifolds $\Sigma_n$ accumulate toward the unstable manifold, and that their distance decreases geometrically with rate $\d$. 

\begin{figure}
\centering
\includegraphics[width=10cm]{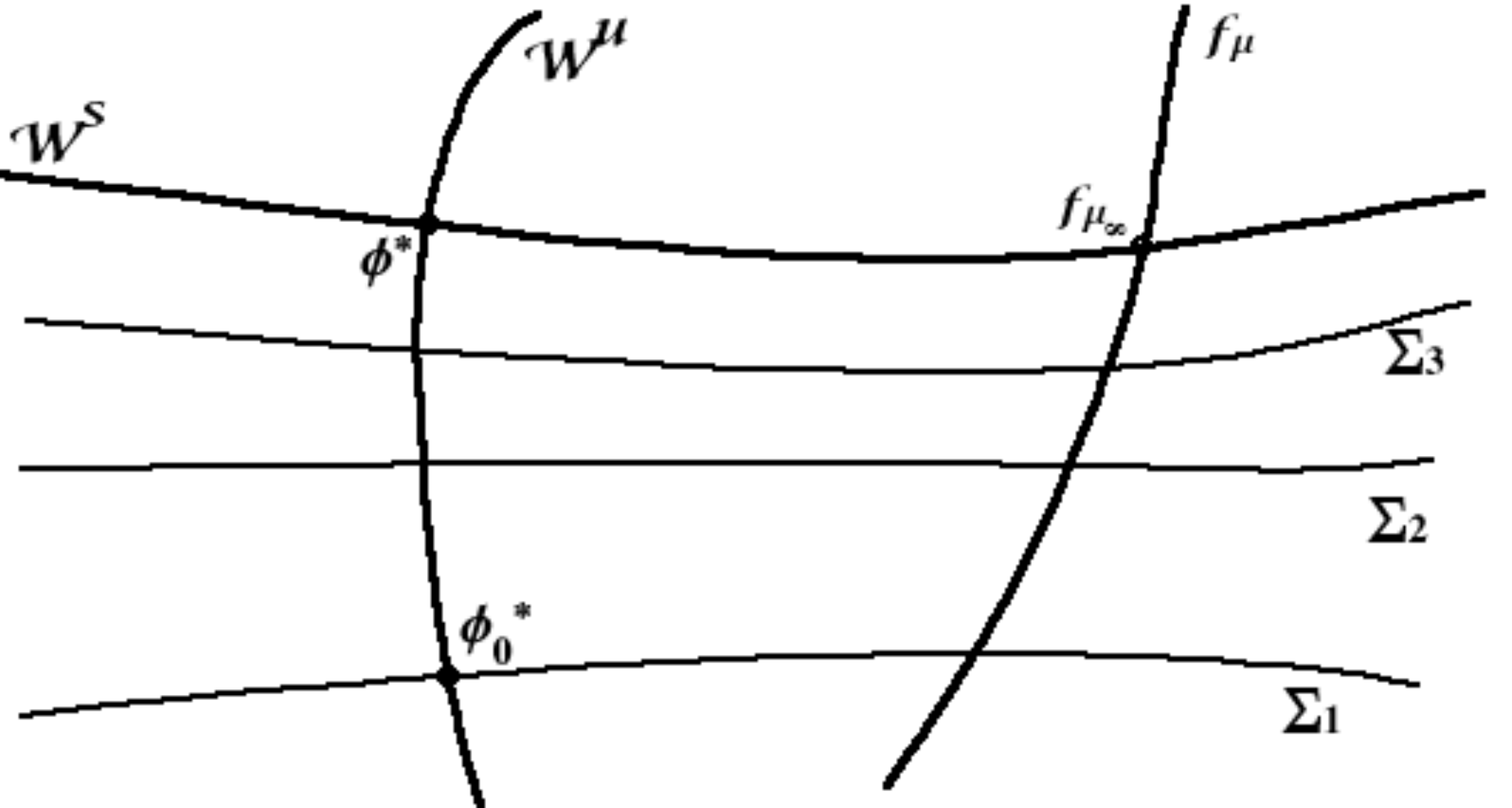}
\caption{The stable and unstable manifold for $\Ren$.}
\label{fig:asymptoticmanifold}
\end{figure}

The whole picture is summarized in Figure \ref{fig:asymptoticmanifold}. This also suggest how to explain the period-doubling cascade in a generic family of maps $\{f_\mu\}\subset\U$. In fact, the sequence of bifurcations occurring at $\mu_n$, or equivalently the sequence of superstable maps at $\tilde{\mu}_n$ can be described in terms of the behavior of the sequence of manifolds $\Sigma_n$. We will return on this later, in order to make the relation with $\d$ more quantitative.

As remarked, this whole discussion has been very qualitative. A rigorous treatment would bring us too far from the points of our discussions; the interested reader is invited to consult e.g. \cite{colleteckmann}. Here it is worth mentioning that, once Feigenbaum's conjectures are accepted, it is not hard to prove that the scenario we described happens. What is much harder is to establish the existence of the hyperbolic fixed point. Remarkably, all this could be done rigorously \cite{CT,colleteckmannlanford, lanford,davie, chandramouli}. Finally, let us stress that had we relaxed the condition that our maps are even, we would still have found a single hyperbolic fixed point $\phi^*$, which turns out to be even.

\begin{exercise}[Renormalization operator and logistic map]
Consider the family of maps on $I=[-1,1]$
\be
f_\mu(x)=1-\mu\,x^2\,,\quad\quad\quad\quad 0\leq\mu\leq2\,,
\ee
which is a suitable rescaling of the logistic map. Work out what the renormalizability conditions (\ref{eq:renormalizability}) imply in terms of $\mu$, and comment on what it means.
\end{exercise}

\begin{exercise}[Domain of the renormalization operator]
We implicitly assumed that superstable maps are in the domain of the renormalization operator $\Ren$, i.e. satisfy  (\ref{eq:renormalizability}). Prove it.
\end{exercise}

\begin{exercise}[Linearizing the renormalization operator]
\label{ex:frechetder}
Compute formally the (Fr\'echet) derivative $\ren=D\Ren$ of $\Ren$ at $\phi$, recalling that this can be found from expanding formally to first order $f=\phi+\eps\,\varphi$, showing that it is
\be
\label{eq:frechetder}
\ren_\phi\cdot\varphi(x)=-\frac{1}{a}\varphi\big(\phi(-a\,x)\big)-\frac{1}{a}\varphi(-a\,x)\,\phi'\big(\phi(-a\,x)\big)\,.
\ee
\end{exercise}

\subsection{Predictions from renormalization group} 
Now that we have formulated the RG ideas for maps on the interval, let us try to obtain some quantitative predictions out of them.

The first step is to find some information of the RG fixed point $\phi^*$. We immediately encounter a difficulty: we must understand the role of $a$ which appears in the definition (\ref{eq:ren}). It has to do with the rescaling of the $x$-direction, and we know that for the fixed point this amounts to shrinking by (negative) $\alpha$. This leads to the identification, at the fixed point
\be
a=-1/\a\,.
\ee

This still leaves $\a$ undetermined. However, it is easy to see how this is fixed by the normalization of the maximum to $\phi^*(0)=1$. Let us consider the ansatz for a symmetric $\phi^*(x)$
\be
\phi^*(x)=1+\sum_{n=1}^N c_n\,x^{2n}+O(x^{2N+2})\,.
\ee
Plugging this into the fixed point equation
\be
\phi^*(x)=\Ren(\phi^*)(x)=\a\,\phi^*(\phi^*(x/\a))\,,
\ee
we find e.g. for $N=3$ the solution
\be
\a\approx-2.479\,,\quad
c_1\approx-1.522\,,\quad
c_2\approx0.073\,,\quad
c_3\approx0.046\,.
\ee
Going to $N=6$ yields $\a\approx-2.502897$, an estimate which turns out to be correct up to order $\sim10^{-6}$. Incidentally, the above procedure illustrates the importance of the order of the maximum, which is a crucial ingredient in our ansatz: a different choice would have lead to a different result for $\phi^*(x)$.

Let us now try to find a more quantitative relation between $\d$ and the sequence $\mu_n\to\mu_\infty$ for a family of maps $\{f_\mu\}\subset\U$. It is convenient to denote introduce the short-hands 
\be
F(x)\equiv f_{\mu_\infty}(x)\,,\quad\quad\quad \varphi(x)\equiv\left.\frac{\partial\,f_\mu}{\partial\,\mu}\right|_{\mu_\infty}\!.
\ee
Then a function $f_\mu$ can be written, when $\mu$ is close to $\mu_\infty$, as
\be
f_\mu(x)\approx F(x)+(\mu-\mu_\infty)\,\varphi(x)\,,
\ee
and similarly we can expand
\be
\Ren(f_\mu)(x)\approx \Ren(F)(x)+(\mu-\mu_\infty)\,\ren_F\cdot\varphi(x)\,.
\ee
Recall that $F$ lies exactly in the intersection $\Ws\cap\{f_\mu\}$. Therefore, $\Ren(F)\in\Ws$ is closer to $\phi^*$ than $F$, and indeed due to the geometric convergence on the stable manifold, $\Ren^n(F)\approx \phi^*$ after a few iterations. We can thus write
\be
\Ren^n(f_\mu)(x)\approx \phi^*(x)+(\mu-\mu_\infty)\,\ren_{\phi^*}^n\cdot\varphi(x)\,.
\ee
Let us expand $\varphi$ on a basis of eigenfunctions of $\ren_{\phi^*}$,
\be
\varphi(x)=c_\delta \,\varphi_\delta(x)+\sum_j c_j\,\varphi_j(x)\,,
\ee
where we have distinguished the eigenfunction pertaining to $\d$. Since all eigenvalues except $\d$ have modulus smaller than one, their eigenvectors $\varphi_j$, $j\neq\d$ are sent to zero by $\ren_{\phi^*}^n$. Only the eigenvector $\varphi_\delta$ of $\delta$ plays a role, so that we can eventually write
\be
\Ren^n(f_\mu)(x)\approx \phi^*(x)+(\mu-\mu_\infty)\,c_\delta\,\delta^n\,\varphi_\delta(x)\,.
\ee

Let us now specialize the above expression to the case $\mu=\tilde{\mu}_n$, i.e. the case where the map is superstable of period $2^n$, and evaluate it at $x=0$. On the one hand, we have
\be
\Ren^n(f_{\tilde{\mu}_n})(0)=(f_{\tilde{\mu}_n})^{2^n}(0)=0\,,
\ee
due to the presence of the $2^n$-cycle and the fact that $x=0$ is a point of the cycle. On the other hand we have
\be
 \phi^*(0)+(\tilde{\mu}_n-\mu_\infty)\,c_\delta\,\delta^n\,\varphi_\delta(0)=\frac{1}{\a}+(\tilde{\mu}_n-\mu_\infty)\,\delta^n\ \,c_\delta\,\kappa_\delta\,,
\ee
where we emphasized that $\kappa_\delta=\varphi_\delta(0)$ does not depend on $n$. Therefore, at least up to higher order terms in $\tilde{\mu}_n-\mu_\infty$ it must be
\be
(\tilde{\mu}_n-\mu_\infty)\,\delta^n\approx -\frac{1}{\a\, c_\d\, \kappa_\d}={\rm const.}\,,
\ee
\emph{for any $\d$}, which means exactly that the rate of convergence $\tilde{\mu}_n\to\infty$ is $\delta$.

The only thing that remains to do is to compute $\d$, using the expression for the differential $\ren$ found in Exercise \ref{ex:frechetder}, and inserting the approximate result for $\phi^*(x)$ found by means of the power series expansion into (\ref{eq:frechetder}). The eigenvalue equation reads
\be
\a\,\varphi_\d\big(\phi(x/\a)\big)+\a\,\varphi_\d(x/\a)\,\phi'\big(\phi(x/
\a)\big)=\d\,\varphi_\d(x)\,,
\ee
and can be solved approximately by using an ansatz for $\varphi_\d(x)$ too. The result, taking $N=6$ in the ans\"atze, is
\be
\delta\approx4.66914\,,
\ee
with an accuracy of order $\sim10^{-5}$ with respect to the known result \cite{briggs}.
\bigskip

We conclude this first part with some comments on what we have seen. We have considered systems that are described by iterated maps on the interval (that is among the simplest dynamics one can imagine) where the experimenter is able to tune one parameter $\mu$; some class of these systems exhibit similar properties as the parameter approaches a critical value $\mu_\infty$. We have explained this by using the properties of the renormalization of operator $\Ren$.

What is the physical interpretation of $\Ren$? When we are looking at the dynamics, $\Ren$ operates a rescaling of the time-scale ($f\mapsto f\circ f$) together with a rescaling of the $x$-scale. In this sense it is similar to Kadanoff's coarse-graining transformation \cite{kadanoff66a,kadanoff66b}: acting with the renormalization operators corresponds to changing the description of the problem, ``zooming out'' in such a way as to preserve the interesting (universal) properties of the dynamics.

In the language of statistical physics, we would say that $\mu$ plays the role of some adjustable ``knob'' (temperature, external magnetic field, etc.), and that at $\mu_\infty$ a phase transition occurs. The divergence of the correlation length in our case is mimicked by the appearance of an infinite-period cycle. All the systems at the phase transitions are points on the stable manifold $\Ws$, and due to this they are very similar. In fact, under the action of $\Ren$, all these points get to the fixed point $\phi^*$, so that the properties invariant under $\Ren$ (the large scale properties, in a statistical system) are common to all of them. We did not investigate at all what happens to our maps at $\mu_\infty$ (to avoid the complications of chaotic systems), but it is indeed possible to single out several universal properties.

The role played by $\d$ is that of a \emph{relevant} eigenvalue, and $\varphi_\delta(x)$ is a relevant direction at the fixed point. This means that if we perturb $\phi^*(x)$ by something proportional to $\varphi_\delta(x)$, acting with $\Ren$ will drive us away from the fixed point. Notice that in Figure \ref{fig:asymptoticmanifold} the stable manifold separates $\U$ in two regions. Maps on either side of the $\Ws$ will have different behavior under $\Ren$ and therefore different properties. We have seen that all the maps ``below'' the stable manifold ($\mu<\mu_\infty$) are periodic, with negative Lyapunov exponent, whereas indeed maps ``above'' $\Ws$ will yield chaotic behavior. Again this is in analogy with what happens in statistical systems, where the position in parameter space of a theory determines its large-scale properties.

\begin{exercise}[Renormalization group for quartic tips]
Work out, on a computer, the computation of $\a$ and $\d$ for maps with \emph{quartic} tip.\\
{\rm (Answer: $\a\approx-1.69030297$ and $\d\approx7.28486622$)}
\end{exercise}

\section*{Part II: Functional renormalization group equations}
In this part we will consider a renormalization group technique for Quantum Field Theory (QFT) that is referred to as Functional Renormalization Group Equation (FRGE) or, quite improperly, as Exact Renormalization Group Equations (ERGE). Before doing so, we will recap some textbook notions about QFT, see e.g. \cite{ramond,weinbergQFT}.

\section{Elements of quantum field theory}
Quantum Field Theory (QFT) arose as an attempt to describe the quantum theory of elementary point-like particles so as to be consistent with the requirements of Special Relativity. One consider the Hilbert space $\mathcal{H}$ of physical states, which bears a unitary representation of the Poincar\'e group, under which only the vacuum state $\left|v_0\right\rangle$ is left invariant and such that its generators have positive norm. In this language fields are distributions which take values in some space of operators on $\mathcal{H}$. Furthermore, since no signal can propagate faster than light, we require that fields smeared by test functions with disjoint supports whose distance is space-like must commute, or anti-commute according to the statistics they obey.

Let us consider a free scalar field in four-dimensional Minkowski space. The state space is a Fock space
\begin{equation}
\mathcal{H}=\bigoplus_{n=0}^{\infty}H_n,
\end{equation}
where $H_n$ is the $n$-particle Hilbert space (the symmetrized $n$-fold tensor product of the single particle one), and $H_0$ is generated by the vacuum $\left|v_0\right\rangle$. In this case, $H_1$ is generated by the complete orthonormal set of positive energy solutions $f_i$ of Klein-Gordon equation\footnote{We take the metric to have signature $(+---)$.}
\begin{equation}
\label{kleingordon}
\left(\eta^{\mu\nu}\partial_\mu\partial_\nu+m^2\right)\,f(x)=0,
\end{equation}
and the scalar product is
\begin{equation}\label{planewave}
\langle f_i,f_j\rangle=\int d\mu(p)\;\hat{f}_i^*(p)\,\hat{f}_j(p)=\delta_{ij},
\end{equation}
where
\begin{equation}
f(x)=\int d\mu(p)\,\hat{f}(p)\,e^{-ip\cdot x},\ \ \ \ \ d\mu(p)=\frac{d^4p}{(2\pi)^4}2\pi\delta(p^2-m^2)\theta(p_0).
\end{equation}

The free scalar field $\phi(x)$ also satisfies Klein-Gordon equation, and can be written in terms of creation and annihilation operators $a^\dagger, a$,
\begin{equation}\label{freescalar}
\phi(x)=\int d\mu(p)\;\left(a(p) e^{-ip\cdot x}+a^\dagger(p) e^{ip\cdot x}\right),
\end{equation}
 which satisfy canonical commutation relations: when $a^\dagger,a$ are smeared by a solution $f$ of Klein-Gordon's equation, i.e. $a_f=\int d\mu \hat{f}a$, one has
 \begin{equation}
 \label{scalarcommrel}
 [a^\dagger_{f_i},a^\dagger_{f_j}]= [a_{f_i},a_{f_j}]=0,\ \ \ \ [a_{f_i},a^\dagger_{f_j}]=\delta_{ij}.
 \end{equation}
 The physical content of the theory, in this simple case, is given by the vacuum expectation value of any (suitably ordered) product of field operators (\ref{freescalar}), which can be explicitly expressed in terms of Green's functions, using (\ref{scalarcommrel}) and Wick's Theorem.
 
The theory of free fields is of little interest for Physics. We want to  consider theories defined by Lagrangian densities of the form e.g.
\begin{equation}
L=L_{\rm free}+L_{int}=\frac{1}{2}\left(\eta^{\mu\nu}\partial_\mu\phi\partial_\nu\phi-m^2\phi^2\right)+L_{\text{int}},
\end{equation}
where the free Lagrangian $L_{\rm free}$ alone would give rise to (\ref{kleingordon}) as its variational equations\footnote{Notice that one would find modified Klein-Gordon equations for the interacting theory. However those would involve products of $\phi$, that is, ill-defined products of distributions, which is a first hint of the presence of ultraviolet pathologies.}. To describe the Physics of interacting QFTs, one considers  two asymptotic state spaces $\mathcal{H}_{\text{in}}$ and $\mathcal{H}_{\text{out}}$, both isomorphic to the space of free particles, related by a unitary operator, the \lq\lq $S$-matrix\rq\rq, which contains the information about the scattering. In this language, the $S$-matrix
 for a free theory will just be the identity.

The predictions of a QFT, such as scattering cross sections, will depend on its $S$-matrix. This, under suitable asymptotic conditions, can be expressed in terms of the time-ordered $n$-points correlation functions of the theory, by virtue of Lehmann-Symanzik-Zimmermann (LSZ) formula. Such quantities, in the path integral formalism invented by Feynman, are the momenta of an (ill-defined) functional measure over the \lq\lq paths\rq\rq, or rather histories, of the fields configurations:
\begin{equation}\label{greenfunctions}
\langle\phi(x_1)\cdots\phi(x_k)\rangle=\frac{1}{\mathcal{N}}\int\mathcal{D}\phi\left(\phi(x_1)\cdots\phi(x_k)\;e^{i\int d^4y L[\phi(y)]}\right)\,,
\end{equation}
where $\mathcal{N}=\int\mathcal{D}\phi\, e^{i\int\! L}$ is an (infinite) normalization constant, such that $\langle1\rangle=1$. The issue in (\ref{greenfunctions}) is that field configurations are weighted by a complex exponential; furthermore the action in Minkowski space involves hyperbolic differential operators, which are rather unpleasant to deal with. A much better setting would be provided by the \lq\lq Wick rotation\rq\rq (the analytic continuation) of (\ref{greenfunctions}), which formally formally yields the \lq\lq Euclidean\rq\rq\ theory, expressed by the Feynman-Kac formula
\begin{equation}\label{schwingerfunctions}
\langle\phi(x_1)\cdots\phi(x_k)\rangle=\frac{1}{\mathcal{N}}\int\mathcal{D}\phi\left(\phi(x_1)\cdots\phi(x_k)\;e^{-\int d^4y L[\phi(y)]}\right)\,.
\end{equation}
The Euclidean correlation functions are called Schwinger functions and, remarkably, they can be rigorously related to the correlations in the physical, Minkowskian theory \cite{OS}. In principle one can construct the QFT in  a Euclidean formalism and \emph{reconstruct} it in Minkowski space. For the analytic continuation to be possible and lead to a physically meaningful relativistic theory, the Schwinger functions have to satisfy a number of conditions, or axioms \cite{streater, wightman}.
However, even Euclidean QFTs are far from being well defined. A number of pathologies arise when one tries to evaluate (\ref{schwingerfunctions}) to extract physical predictions. In what follows, we will always deal with Euclidean theories and forget about the reconstruction problem.

\subsection{Generating functionals}
Let us consider the Euclidean theory of a free scalar field; this amounts to studying the (normalized) functional measure formally defined as
\begin{equation}
d\mu_{C}(\phi)=\frac{1}{\int \mathcal{D}\phi e^{-\frac{1}{2}\int \phi(-\Delta+m^2)\phi}}\;\mathcal{D}\phi e^{-\frac{1}{2}\int d^4x\, \phi(x)\,(-\Delta+m^2)\,\phi(x)},
\end{equation}
where we have written $\mu_C$ to remember that the measure depends on a certain operator $C=[-\Delta+m^2]^{-1}$.\footnote{A motivation to introduce $d\mu_C$ is that one can rigorously define it as the conditional Wiener measure of covariance $C$ on the space of \lq\lq paths\rq\rq\ $\phi$, whereas $\mathcal{D}\phi$ is just  a formal measure.} Since the integral appearing as a normalization is Gaussian  and can be performed exactly, we find that the normalization constant is given by $\int \mathcal{D}\phi e^{-\frac{1}{2}\int \phi(-\Delta+m^2)\phi}=(\mathrm{det}[-\Delta+m^2])^{-1/2}$. We want now to compute the quantities
\begin{equation}
\label{eq:correlfunct}
\langle\phi(x_1)\cdots\phi(x_k)\rangle=\int d\mu_C\left(\phi(x_1)\cdots\phi(x_k)\right),
\end{equation}
and this can be done by introducing the generating functional $Z[J]$
\begin{equation}
Z[J]\equiv \int d\mu_C\, e^{\int d^4x\,J(x)\phi(x)},
\end{equation}
such that 
\begin{equation}\label{logcorr}
\langle\phi(x_1)\cdots\phi(x_k)\rangle=\left.\frac{\delta}{\delta J(x_k)}\cdots\frac{\delta}{\delta J(x_1)} \log Z[J]\right|_{J=0}.
\end{equation}

As is well known, the generating functional can be evaluated exactly in this case, by performing the Gaussian integral (by a simple change of variables), yielding
\be
\label{explicitfreeZ}
Z[J]=e^{\frac{1}{2}\int d^4xd^4y J(x)\,C(x-y)\,J(y)},
\quad\quad
C(z)=\frac{1}{(2\pi)^4}\int d^4p \frac{e^{ip\cdot z}}{p^2+m^2}\, ,
\ee
from which it follows that the two-points Schwinger function is just the free Euclidean propagator\footnote{Of course, if we wanted to consider the Minkowskian theory, we would have to worry about a prescription to obtain the correct Green's function by analytic continuation.}. Similarly, we can find the explicit expression for all correlation (\ref{eq:correlfunct}), which can be expressed by the familiar Feynman diagrams. We could introduce a different functional of the sources $W[J]$, defined by
\begin{equation}
e^{-W[J]}\equiv Z[J],
\end{equation}
so that we do not have to take the logarithm in (\ref{logcorr}). This  can be proven to generate all the connected correlation functions, see e.g. \cite{zinnjustin} \S 6. For the free theory, the only non-vanishing one is the two-points function. Let us  introduce another generating functional, which will be useful later on. Observe that in the free case we have, using (\ref{explicitfreeZ}),
\begin{equation}
\frac{\delta Z[J]}{\delta J(x)}=\int d^4y C(x-y)J(y)\;Z[J],\ \ \ \ \frac{\delta W[J]}{\delta J(x)}=-\int d^4y C(x-y)J(y),
\end{equation}
so that if we define the effective field as the expectation value $\varphi(x)\equiv\langle\phi(x)\rangle=-\frac{\delta W[J]}{\delta J(x)}|_{J}$,  this satisfies the equations of motion
\begin{equation}\label{effectivefreefield}
(-\Delta+m^2)\varphi(x)=J(x).
\end{equation}
We can invert the above relation to obtain $J[\varphi]$, and substitute it in the generating functional $W[J]$. This amounts to defining the new functional $\Gamma[\varphi]=W[J[\varphi]]$, which, in a generic interacting theory, can be done by Legendre transform
\begin{equation}\label{legendretransform}
\Gamma[\varphi]=\inf_{J} \left(W[J]+\int d^4x J(x)\varphi(x)\right),
\end{equation}
from which we find the relation
\begin{equation}\label{JphiGamma}
\frac{\delta\Gamma[\varphi]}{\delta \varphi(x)}=J(x)\,.
\end{equation}
This is a constraint for the effective field $\varphi(x)$ which in absence of currents must solve a variational equation similar to the one for $S[\phi]$ in the classical theory. For this reason, we can consider $\Gamma[\varphi]$ as an \emph{quantum effective action}. We have two complementary interpretations for $\Gamma$: on the one hand, in a diagrammatic approach, it can be proven that $\Gamma$ can be obtained from the one particle irreducible\footnote{Recall that a Feynman graph is one particle irreducible if it cannot be disconnected by cutting any single internal line.} (1PI) Feynman diagrams:
\begin{equation}
\Gamma[\varphi]=\int_{\text{1PI}}\!
\mathcal{D}\phi\; e^{-S[\varphi+\phi]},
\end{equation}
where by this formal notation (borrowed from \cite{weinbergQFT}) we indicate that the path integral is restricted in such a way that, when expanded in a perturbative series, only 1PI graphs appear.\footnote{This was actually the way the effective action was originally defined \cite{goldstone}, whereas the functional definition (\ref{legendretransform}) was given in \cite{dewitt63,jonalasinio64}; see also \cite{jackiw} for a derivation of the $n$-loop expansion of $\Gamma$.} We can also think of $\Gamma$ as given by an infinite series in $\varphi^n$, whose coefficients depend on the loop integrals\footnote{Actually, on the renormalized loop integrals, as it will be clearer later.}. Such an expansion in effective vertices will be\lq\lq semi-local\rq\rq, since we are not just summing tree diagrams:
\begin{equation}\label{1PIexpansion}
\Gamma[\varphi]=\sum_{n}\frac{1}{n!}\int d^4x_1\cdots d^4x_N\,\Gamma^{(n)}[x_1,\dots,x_N]\varphi(x_1)\cdots\varphi(x_n).
\end{equation}
This gives a second interpretation: the full quantum theory generated by the action $S$ is equivalent to the classical (tree-level) theory for an action taken to be $S_{new}=\Gamma$, computed in terms of the effective vertices. As a result we can write
\begin{equation}
\Gamma[\varphi]=S[\varphi]+\mathrm{quantum\ corrections}.
\end{equation}
For more on this, see \S 16 in \cite{weinbergQFT} and \S 8 in \cite{toms}.

If we go back to the free theory, we can explicitly substitute (\ref{effectivefreefield}) in the definition of $W[J]$ to find that
\begin{equation}
\Gamma[\varphi]=\int d^4x \frac{1}{2}\varphi(x)\left(-\Delta+m^2\right)\varphi(x)\,.
\end{equation}
This once again shows that a free theory, even if we take into account all possible quantum corrections, just describes the propagation of noninteracting particles, as was already clear from (\ref{explicitfreeZ}), and the expansion (\ref{1PIexpansion}) is just given by the two point function. Furthermore, we see that
\begin{equation}\label{GammaWrel}
\frac{\delta^2\Gamma[\varphi]}{\delta\varphi(x)\,\delta\varphi(y)}=-\left[\frac{\delta^2 W[J]}{\delta J(x)\,\delta J(y)}\right]^{-1},
\end{equation}
which, as can be checked using (\ref{JphiGamma}), is true also in a generic theory. Furthermore, one also obtains from  (\ref{legendretransform}) that\footnote{For a proof of this last expression, see \cite{weinbergQFT} \S 18, or \cite{salmhofer} \S 2.5 where a different strategy is used: the above formula is taken as definition, and the familiar properties of $\Gamma$ are derived from it.}
\begin{align}\nonumber
\exp\left(-\Gamma[\varphi]\right)&=\frac{1}{\mathcal{N}}\int \mathcal{D}\rho\,\exp\left\{-S[\rho+\varphi]+\int d^4x\frac{\delta\Gamma[\varphi]}{\delta\varphi(x)}\,\rho(x)\right\}=\\
\label{gammafull}
&=e^{-\frac{1}{2}\int\! d^4x\,d^4y\varphi(x)C(x-y)\varphi(y)}\int d\mu_C[\rho]e^{-S_{\text{int}}[\rho]+\int\! d^4x \rho(x)(-\Delta+m^2)\varphi(x)},
\end{align}
that can be taken as an alternative definition of $\Gamma[\varphi]$, where we replace the current $J(x)$ by $(-\Delta+m^2)\varphi(x)$. The insertion of the inverse propagator has the effect of \lq\lq amputating\rq\rq the external propagators, see \S 2.5 in \cite{salmhofer}.

\begin{exercise}[The $\Gamma$ and $W$ functionals]
Check (\ref{GammaWrel}) for an interacting theory.
\end{exercise}

\begin{exercise}[The saddle point approximation]
Consider the interacting theory
\begin{equation}
S[\phi]=\int d^4x\left[\phi (-\Delta+m^2)\phi +\frac{\lambda}{4!}\phi^4\right]\,,
\end{equation}
and work out the saddle-point approximation of $\Gamma[\varphi]$ using (\ref{gammafull}).
\end{exercise}

\subsection{Perturbative expansion of $\lambda\phi^4$}
As we know, the difficulties and the wonders lie in the interacting theories. We will consider a specific theory for our examples, which is the \textsl{drosophila} of QFT: the $\lambda\phi^4$ theory. Its Euclidean action is
\begin{equation}
\label{eq:phi4action}
S[\phi]=\int d^4x\left[\phi (-\Delta+m^2)\phi +\frac{\lambda}{4!}\phi^4\right],
\end{equation}
where $\lambda>0$. Its generating functional will have the form
$Z=\int d\mu_C e^{-S_{\text{int}}}$.
For such an interacting theory, we will not be able to exactly evaluate the path integral in general. Instead, we can perform a \emph{perturbative expansion}, i.e. a formal expansion in powers of $\lambda$. Supposing that $\lambda\ll1$, we have
\begin{align}\nonumber
d\mu_C e^{-\frac{\lambda}{4!}\,\int d^4x \phi(x)^4}&=\\
&=d\mu_C\left[1-\frac{\lambda}{4!}\,\int d^4x \phi(x)^4+\frac{\lambda^2}{(4!)^22!}\left(\int d^4x \phi(x)^4\right)^2-\dots\right].
\end{align}
In the same way,  $Z[J]$ is expanded as a formal series in $\lambda$:
\begin{equation}
Z[J]=e^{-\frac{\lambda}{4!}\frac{\delta^4}{\delta J^4}}\int d\mu_C e^{\int J\phi}=e^{-\frac{\lambda}{4!}\frac{\delta^4}{\delta J^4}}e\,^{\frac{1}{2}\int d^4xd^4y J(x)\,C(x-y)\,J(y)},
\end{equation}
 and the Schwinger functions are computed order by order in the coupling parameter. However several difficulties arise:
\begin{enumerate}
\item This procedure can work only when $\lambda$ is small, which makes it very hard to address strong-coupling issues.
\item It is not clear \emph{how small} the coupling has to be. In fact it is not even clear if there is any nonzero radius of convergence (even in some weak sense such as Borel summation) for the perturbative series.
\item As we will see, not even the individual terms in the series are well defined, as many of them are given by divergent integrals.
\end{enumerate}
We will  focus only on the last point. The techniques we will discuss are suitable for perturbative renormalization, even to all orders; if we wanted to address the issue of convergence of the perturbative series we would need more refined tools.

\begin{figure}[!t]
  \begin{center}
    \subfigure[The propagator.]{\label{fig:propagator}\includegraphics[width=3cm, angle=-90]{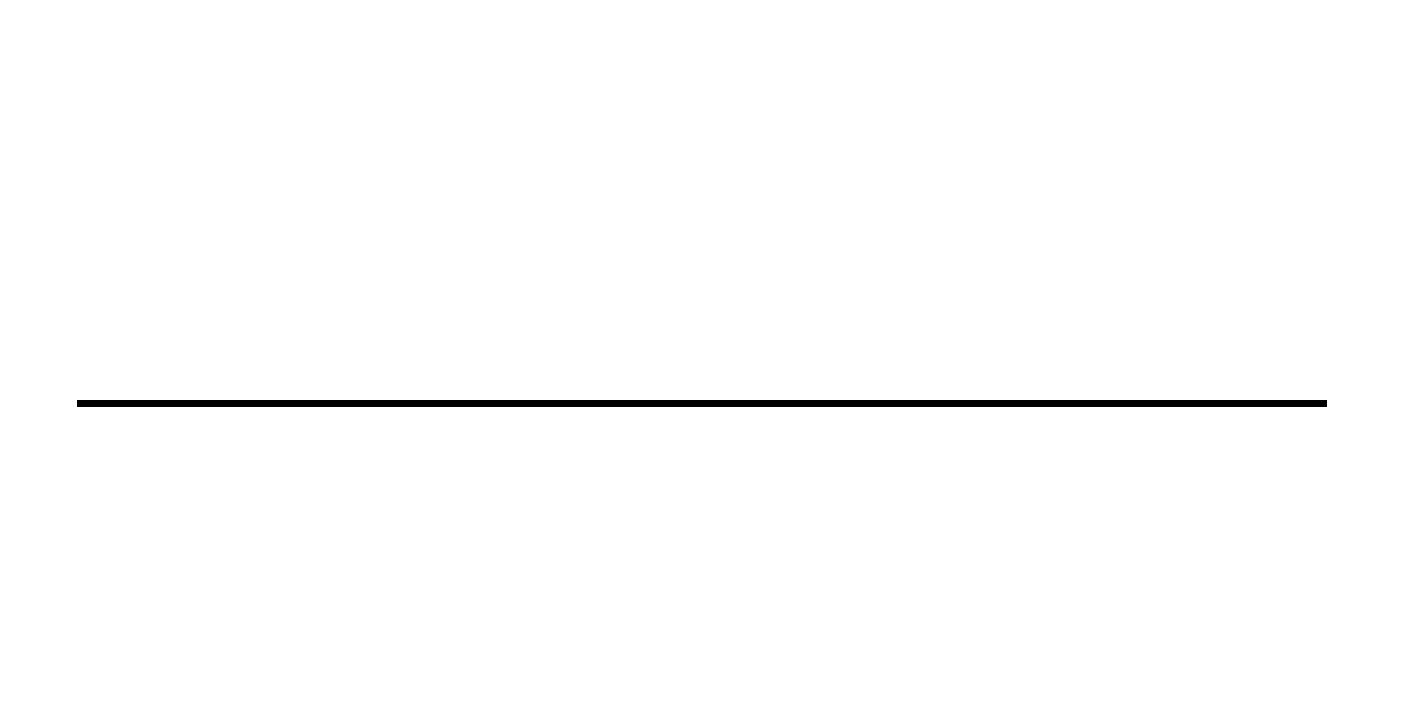}}
    \subfigure[The \textsl{tadpole}.]{\label{fig:tadpole}\includegraphics[width=3cm, angle=-90]{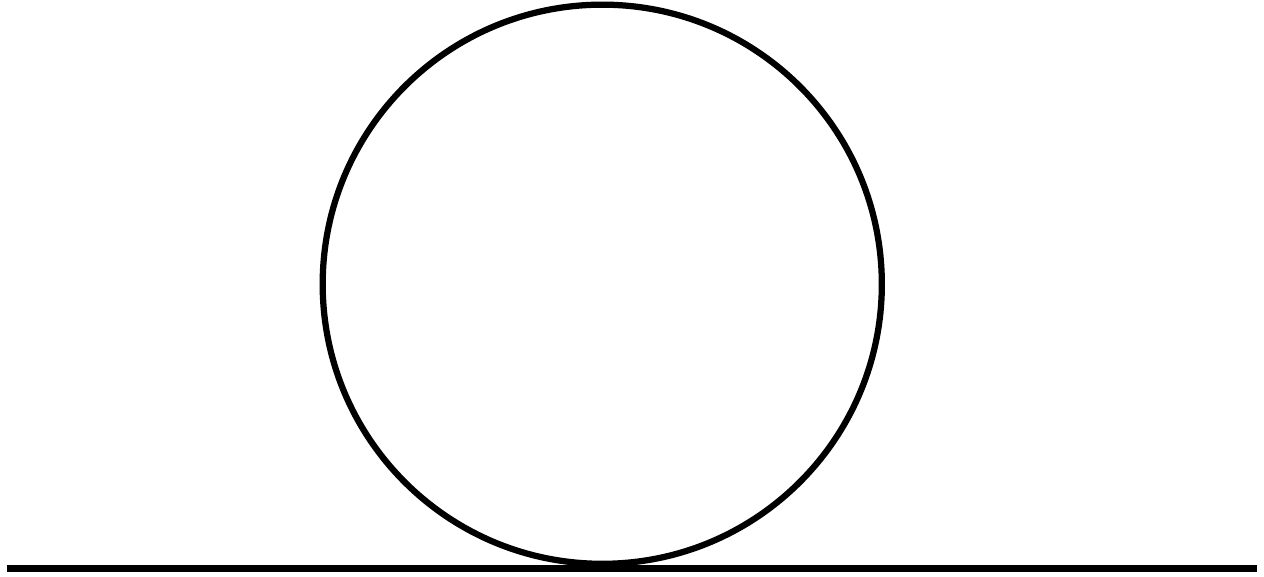}}\\
     \subfigure[The \textsl{setting sun}.]{\label{fig:settingsun}\includegraphics[width=3cm, angle=-90]{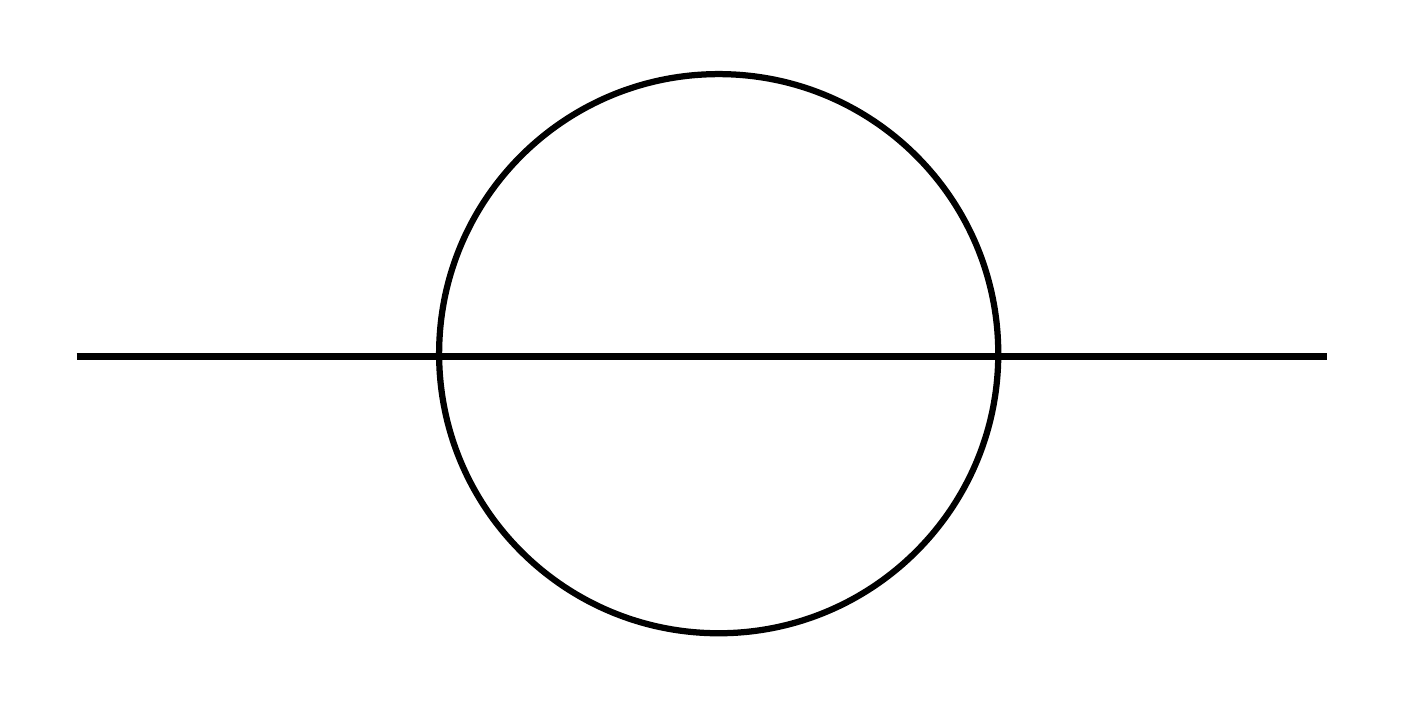}}
          \subfigure[The \textsl{double scoop}.]{\label{fig:doublescoop}\includegraphics[width=3cm, angle=-90]{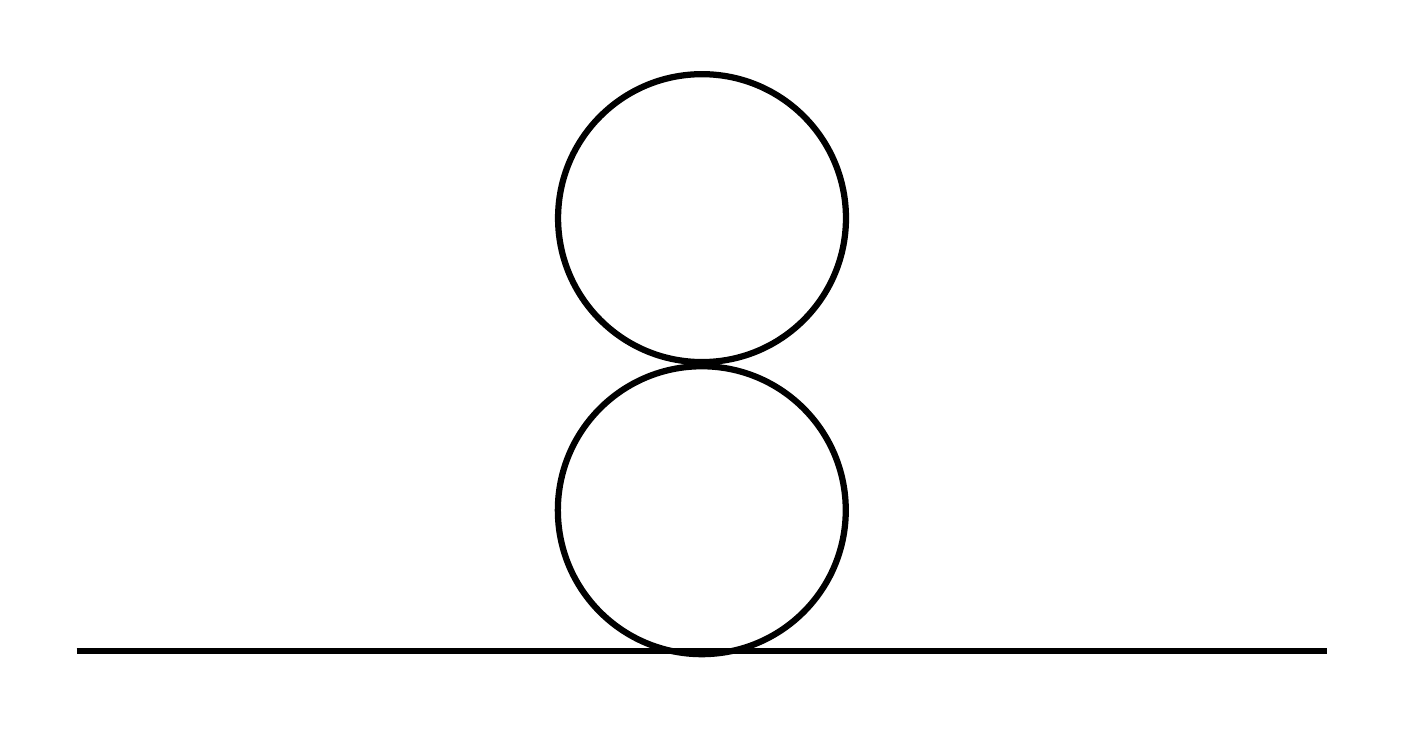}}
  \end{center}
  \caption{One particle irreducible two-points graphs for $\lambda\phi^4$, up to two loops.}
  \label{fig:twopoint}
\end{figure}

\begin{figure}[!t]
  \begin{center}
\includegraphics[width=3.3cm, angle=-90]{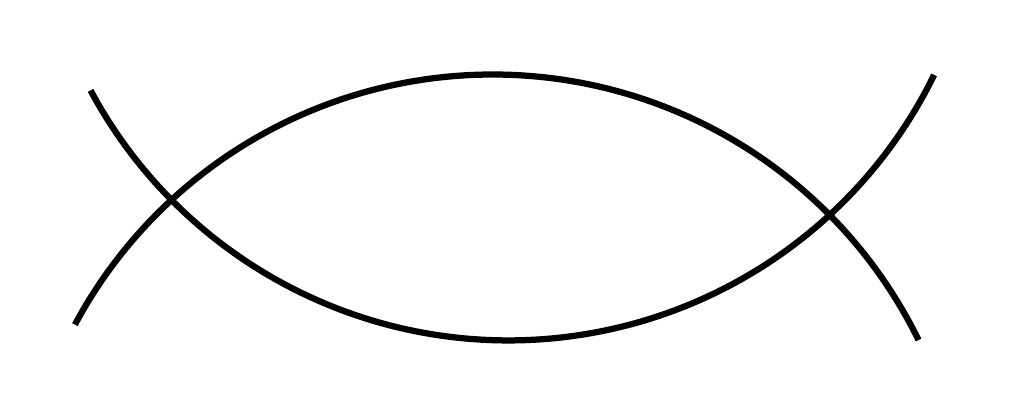}
  \caption{The \textsl{bubble} is a one-loop divergent graph with four external legs.}
  \label{fig:bubble}
  \end{center}
\end{figure}

It is simple (see e.g. \cite{ramond}) to derive the first terms of the expansion of, say, the two-points Schwinger function\footnote{Should we consider the amputated two-points function, we would not find the factor of $(p^2+m^2)^{-2}$ in front of the integral.} in momentum space:
\begin{equation}
\int\langle\phi(0)\phi(x)\rangle\,e^{-ip\cdot x}=\frac{1}{p^2+m^2}-\frac{1}{(p^2+m^2)^2}\int \frac{d^4q}{(2\pi)^4}\frac{\lambda/2}{q^2+m^2}+O(\lambda^2).
\end{equation}
 The latter term  corresponds to the well known \textsl{tadpole} graph (Figure \ref{fig:tadpole}), that diverges for large internal momenta $q$.
Let
\be
T=\frac{\lambda}{2}\frac{\Omega_4}{(2\pi)^4} \int_0^{+\infty} \frac{q^3dq}{q^2+m^2}\,.
\ee
This divergent expression can be regularized for instance by means of  \emph{dimensional regularization}. The result is then a function of the dimension $D$
\be
\label{tadpole}
T(D)=\frac{\tilde{\lambda}\, ( \mu\tilde{m})^2}{2(4\pi)^2}\left(\frac{4\pi}{\tilde{m}^2}\right)^{2-D/2}\,\Gamma\left(1-\frac{D}{2}\right)\;,
\ee
where $\tilde{m},\tilde{\lambda}$ have been rescaled so that they are dimensionless in $D$ dimensions, and $\mu$ is a mass scale.  By treating the dimension as a complex variable, it is easy to isolate a divergent contribution as $D\to4$, which is just given by a simple pole.  
A similar calculation can be done for the \textsl{bubble} (Figure \ref{fig:bubble}), which instead gives
\begin{align}\label{bubble}
B(D)=\mu^{4-D}\frac{\tilde{\lambda}^2}{32\pi^2}\sum_{\xi\in\{\tilde{s},\tilde{t},\tilde{u}\}}&\left[\frac{2}{4-D}+2+\psi(1)+\ln\frac{4\pi}{\tilde{m}^2}+\right.\\
\nonumber
&\left.-\sqrt{1+\frac{4\tilde{m}^2}{\xi}}\ln\frac{\sqrt{1+\frac{4\tilde{m}^2}{\xi}}+1}{\sqrt{1+\frac{4\tilde{m}^2}{\xi}}-1}+O(D-4)\right].
\end{align}
The variables $\tilde{s},\tilde{t},\tilde{u}$ are Mandelstam's invariant momenta, scaled to be dimensionless: $\tilde{s}=s/\mu^2$, etc. The finite part of the diagram has also a complicated dependence on the momenta entering the loop (which did not happen for the \textsl{tadpole} due to momentum conservation).

The trick now is that the original action (\ref{eq:phi4action}) can be modified by adding term that cancel the divergent contributions of (\ref{tadpole}) and (\ref{bubble}) at $D\to4$. This simply amounts to adding \emph{counterterms} that are the quadratic and and quartic in $\phi(x)$ and therefore just add up to the already existing couplings.

Now, given an arbitrary divergent graph, we can consider its one particle irreducible (1PI) components, in which the momentum integration factorizes, and renormalize each of the divergent ones. Of course the above trick could not work if we were dealing with, for instance, a six-legs 1PI divergent graph, since there is no six-points term in the original Lagrangian. Fortunately, as it is well known, this never happens in the four-dimensional $\lambda\phi^4$ theory, as can be established by counting the superficial degree of divergence.

However, it is still not clear what happens when we consider a generic 1PI diagram containing several divergent sub-diagrams. Is it still possible to factor the divergent contributions and absorb them in a redefinition of the couplings, or are new terms generated? Already at two loops there is a divergent 1PI graph whose internal \textsl{bubbles} share a line (Figure \ref{fig:settingsun}); allowing more loops, the situation gets worse. Fortunately, it is possible to show that such overlapping divergences cancel out at any order, and that the theory can be indeed regularized by modifying only the couplings appearing in the original lagrangian. The proof of this remarkable fact relies on ideas due to Bogolubov, Parasiuk, Hepp and Zimmermann \cite{bogolubovparasiuk,hepp,zimmermann69,zimmermann73}, and is beyond the scope of this discussion.

\begin{exercise}[Dimensional regularization]
Compute (\ref{tadpole}) in dimensional regularization.
\end{exercise}

\subsection{Renormalization and anomalous scaling}
The previous discussion hinted that $\lambda\phi^4$ in four dimensions is \emph{perturbatively renormalizable to all orders}. With this in mind, let us go back to the idea of redefining the couplings. We use as an input in the path integral the bare Lagrangian $L_{\text{bare}}$. We have
\begin{equation}
\label{eq:bareact1}
L_{\text{bare}}= \frac{1}{2} \mathcal{Z}_{\text{bare}}\phi (-\Delta)\phi + \frac{1}{2} m^2_{\text{bare}}\mathcal{Z}_{\text{bare}} \phi^2+\frac{1}{4!}\lambda_{\text{bare}}\,\mathcal{Z}_{\text{bare}}^2\phi^4\,,
\end{equation}
where we have introduced a wave-function renormalization $\mathcal{Z}$, such that $\phi_{\text{bare}}=\phi \mathcal{Z}_{\text{bare}}^{1/2}$. The bare Lagrangian is not physical, i.e. its couplings cannot be determined by any experiment. However, we want it to satisfy 
\begin{equation}
\label{eq:bareact2}
L_{\text{bare}}=\frac{1}{2}\,(1+P) \phi (-\Delta)\phi + \frac{1}{2} m^2\, (1+Q) \phi^2+\frac{1}{4!}\lambda\,(1+R)\phi^4\,,
\end{equation}
so that the physical quantities, the $n$-points Schwinger functions, are finite. Here $P,Q,R$ are functions$-$counterterms$-$of the dimension displaying the correct pole structure as $D\to4$. While this makes the observables finite, there is an ambiguity in fixing the exact value of the physical couplings, which comes from putting a finite, regular (in $D$) part in the counterterms. This has to be done through measuring them in an experiment, but we notice form the explicit expressions of $T(D), B(D)$ that the prediction will depend (logarithmically) on the scale $\mu$:  we have to measure the couplings at a given energy scale and set conveniently the bare quantities. Then \emph{we have to expect the physical couplings to depend on the scale}.

Let us consider the $n$-points 1PI correlation function $\Gamma^{(n)}$; observing that $\phi_{\text{bare}}=\mathcal{Z}_{\text{bare}}^{1/2}\phi$, and rewriting $\mathcal{Z}_{\text{bare}}$ as function of $\lambda(\mu),\mu$ as $\mathcal{Z}$, we have\begin{equation}
\Gamma_{\text{bare}}^{(n)}(p_1,\dots,p_n;m^2_{\text{bare}},\lambda_{\text{bare}})=\mathcal{Z}^{-n/2}\Gamma^{(n)}(p_1,\dots,p_n;\mu,m^2(\mu),\lambda(\mu))\,.
\end{equation}
Notice that the right hand side depends only on physical coupling, and since the l.h.s. is independent of $\mu$, $\frac{d\Gamma_{\text{bare}}}{d\mu}=0$, we have the equation for the physical quantities
\begin{equation}
\left(\frac{\partial }{\partial \mu}+\frac{\partial m^2}{\partial \mu}\frac{\partial }{\partial m^2}+\frac{\partial \lambda}{\partial \mu}\frac{\partial }{\partial \lambda}-\frac{n}{2}\frac{\partial \ln \mathcal{Z}}{\partial \mu}\right)\Gamma^{(n)}(\vec{p};\mu,m^2(\mu),\lambda(\mu))=0\,.
\end{equation}
The same is true if we consider logarithmic derivatives. Call $\varepsilon=4-D$, and define
\begin{align}\label{betafunctions}
\beta_\lambda(\tilde{\lambda},\tilde{m}^2,\varepsilon)&=\mu\frac{\partial \tilde{\lambda}}{\partial \mu}\,,\\
\beta_{m}(\tilde{\lambda},\tilde{m}^2,\varepsilon)&=\mu\frac{\partial \tilde{m}^2\mu^2}{\partial \mu}\,,\\
\eta_\phi(\tilde{\lambda},\tilde{m}^2,\varepsilon)&=\mu\frac{\partial \ln \mathcal{Z}}{\partial \mu}\,,
\label{betafunctionsend}
\end{align}
recalling that $\tilde{\lambda}=\lambda$ only when $\varepsilon\to0$. Remark that once the l.h.s. (the \emph{$\b$-functions}) are known, these equations determine the evolution of the couplings.\\
Let us introduce a dummy parameter $s$ to count the powers of momenta appearing in $\Gamma^{(n)}$, $\vec{p}\to s\,\vec{p}$. On dimensional grounds
\begin{equation}
\left(\mu\frac{\partial }{\partial \mu}+2m^2\frac{\partial }{\partial m^2}+s\frac{\partial }{\partial s}-[4\!-\!n\!+\!\varepsilon(n-2)]\right)\Gamma^{(n)}(s\vec{p};\mu,m^2(\mu),\lambda(\mu))=0\,.
\end{equation}
Putting everything together, we have the scaling equation (in the limit $\varepsilon\to0$)
\begin{equation}\label{scalingequation}
\left(-s\frac{\partial}{\partial s}+\beta_\lambda \frac{\partial}{\partial \lambda}+\left[4\beta_m\!-\!2\right]m^2\frac{\partial}{\partial m^2}-\frac{n}{2}\eta_\phi+\!4\!-\!n\right)\Gamma^{(n)}(s\vec{p};\mu,m^2,\lambda)=0\,,
\end{equation}
supplemented by (\ref{betafunctions}-\ref{betafunctionsend}). This is \emph{Callan-Symanzik equation}. The interpretation is that the same $n$-points function, considered at different energy scales, changes only by a rescaling. In a classical theory, we could have written immediately  the solution:
\be
\Gamma^{(n)}(s\vec{p},\tilde{m},\tilde{\lambda},\mu)=s^{\rm classical\ dim.}\;\Gamma^{(n)}(\vec{p},\lambda,\tilde{m},\mu)\;,
\ee
Remarkably a similar scaling holds in the quantum case. Once we find the $\b$-functions perturbatively from the counterterms by imposing the equality of (\ref{eq:bareact1}) and (\ref{eq:bareact2}) \cite{thooft73,weinberg73}, we can integrate (\ref{betafunctions}), and finally use them to solve (\ref{scalingequation}) by the method of characteristics (e.g. \cite{collins} \S 7.3), yielding the scaling
\begin{equation}
\label{eq:CSsolution}
\Gamma^{(n)}(s\vec{p},\tilde{m},\tilde{\lambda},\mu)=s^{4-n}e^{-\frac{n}{2}\int_1^sdt\frac{\eta_\phi(\lambda(t))}{t}}\;\Gamma^{(n)}(\vec{p},\lambda(s),\tilde{m}(s),\mu)\;,
\end{equation}
where the extra exponent gives the anomalous dimension, as is immediate to see in the simple case when $\eta_\phi$ is constant. Observe that, for a scalar theory in four dimension, the presence of an anomalous scaling is particularly relevant for the $4$-points function, which has zero classical dimension.

\subsection{Renormalizable and non-renormalizable theories}\label{sec:powercounting}
Let us briefly recall a criterion for perturbative renormalizability, which we mentioned to justify the absence of six-legged 1PI divergent graphs in  $\lambda\phi^4$. Consider a scalar theory in arbitrary dimension $D$, of the form
\begin{equation}
S=\int d^Dx \left[\frac{1}{2}\phi\left(-\Delta+m^2\right)\phi+g\, \phi^{N} \right]\;,
\end{equation}
and a 1PI graph with $L$ loops, $V$ vertices, $I$ internal lines and $E$ external lines. Let the \emph{superficial degree of divergence} $\delta$ of a graph be given by $D$ (from the integral $\int d^Dp$) times the number of loops, minus two  (in the scalar case) times the number of propagators appearing in the integration. It is not hard to see that then
\begin{equation}\label{primitivedivergence}
\delta=D-\frac{1}{2}(D-2)E+V\left(\frac{N-2}{2}D-N\right)\;.
\end{equation}

It can happen that a graph converges better than what its \emph{superficial} degree of divergence would suggest\footnote{This frequently happens in presence of symmetries, an example being the light-by-light scattering in Quantum Electrodynamics.}. However, by a theorem of Weinberg, if $\d<0$ the graph is convergent. Therefore we can give the following \emph{perturbative and superficial} classification of quantum field theories, based on proliferation of divergent graphs:
\begin{enumerate}
\item  Non-renormalizable, when $\frac{N-2}{2}D-N\geq0$.
\item Renormalizable, when  $\frac{N-2}{2}D-N=0$, so that $\delta$ does not depend on $V$; this is the familiar case of $\lambda\phi^4$ in four dimensions.
\item Super-renormalizable if $\frac{N-2}{2}D-N<0$; in this case the UV behavior gets better and better in higher loops diagrams, and there are only a finite number of divergent ones.
\item Finite, if there are no divergent diagrams at all.
\end{enumerate}

What is special about renormalizable theories? Let us rephrase (\ref{primitivedivergence}) in terms of the mass dimension of the couplings. The coupling $g^{(N)}$ for the $g^{(N)} \phi^N$ term has mass dimension\footnote{It is convenient to define the operators $\mathcal{O}$ to include the $x$-integral, so that $\mathcal{O}_{\rm kin}=\frac{1}{2}\int d^4x \partial_\mu \phi\partial^\mu\phi$ is dimensionless.}
\begin{equation}
\Delta_N=[g^{(N)}]=D-\frac{D-2}{2}N\;,
\end{equation}
and it is immediate to see that if $\Delta_N<0$ the theory is non-renormalizable. This can easily be extended to more physical theories involving fermions and gauge fields, see \cite{weinbergQFT} \S 12.1. Therefore if we consider a general theory, allowing for any term compatible with the symmetries, we will have an action of the form
\begin{align}
\label{eq:gaussFP}
S=S_ \mathrm{ren}+S_ \mathrm{nonren}=\sum_{i\in\mathrm{ren}} g_{\mathrm{R}}^{(i)} \mathcal{O}_i[\phi]+\sum_{j\in\mathrm{nonren}} g_{\mathrm{NR}}^{(j)}\mathcal{O}_j[\phi]\;,
\end{align}
where $\Delta_j=[g_{\mathrm{NR}}^{(j)}]<0$. If we now assume that there is a common (large) energy scale $M$ such that 
$g_{\mathrm{NR}}^{(j)}\approx M^{\Delta_j}$,
we automatically get that, when considering processes of energy $s^2\ll M^2$, the effect of irrelevant operators on physical observables is suppressed by $(M^2/s^2)^{\Delta_i/2}\ll1$. Therefore a first reason for the importance of renormalizable theories is that at sufficiently low energy they capture the leading physical effects to a good accuracy. Furthermore, they can be defined by measuring a small number of physical parameters $g_{\mathrm{R}}^{(i)}$, and hence are very predictive. Irrelevant operators can be taken into account as higher order effects, which however require fixing additional parameters \cite{burgess07,burgess09}.

This does not mean that non-renormalizable theories are useless. In fact, when the number of renormalizable couplings is reduced by symmetries, even to zero, it can be important to consider the lowest order non-renormalizable ones, as it happens in chiral perturbation theory, which has a remarkable experimental success describing quark bound states (e.g. pions) at low energies \cite{leutwyler}. However, when we reach the threshold energy $\sim M$, any possible coupling has to be considered, and the theory is no longer predictive. Usually this is regarded as a breakdown of the theory, meaning that it has to be replaced with a more fundamental one. For the effective theory of pions, the fundamental theory is Quantum Chromodynamics, but in principle nothing forbids that the fundamental theory cannot even be expressed as a Quantum Field Theory.

\begin{exercise}[Superficial degree of divergence]
Using the topological properties of Feynman graphs for a scalar theory, prove~(\ref{primitivedivergence}). Drawing some examples will help you.
\end{exercise}

\subsection{Behavior at fixed points}
\label{sec:betafunctions}

The two previous sections indicated that the renormalization techniques in QFT have similar features to the dynamical systems we described earlier. In fact, equation (\ref{eq:gaussFP}) looks a lot like the subdivision of eigenvectors of a hyperbolic fixed point into relevant and irrelevant, where $\Delta_j$ plays the role of critical exponents, specifying how the corresponding operator behave as the energy scale $s$ is varied, depending on their sign.

On top of that, Callan-Symanzik equation  (\ref{scalingequation}) can be added to the previous qualitative discussion to specify the case where the classical dimension is zero. In that case it reads (if we also neglect the mass)
\bea
0=\left(-s\frac{\partial}{\partial s}+\beta_\lambda \frac{\partial}{\partial \lambda}-2\eta_\phi\right)\Gamma^{(4)}=\left(-s\frac{\partial}{\partial s}+\beta_{\hat{\lambda}} \frac{\partial}{\partial \lambda}\right)\Gamma^{(4)}\,.
\eea
where we have introduced a $\mathcal{Z}_\phi$-rescaled coupling $\hat{\lambda}$.\footnote{This actually is not important for $\lambda\phi^4$, because at small coupling $\eta_\phi$ is sub-leading in $\lambda$.}  At vanishing coupling the $\b$-function is zero. We see that the stability property of the fixed point as $s$ increases depend on the sign of the $\b$-function.
The fixed point  (FP) that we are talking about is obviously the free theory that as we have checked explicitly does not get any quantum correction, and therefore does not renormalize. Of course the stability analysis could in principle be done at \emph{any} fixed point, but our perturbative techniques allow us only to deal with the free (or Gaussian) one.

\begin{figure}[!t]
  \begin{center}
    \subfigure[$\beta(g)>0$ and increasing.]{\label{fig:betalandau}\includegraphics[width=4cm, angle=-90]{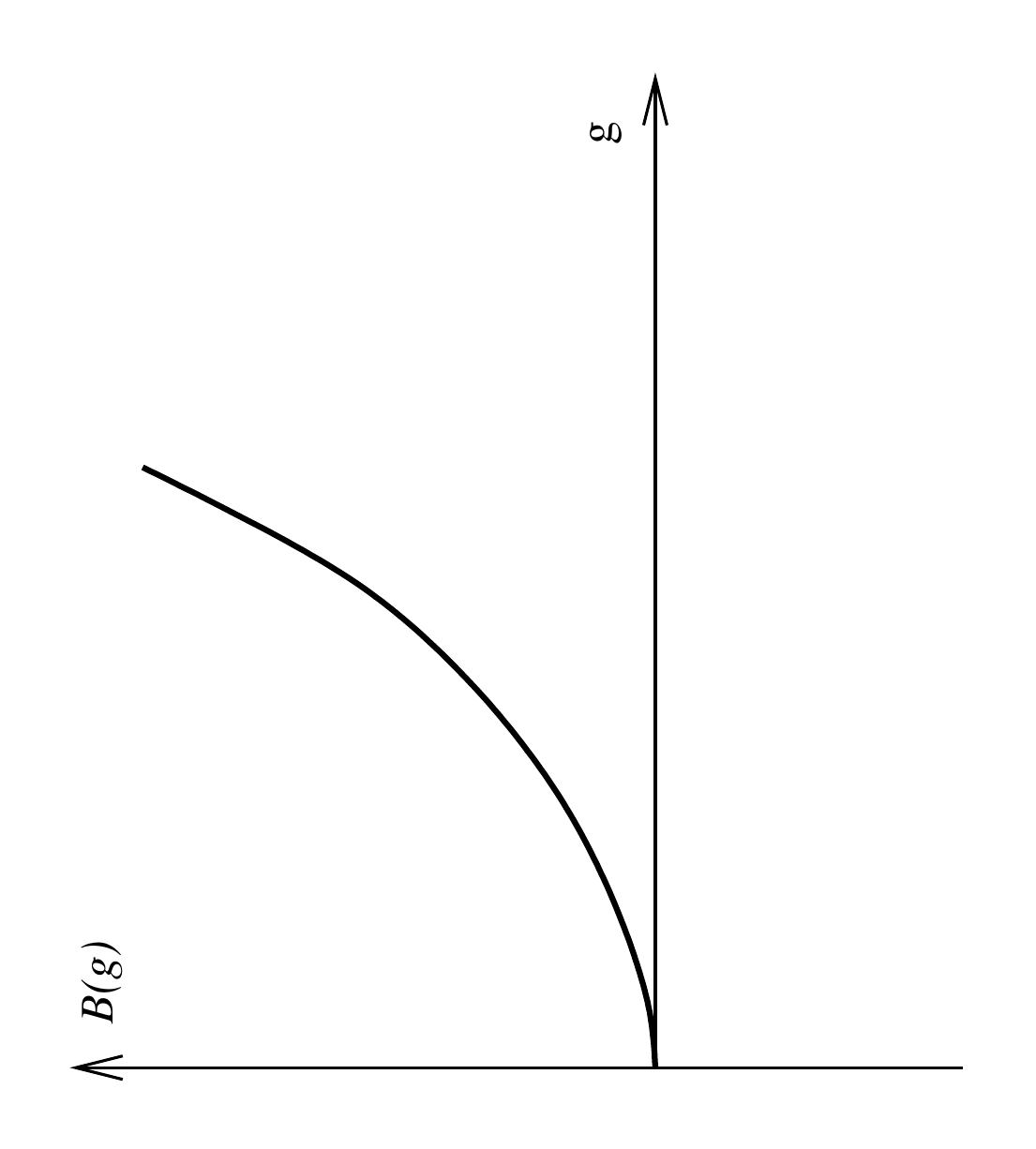}}
    \subfigure[$\beta(g)<0$ and decreasing.]{\label{fig:betafree}\includegraphics[width=4cm, angle=-90]{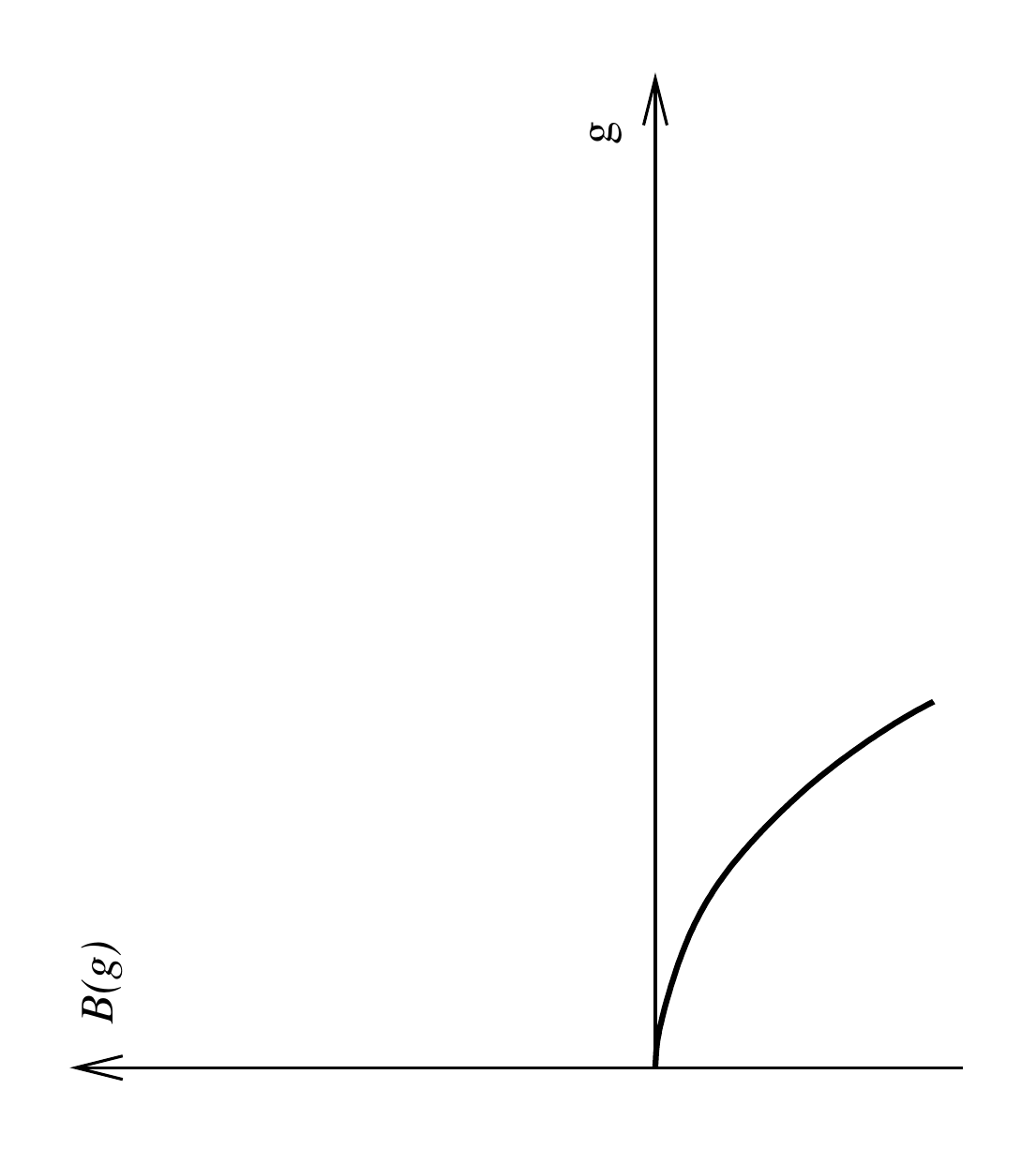}}
     \subfigure[Non Gaussian FP.]{\label{fig:betasafe}\includegraphics[width=4cm, angle=-90]{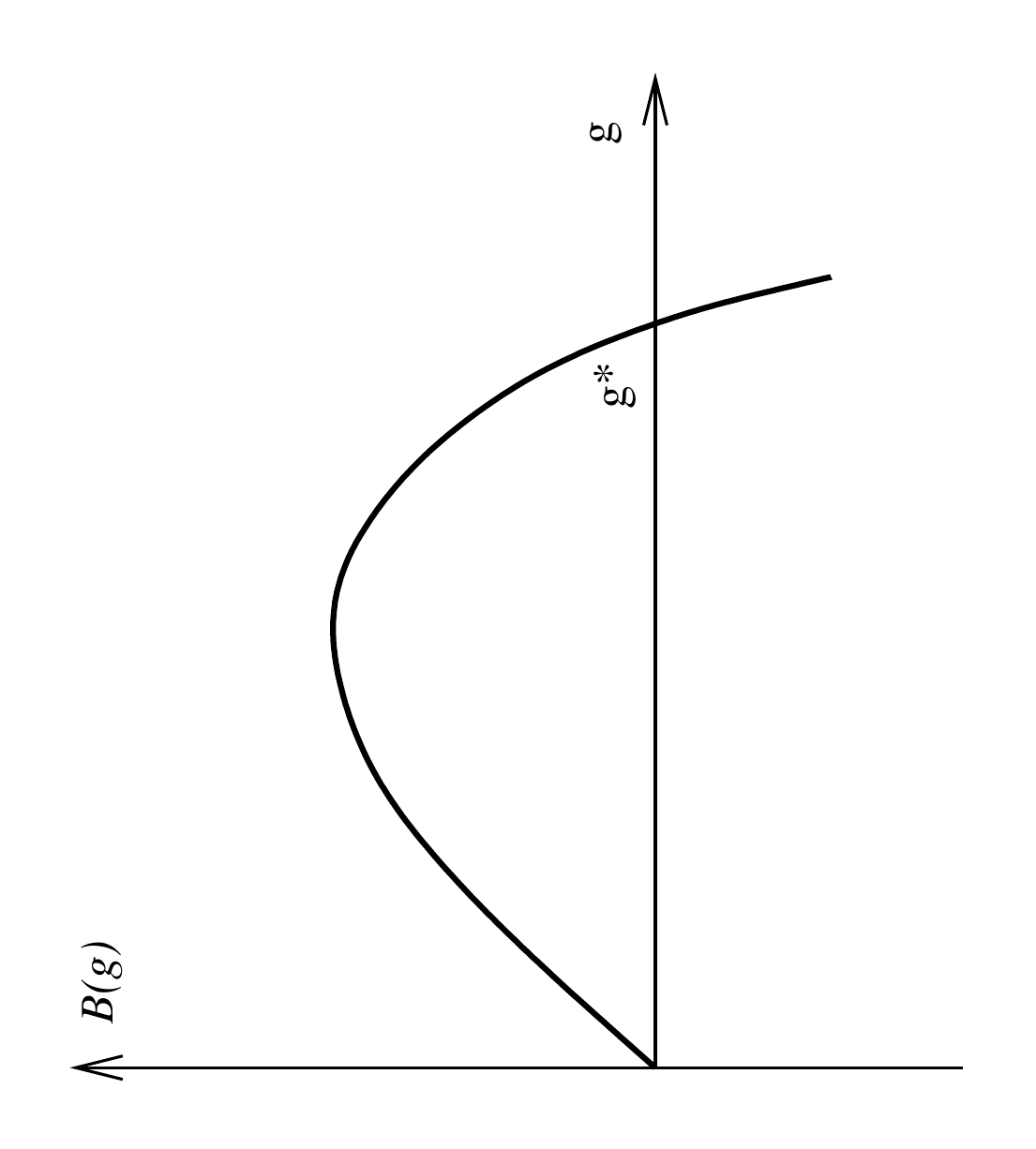}}

    \end{center}
  \caption{Possible qualitative behaviors of $\beta(g)$.}
  \label{fig:betaqualitative}
\end{figure}

Let us comment a bit on the possible behaviors, based on the sign of the $\b$-functions. The most interesting couplings to study perturbatively are the ones having zero mass dimension at the Gaussian fixed point (GFP), the \emph{marginal couplings} for which quantum corrections are crucial. Suppose to consider a theory with some renormalizable couplings and a single marginal one, $g=\tilde{g}$. To classify behavior of the theory close to the fixed point we can assume $\frac{d}{dt}g=\beta(g)$, provided that we restrict to energy scales much larger than the mass. We assume that the physical regime for the coupling is $g\geq0$. Qualitatively, we have
\begin{enumerate}
\item Figure \ref{fig:betalandau}, $\beta(g)>0$. In this case, $g(t)$ increases toward high energies. Even if we measure the coupling to be small at low energies (as happens for the fine structure constant $\alpha_{\rm em}$), we have to expect a UV regime where perturbation theory is not be applicable. On the other hand, in the IR the theory would just sink in the Gaussian fixed point. Remark that if $\beta(g)$ grows with $g$, it is possible that the solution of the differential equation does not exists for any $t$, i.e. the coupling blows up to infinity in a finite time.

\item Figure \ref{fig:betafree}, $\beta(g)<0$. Since it is the same scenario up to one sign, the preceding discussion is valid up to swapping the IR and the UV limits. In this case, the coupling would become smaller and smaller at high energies, and the theory is \emph{asymptotically free}, which is the case of e.g. QCD.

\item Figure \ref{fig:betasafe}. $\beta(g^*)=0$, $g^*\neq0$. We have two possibilities, depending on the sign of $\beta(\lambda)$. Let us take it to be positive as in figure. In this case, taking an initial condition $g_0<g^*$ the coupling would grow at high energies, approaching $g^*$. The theory would be well behaved both in the UV and the IR, even if, depending on how large $g^*$ is, it may not be possible to describe it just perturbatively.
The same picture holds when $\beta(g)<0$, up to swapping IR and UV.  We call these \emph{asymptotically safe theories}.

\end{enumerate}

A generic dynamical system can have a much richer behavior. If we consider many coupled $\beta$-functions, limit cycles or even chaotic behavior may arise. For certain classes of QFTs it is possible to constrain these behaviors \cite{zamolodchikov,komargodski}, at least to some extent \cite{curtright}; this is an active topic of research \cite{glazek02,glazek04,fortin}, see also \cite{moskovic}.

\section{Wetterich's FRGE for scalar fields}
Here we will introduce the Functional Renormalization Group Equation (FRGE) due to Wetterich \cite{wetterich91,wetterich93},  in the case of a scalar field. While it is not hard to formally derive the FRGE, it is more involved to understand how to apply it.

Roughly speaking, the idea is to obtain the $\beta$-functions from the change in the effective action $\Gamma$ computed up to two different scales. Rather than considering all quantum fluctuations, we want to integrate over the ultraviolet (UV) degrees of freedom down to the scale $k$:
\begin{equation}
Z_k\approx\int \prod_{|p|>k}d\phi(p)\;e^{-S[\phi]}\,,
\end{equation}
which is however ill-defined in the UV, so that we must put a UV cutoff at $\Lambda$
\begin{equation}
Z_{k,\Lambda}\approx\int \prod_{k<|p|<\Lambda}d\phi(p)\;e^{-S[\phi]}\,.
\end{equation}
While the above expression depends on the unphysical cutoff $\Lambda$, its variation with respect to $k$ does not
\be
\partial_k Z_k \approx\; Z_{k+\delta k,\,\Lambda}-Z_{k,\,\Lambda}
\approx\int\! \prod_{k<|p|<k+\delta k}\!\!d\phi(p)\;e^{-S[\phi]}\,,
\ee
and could be used to investigate physical properties.

The above argument is only qualitative, since working directly on the path integral measure is unpractical, and we would also need to consider the presence of a current to generate the whole theory. However, it is possible to find a differential equation for the scale dependent family of functionals $\Gamma_k[\varphi]$, i.e. to express the equivalent of the right hand side of the above equation in presence of a current in a nice way. More extensive reviews on this subject are \cite{bergestetradis, bagnuls, gies06, pawlowski}; interesting complementary reading on somewhat different approaches is \cite{wilsonkogut,cardy} and \cite{rosten10}.

\subsection{Derivation of the FRGE}\label{sec:scalarGWderivation}
\begin{figure}[!b]
  \begin{center}
    \subfigure[The regulator $R_k(p^2)$ for different shape functions. Dotted line: polynomial, $\alpha=1$. Light solid: exponential, $\beta=2$. Dark solid: exponential, $\beta=1$. Dashed: optimized.]{\label{fig:regulatorRG}\includegraphics[width=0.48\textwidth]{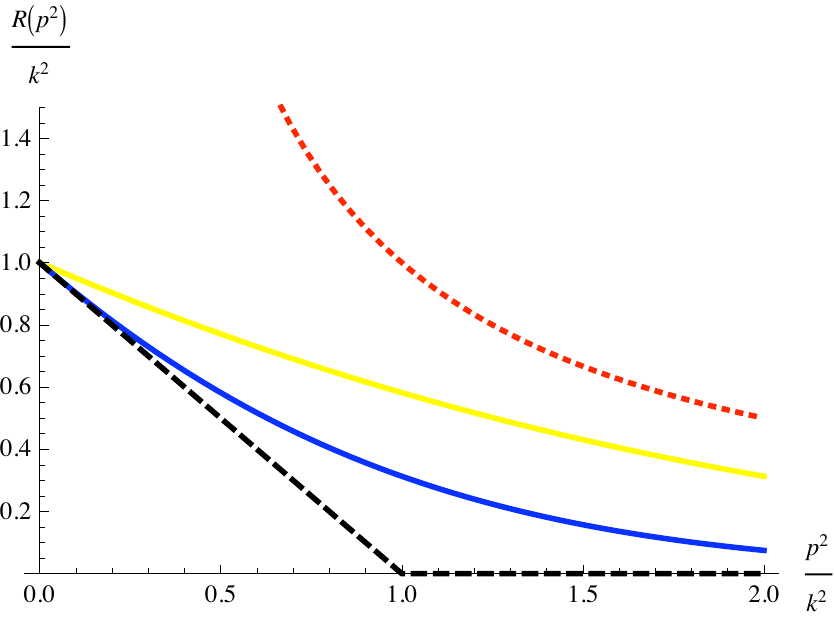}}
    \subfigure[The (normalized) modified propagator $F(p^2)$ for different shape functions, as in the left panel. The grey thin line is the unmodified propagator.]{\label{fig:propagatorRG}\includegraphics[width=0.48\textwidth]{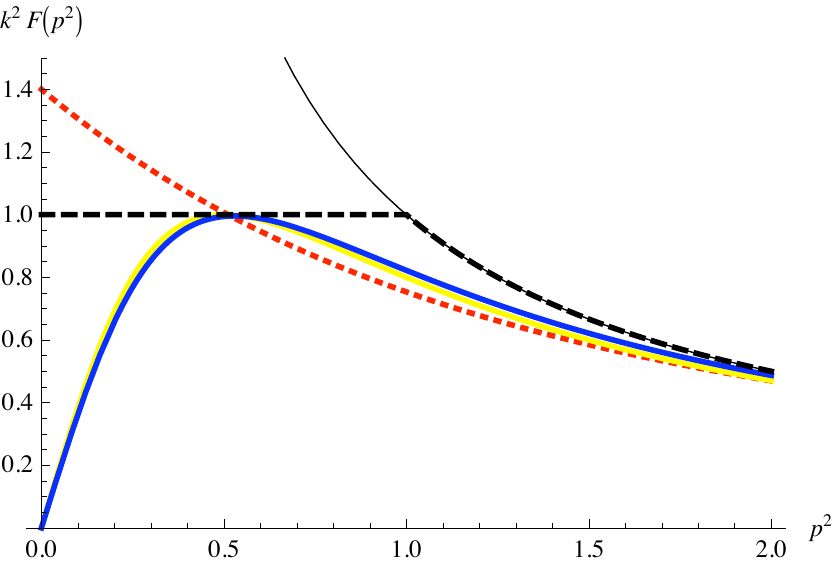}}
  \end{center}
  \caption{Regulator and regularized propagator for different shape functions.}
  \label{fig:regulators}
\end{figure}

In all that follows, we will deal with a single scalar field, but everything can be straightforwardly extended to multiplets of fields. We start from the UV cut-off generating functional
\begin{equation}
e^{-W_\Lambda[J]}=\frac{1}{\mathcal{N}_\Lambda}\int\mathcal{D}_\Lambda \phi\, e^{-S[\phi]+J\cdot\phi}\;,
\end{equation}
where $\mathcal{D}_\Lambda \phi$ is the functional measure with a formal UV cut-off at energy scale $\Lambda$, $\mathcal{N}_\Lambda=\int\mathcal{D}_\Lambda \phi e^{-S_{\rm free}[\phi]}$ and we introduced the short hand $J\cdot \phi=\int d^Dx\,J(x)\,\phi(x)$. We obtain from it a one parameter family of generating functionals
\begin{equation}\label{Wk}
e^{-W_{k,\Lambda}[J]}=\frac{1}{\mathcal{N}_{k,\Lambda}}\int\mathcal{D}_\Lambda \phi\, e^{-S[\phi]+J\cdot\phi-\Delta S_k[\phi]}\;,
\end{equation}
where now $\mathcal{N}_{k,\Lambda}=\int\mathcal{D}_\Lambda \phi e^{-S_{\rm free}[\phi]-\Delta S_k[\phi]}$. $\Delta S_k[\phi]$ is a modification of the action which suppresses low energy modes ($E\lesssim k$) in the path integral. Instead of using a sharp cutoff in the functional measure, this can be done by setting
\begin{equation}
\Delta S_k[\phi]=\!\int\! \frac{d^Dp}{2(2\pi)^D}\,\phi(p)\, \mathcal{R}_k(p)\,\phi(-p)=\!\int\! \frac{d^Dp}{2(2\pi)^D}\,\phi(p)\; k^2\, r(p^2/k^2)\,\phi(-p).
\end{equation}
The regulator $\mathcal{R}_k(p)$ is determined by the choice of the shape function $r(z)$. The requirements on $r(z)$ are to be monotonic and to satisfy
\begin{align}
r(0)>0,\ \ \ \ \ &\lim_{p^2/k^2\to0}\mathcal{R}_k(p)>0;\\
\lim_{z\to\infty}r(z)=0,\ \ \ \  & \lim_{k^2/p^2\to0}\mathcal{R}_k(p)=0;\\
r(z)>0,\ \ 0\leq z\lesssim 1,\ \ \ &  \lim_{k\to\Lambda\to\infty}\mathcal{R}_k(p)=\infty.
\end{align}
This is to ensure that $\mathcal{R}_k$ implements a IR cutoff, which is a sort of mass term if $r(z)<\infty$; then we impose that the regulator vanishes when $p^2$ is in the UV; it is also intended that not only $r(z)$ has to become very small when $z\gtrsim1$, but also the decay to zero should be fast. Finally, when we send $k\to\Lambda\to\infty$, we require the path integral to be dominated by the quadratic part; this can allow to set a sort of initial condition on the flow equation.
Particular choices of $r(z)$ can be of the form
\begin{align}
\mathrm{polynomial}\ \ \ \ &r(z)=z^{-\alpha},\ \ \alpha\geq0\;,\\
\mathrm{exponential}\ \ \ \ &r(z)=\frac{z}{e^{z^\beta}-1},\ \ \beta\geq1\;,\\
\label{litim}
\mathrm{optimized}\ \ \ \ &r(z)=(1-z) H(1-z)\;,
\end{align}
where $H$ is Heaviside's step function. The last cutoff, due to Litim is called \lq\lq optimized\rq\rq\ in a sense described in \cite{litim00,litim01,litim01b}, and we will be using it extensively. The different cutoffs are plotted in Figure \ref{fig:regulators}, together with the respective shapes of the regularized propagators for $k=1$
\begin{equation}
F(p^2)=\frac{1}{\mathcal{R}_{k=1}(p^2)+p^2}\;.
\end{equation}
However, provided that the regulators meet the mentioned requirements, the derivation of Wetterich equation does not depend on details of $r(z)$.

From now on, we will always assume the presence of an UV cutoff $\Lambda$, but just write $W_k\equiv W_{k,\Lambda}$ (and similarly $\mathcal{N}_k$, $\Gamma_k$, etc). We consider the Legendre transform of $W_k[J]$, which is written in term of the $k$-dependent mean field $\varphi(x)=-\frac{\delta W_k[J]}{\delta J(x)}$:
\begin{equation}
\tilde{\Gamma}_k[\varphi]=\inf_{J}\left[W_k[J]+\int d^Dx\; \varphi(x)J(x)\right]\;.
\end{equation}
To understand what this is, we can evaluate it exactly in the free case, finding 
\begin{align}
W_k[J]&=-\frac{1}{2}\int d^Dx\,d^Dy\; J(x)\, C_{k}(x-y)\,J(y)\;,\\
C_{ k}(z)&=\frac{1}{(2\pi)^D}\int d^Dp \frac{e^{ip\cdot z}}{p^2+m^2+k^2\;r(p^2/k^2)}\;,\\
\tilde{\Gamma}_k[\varphi]&=\frac{1}{2}\int d^Dx\,\varphi(x)\left(-\Delta+m^2+k^2r(-\Delta/k^2)\right)\varphi(x)\;.
\end{align}
To cancel the explicit $k$-dependence from $\tilde{\Gamma}_k[\varphi]$ (which in the free case is the only dependence), we define
\begin{equation}
\Gamma_k[\varphi]=\tilde{\Gamma}_k[\varphi]-\Delta S_k[\varphi]\;.
\end{equation}
We retrieve the full quantum effective action when $k\to0$, when $\Delta S_k$ vanishes. In the opposite limit,  by saddle-point method  we have that $\Gamma_\infty[\varphi]\approx S[\varphi]$.

With this definition, we can find the one loop effective action $\Gamma_k^\mathrm{1loop}[\varphi]$ for a generic interacting theory by saddle point approximation:
\begin{equation}
\Gamma_k^\mathrm{1loop}[\varphi]=S[\varphi]+\frac{1}{2}\mathrm{Tr}\left[\log
\left.\frac{\delta^2 (S[\phi]+\Delta S_k[\phi])}{\delta\phi\,\delta\phi}\right|_{\phi=\varphi}\right]\;,
\end{equation}
which is an infrared modified version of the familiar expression and reduces to that when $k\to0$. Its differential form is inspiring:
\begin{equation}\label{wetterich1loop}
k\frac{d}{d k}\Gamma_k^\mathrm{1loop}[\varphi]=\frac{1}{2}\mathrm{Tr}\left[k\frac{d \mathcal{R}_k}{d k}\,
\left(\left.\frac{\delta^2 (S[\phi]+\Delta S_k[\phi])}{\delta\phi\,\delta\phi}\right|_{\varphi}\right)^{-1}\right]\;.
\end{equation}
From this we can extract the one-loop $\beta$-functions of the theory similarly to what one does in Wilsonian renormalization. We must expand the one-loop effective action on a basis of operators, assigning a coupling to each of them:
\begin{equation}
\Gamma_k^\mathrm{1loop}[\varphi]=\sum g^{(i)}_k \mathcal{O}^{(i)}[\varphi]\;.
\end{equation}
Some of them will be the ones already appearing in the action, but most of them will be new. Then we have to compute the trace on the r.h.s., and match the terms appearing in the action with the ones coming from the trace, obtaining
\begin{equation}
k\frac{d g^{(i)}_k}{d k}=\beta_0^{(i)}(g_k),\ \ \ \ k\frac{d \tilde{g}^{(i)}_k}{d k}=-\Delta_i\tilde{g}^{(i)}_k+\beta_0^{(i)}(\tilde{g}_k)\equiv\beta^{(i)}(\tilde{g}_k)\;,
\end{equation}
where in the latter expression we consider dimensionless couplings $\tilde{g}^{(i)}_k$\footnote{A more physical analysis would require appropriately scaling each coupling by powers of the field strenght renormalization $\mathcal{Z}^{1/2}$.}, where $\Delta_i$ is the mass dimension. Remarkably, this can be extended to all-loops.

We will introduce the renormalization time $t=\log(k/\Lambda)$, so that $\frac{d}{dt}=k\frac{d}{dk}$. Differentiating (\ref{Wk}), we have the identities
\begin{equation}
\left.\frac{\partial W_k[J]}{\partial t}\right|_{J\ \mathrm{fixed}}=\left\langle\frac{d \Delta S_k\left[\phi\right]}{dt}\right\rangle_J=\frac{1}{2}\mathrm{Tr}\left[\langle \phi\phi\rangle_J \frac{d \mathcal{R}_k}{dt}\right]\;,
\end{equation}
where the trace stands for integration and, if we were to consider multiplets of fields, summation on internal indexes. Then, recalling that $\varphi=\langle\phi\rangle$ and using the straightforward generalization of (\ref{GammaWrel})
\begin{equation}
\frac{\delta^2\tilde{\Gamma}_k[\varphi]}{\delta\varphi(x)\,\delta\varphi(y)}=-\left.\left[\frac{\delta^2 W_k[J]}{\delta J(x)\,\delta J(y)}\right]^{-1}\right|_{J=J_\varphi},
\end{equation}
where $J_\varphi$ is the current used in the Legendre transform, we have
\begin{align}%\label{beginwetterich}
\nonumber
\frac{d \Gamma_k[\varphi]}{dt}&=\frac{\partial W_k[J_\varphi]}{\partial t}+\int d^Dx \frac{\delta W[J]}{\delta J(x)}\frac{dJ_\varphi(x)}{dt}+\int d^Dx \varphi(x)\frac{dJ_\varphi(x)}{dt}-\frac{d\Delta S_k[\varphi]}{dt}=\\
\nonumber
&=\frac{\partial W_k[J_\varphi]}{\partial t}-\frac{d\Delta S_k[\varphi]}{dt}=\frac{\partial W_k[J_\varphi]}{\partial t}-\frac{1}{2}\mathrm{Tr}\left[\varphi \frac{d\mathcal{R}_k}{dt}\varphi\right]=\\
\nonumber
&=\frac{1}{2}\mathrm{Tr}\left[(\langle \phi\phi\rangle_{J_\varphi}-\langle\phi\rangle_{J_\varphi}^2) \frac{d \mathcal{R}_k}{dt}\right]=-\frac{1}{2}\mathrm{Tr}\left[\left.\frac{\delta^2 W[J]}{\delta J\delta J}\right|_{J_\varphi} \frac{d \mathcal{R}_k}{dt}\right]=\\
\nonumber
%\label{minussign}
&=\frac{1}{2} \mathrm{Tr}\left[\left(\frac{\delta^2\tilde{\Gamma}[\varphi]}{\delta\varphi\delta\varphi}\right)^{-1} \frac{d \mathcal{R}_k}{dt}\right]=\frac{1}{2} \mathrm{Tr}\left[\left(\frac{\delta^2\Gamma[\varphi]}{\delta\varphi\delta\varphi}+\mathcal{R}_k\right)^{-1} \frac{d \mathcal{R}_k}{dt}\right]=\\
&=\frac{1}{2} \mathrm{Tr}\left[\frac{\frac{d }{dt}\mathcal{R}_k}{\mathcal{R}_k+\Gamma^{(2)}[\varphi]}\right]\;.\label{wetterich}
\end{align}
The above Functional Renormalization Group Equation (FRGE), due to Wetterich, is a differential equation for the one-parameter family of functionals $\Gamma_k$. Its solution $\Gamma_k$ describes the flow of the effective action (hence the name \emph{flow equation}) in the \emph{theory space} under changes of the cutoff scale. The theory space is the space of all functionals of the fields compatible with the symmetries of the theory (the analog of the space $\U$ for maps on the interval). For instance, if we assume the theory of a single real scalar field has a symmetry for $\phi\leftrightarrow-\phi$ a sensible definition of a theory space could be
\begin{equation}\label{scalarT}
\mathcal{T}=\mathrm{span}\left\{(-\Delta)^p\phi^{2n}:\;p\geq0,n\geq1\ \ \mathrm{and\ permutations}\right\}.
\end{equation} 

Some further remarks are in order: first, the above equation is formally \lq\lq exact\rq\rq (in the sense discussed), and in particular accounts for arbitrarily high loop effects. This may seem a bit surprising, because (\ref{wetterich}) has a one-loop structure. However, the equation is exact \emph{in the full theory space}. Let us illustrate what this means with an example: in the theory defined by (\ref{scalarT}), we denote by $\lambda^{(n,p,\sigma)}$ the couplings corresponding to $n$ powers of $\phi$ and $p$ laplacians (the permutations are labelled by $\sigma$)\footnote{To make contact with the familiar $\lambda\phi^4$ notation, $\lambda\mathcal{Z}^2=\tilde{g}^{(4,0)}$, $\tilde{m}^2\mathcal{Z}=\tilde{g}^{(2,0)}$ and $\mathcal{Z}=\tilde{g}^{(2,1)}$. This will become hopefully clearer in the next section.}. Then we would have an infinite system of $\beta$-functions 
\begin{equation}\label{manybetafunctions}
\left\{\begin{array}{l}
\frac{d}{dt}\lambda^{(2,0)}=\beta_{2}(\lambda^{(2,p,\sigma)},\lambda^{(4,p,\sigma)})\;,\\
\frac{d}{dt}\lambda^{(4,0)}=\beta_{4}(\lambda^{(2,p,\sigma)},\lambda^{(4,p,\sigma)},\lambda^{(6,p,\sigma)})\;,\\
\dots\ ,
\end{array}\right.
\end{equation}
so that even if the beta function for the $\varphi^2$ coupling does not explicitly depend on the coupling of $\varphi^6$, it does implicitly via e.g. $\lambda^{(4)}$. Clearly the exact solution of (\ref{manybetafunctions}) is very difficult$-$actually as difficult as the exact solution of the quantum field theory, since we can imagine the quantum theory to be implicitly defined by (\ref{wetterich}), since $\Gamma$ contains all the 1PI diagrams. Some approximation will be needed,  but \emph{this needs not to be perturbation theory}.

We completely lost track of the cutoff $\Lambda$. Indeed the flow equation does not contain it and appears well defined, as the trace converges both at large and small momenta due to the regulators. However when we integrate it with respect to $t$, we have no guarantee that the resulting flow does not blow up. If, however, we find that the flow equation can be integrated up to arbitrary high scales, no problem arises. This is the case of what we have defined in Section \ref{sec:betafunctions} as \emph{asymptotically safe} theories. In that case, in $\mathcal{T}$ there will be two fixed points, and a certain number of theories that flow from one to another under the RG flow. These span a critical manifold, whose dimension counts the physical parameters that  must be measured to make a prediction.
When the theory is defined at the scale $k_0$ by setting the physical values of couplings $g^{(1)}(k_0),\dots g^{(m)}(k_0)$ corresponding to operators $\mathcal{O}_1,\dots \mathcal{O}_m$, we have to expect that at a different scale the theory is defined in terms of a possibly infinite number of operators $\mathcal{O}_i$, whose couplings are determined from $g^{(1)}(k_0),\dots g^{(m)}(k_0)$.\footnote{In a geometrical picture, this corresponds to saying that the integral curve $\Gamma_k\in\mathcal{T}$ needs not to be contained in any finite linear subspace of $\mathcal{T}$.}

Therefore, Wetterich equation can be useful both as an alternative method to find approximate expression for the $\b$-functions and as a tool to study the RG properties of functions that seem not to be renormalizable perturbatively.

\subsection{Expansions and truncations}
In practice it is not possible to address the full equation (\ref{wetterich}), and one has to adopt some approximation scheme, which generally relies on the choice of an ansatz. A natural way to do so is to expand it on a basis of the theory space, for instance taking the monomials appearing in the definition (\ref{scalarT}) as basis vector, and retain only some of those in the ansatz, which \emph{will not be stable under the RG flow}. There are two natural strategies:
\begin{enumerate}
\item A derivative expansion. This amounts to retaining terms involving up to a certain number of derivatives. The rationale for doing this is that derivative operators make the explicit computations of $\beta$-functions much messier, and often already the \lq\lq local potential approximation\rq\rq, which basically amounts to the zeroth order of the derivative expansion together with the kinetic term,
\begin{equation}
\Gamma_k[\varphi]=\int d^Dx \left[-\frac{1}{2}\mathcal{Z}_k\varphi \Delta\varphi+ \sum_{n=1}^\infty \mathcal{Z}^n_k\frac{\lambda^{(2n)}_k}{(2n)!}\,\varphi^{2n}\right]\,,
\end{equation}
can be used to make quite accurate physical predictions. Observe that we have replaced the coupling constants $g_k^{(2n)}\to  \mathcal{Z}^n_k\lambda_k^{(2n)}$. The lowerscript $k$ reminds us that the couplings are scale dependent.

\item Expansion according to mass dimension. In this case, one fixes a mass dimension $-\bar{\Delta}\leq0$, and considers all the terms whose couplings have $[g]\geq\bar{\Delta}$. For instance, the truncation corresponding to  $\bar{\Delta}=2$ in $D=4$ is
\begin{align}\nonumber
\Gamma_k[\varphi]=&\int d^4x \left[-\mathcal{Z}_k\frac{1}{2}\varphi \Delta\varphi+\mathcal{Z}_k\lambda^{2,2}\frac{1}{2}\varphi \Delta^2\varphi-\mathcal{Z}^2_k\lambda^{(4,1)}_k\frac{1}{3!}\varphi^3 \Delta\varphi+\right.\\
\label{gammadim-2}
&\ \ \ \ \left. +\mathcal{Z}_k^2m_k^2\frac{1}{2}\,\varphi^{2}+\mathcal{Z}_k^2\lambda^{(4)}_k\frac{1}{4!}\,\varphi^{4}+\mathcal{Z}^3_k\lambda^{(6)}_k\frac{1}{6!}\,\varphi^{6}\right]\;.
\end{align}
This can be convenient in the vicinity of the Gaussian fixed point,  where the discarded couplings are irrelevant and should have little effect on the flow.
\end{enumerate}

In any of the above cases, however, \emph{the equation is no longer exact}. In particular, any prediction of the flow equation (fixed points, critical exponents, \dots) will depend both on the truncation and on the explicit form of the regulator $\mathcal{R}_k$. A  way to make sense of the predictions is to consider larger and larger truncations, and different regulator schemes, and check which predictions are robust under this changes. Only in exceptional cases it is possible to make statements that take into account the flow in the whole theory space \cite{sfondrini}, and even in such cases it is very hard to provide a rigorous mathematical formalization of the flow equation in $\mathcal{T}$. 

Let us also mention that not all couplings are physical, since changes of variable in the path integral do not modify the physical prediction, but lead to redefinitions of the couplings. Such couplings are called \emph{redundant}. 

\begin{exercise}[FRGE for $n$-points functions]
We can use (\ref{wetterich}) to retrieve the behavior of the $n-$points functions, allowing for a graphical representation in Feynman diagrams. For instance, show that the two-points function is given by
\begin{align}\nonumber
\frac{d}{dt}\Gamma_k^{(2)}=&\mathrm{Tr}\left\{\frac{d\mathcal{R}_k}{dt}\left[\Gamma_k^{(2)}+\mathcal{R}_k\right]^{-1}\Gamma^{(3)}\left[\Gamma_k^{(2)}+\mathcal{R}_k\right]^{-1}\Gamma^{(3)}\left[\Gamma_k^{(2)}+\mathcal{R}_k\right]^{-1}\right\}+\\
&-\frac{1}{2}\mathrm{Tr}\left\{\frac{d\mathcal{R}_k}{dt}\left[\Gamma_k^{(2)}+\mathcal{R}_k\right]^{-1}\Gamma^{(4)}\left[\Gamma_k^{(2)}+\mathcal{R}_k\right]^{-1}\right\}\,.
\end{align}
\end{exercise}

\section{Applications of the FRGE}
We will show how the FRGE we have derived can be used to derive $\b$-functions, comparing those from the ones found by a perturbative approach, and then applying them to non-perturbative questions.

\subsection{Revisiting $\lambda\phi^4$}
Let us consider the ansatz
\begin{equation}
\label{eq:phi4ansatz}
\Gamma_k[\varphi]=\int d^Dx \left[-\frac{1}{2}\mathcal{Z}_k\varphi \Delta\varphi+ \mathcal{Z}_k \frac{m_k^2}{2}\varphi^2 +\mathcal{Z}^2_k\frac{\lambda_k}{4!}\,\varphi^{4}\right]\;.
\end{equation}
Then, equation (\ref{wetterich}) almost coincides with (\ref{wetterich1loop}). The only difference is that on the r.h.s. of the latter we have $S^{(2)}$, whose coupling constants have no explicit $k$-dependence. We now define the regulator in momentum space, in terms of (\ref{litim}):
\begin{align}\label{R_k}
\mathcal{R}_k(p^2)&=\mathcal{Z}_k \left(k^2-p^2 \right)\; H(k^2-p^2)\,.
\end{align}
The flow equation takes the explicit form
\be\label{lpaphi4}
\frac{d}{dt}\Gamma_k[\varphi]
=\frac{1}{2}\mathrm{Tr}\left[\frac{2k^2+\eta(k^2+ \Delta) }{k^2+m^2_k+\mathcal{Z}_k\frac{\lambda}{2}\varphi^2}\; H(k^2+ \Delta)\right]\,,
\ee
where $\eta=\frac{d}{dt}\log\mathcal{Z}_k$ and $\Delta$ disappears from the denominator due to our choice of $\mathcal{R}_k$. For the moment, let us not consider the running of the kinetic term, so that we can put $\eta\approx0$. To recover the field operators of the potential, we can restrict ourselves to constant fields, $\varphi(x)\equiv\varphi_o$. In this way, the trace can be easily computed as a momentum integral. In fact, if $\mathcal{V}_D$ is a $D$-dimensional volume element, we have
 \begin{align}\nonumber
\frac{d}{dt}\Gamma_k[\varphi]&=\frac{1}{2}\int d^Dx\int_{p^2<k^2}\!\! \frac{d^Dp}{(2\pi)^D}\left[\frac{2k^2}{k^2+m^2_k+\mathcal{Z}_k\frac{\lambda}{2}\varphi_o^2}\right]=\\
&=\mathcal{V}_D\frac{1}{(4\pi)^{D/2}\, \Gamma(1+D/2)} \frac{k^{D+2}}{k^2+m^2_k} \left[1-\frac{\mathcal{Z}_k\frac{\lambda}{2}\varphi_o^2}{k^2+m^2_k}+\frac{\mathcal{Z}^2_k\frac{\lambda^2}{4}\varphi_o^4}{(k^2+m^2_k)^2}+O(\varphi_o^6)\right]\,,
\end{align}
where in the last line we have expanded the fraction in powers of the fields.\footnote{Accidentally in this case such an expansion equals one in powers of $\lambda$, but the rationale for the expansion is to be able to match the monomials on the left and right hand side.} Notice how terms of the form $\varphi_o^{2n}$ are generated from any $n$, reminding us that we are only in a small sector of $\mathcal{T}$. From the above we find the beta functions for the dimensionless couplings $\tilde{m}^2_k=m^2_k k^{-2}$, $\tilde{\lambda}_k=\lambda_k k^{D-4}$, where the explicit $k$-dependence has washed out.
\begin{align}
\label{mass4}
\frac{d }{dt}\tilde{m}^2_k=&\beta_{\tilde{m}^2}=-2\tilde{m}^2_k-\frac{1}{(4\pi)^{D}\Gamma(1+\frac{D}{2})} \frac{\tilde{\lambda}}{(1+\tilde{m}^2)^2}\,,\\
\label{lambda4}
\frac{d }{dt}\tilde{\lambda}_k=&\beta_{\tilde{\lambda}}=\,(D-4)\tilde{\lambda}+\frac{3!}{(4\pi)^{D/2}\Gamma(1+\frac{D}{2})} \frac{\tilde{\lambda}^2}{(1+\tilde{m}^2)^3}\,,
\end{align}
which in four dimension gives
\be
\frac{d }{dt}\tilde{m}^2_k\approx(-2+\frac{\tilde{\lambda}}{16\pi^2})\tilde{m}^2_k-\frac{\tilde{\lambda}_k}{32\pi^2}\,, \\
\quad\quad
\frac{d }{dt}\tilde{\lambda}_k\approx\frac{3\tilde{\lambda}^2}{16\pi^2} \,.
\ee
This has to be compared with the familiar one-loop result (see e.g. \cite{ramond}, \S 4.7) from perturbation theory with the mass independent renormalization. There are two manifest differences: first, we find higher order contributions in $\tilde{m}^2=m^2/k^2$ to both $\beta$-functions. Second, there is a discrepancy in the running of the mass term at zeroth order in $\tilde{m}$.

The former difference can be understood as a scheme dependence due to the infrared cutoff, which however does not alter the qualitative behavior of the flow, since $m^2/k^2\ll 1$. In fact, it is easy to check that such depend on the choice of the regulator. The latter discrepancy amounts to a quadratic divergence, as it can be seen in terms of the dimensional quantities, $k\frac{d}{dk} m^2_k\approx {\tilde{\lambda}_k\;k^2}/{16\pi^2}\,.$
 Recall that these do not appear directly in dimensional regularization.

We still have to extract the running of the wavefunction. For this purpose it is not enough to limit ourselves to constant fields. We need an $x-$dependent fluctuation term, such that $\varphi(x)=\varphi_o+\tilde{\varphi}(x)$. Then we cannot straightforwardly perform the momentum integration, because there are a number of differential operators, acting to the right on the $\tilde{\varphi}(x)$. However, we just need to use the commutator
\begin{equation}\label{commrel}
\left[-\Delta,\,\tilde{\varphi}(x)\right]=-\Delta\left(\tilde{\varphi}(x)\right)+2i \partial_\mu \left(\tilde{\varphi}(x)\right)\;i\partial^\mu\,,
\end{equation}
and the cyclic property of the trace to sort them to one side and write formally
\be
\mathrm{Tr}\left[\mathcal{A}(x)\,\mathcal{B}(i\partial)\right]=\sum_{x,p}\langle x| \mathcal{A}(x)|x\rangle\,\langle x|p\rangle\,\langle p|\mathcal{B}(p)|p\rangle\, \langle p|x\rangle\,.
\ee

In the case of a local potential approximation for a scalar field (\ref{lpaphi4}) we can sort all the derivative operators to the right just by the ciclic property of the trace for the terms contributing to the flow of $\eta$, i.e. for the ones quadratic the field, which means that it was actually consistent to set $\eta=0$, consistently with what we know form one-loop perturbation theory.

The higher loops effect come into play through the flow of irrelevent couplings. For instance, it is easy to see that including a term such as  $\int -\varphi^3\Delta\varphi=3\int\varphi^2\,\partial_\mu\varphi\partial^\mu\varphi$ into the ansatz would yield a nonzero $\b$-function for $\eta$. Let us sketch the computation. We consider
\begin{equation}
\Gamma_k[\varphi]=\int d^Dx \left[-\mathcal{Z}_k\frac{1}{2}\varphi \Delta\varphi-\mathcal{Z}^2_k\frac{\lambda^{(4,1)}_k}{3!}\varphi^3 \Delta\varphi+\mathcal{Z}_k^2\frac{m^2}{2}\,\varphi^{2}+\mathcal{Z}_k^2\frac{\lambda^{(4)}_k}{4!}\,\varphi^{4}\right]\;.
\end{equation}
To be completely consistent in $D=4$, we would have to consider a more general action such as (\ref{gammadim-2}), but let us stick to the simplest case for simplicity. Then
\begin{align}\label{floweta}
\frac{d}{dt}\Gamma_k[\varphi]=\frac{1}{2}\mathrm{Tr}\left[\frac{2k^2+\eta(k^2+ \Delta) }{k^2+m^2_k-\frac{1}{2}\mathcal{Z}_k\lambda_k^{(4,1)}\Delta\varphi^2+\mathcal{Z}_k\frac{\lambda_k}{2}\varphi^2}\; H(k^2+ \Delta)\right]\;,
\end{align}
where all the derivative operators act to the right and the cyclicity of the trace is understood. To find the $\beta$-functions of the potential we restrict again to constant configurations $\varphi_o$ (without setting $\eta=0$):
 \begin{align}\nonumber
\frac{d}{dt}\Gamma_k[\varphi]=\frac{1}{2}\int d^Dx& \frac{D\pi^{D/2}}{(2\pi)^D\Gamma(1+\frac{D}{2})}\int_0^k\!dp\,p^{D-1}\times\\
&\times\left[\frac{2k^2+\eta(k^2-p^2)}{k^2+m^2_k+\frac{1}{2}\mathcal{Z}_k\,\lambda_k^{(4,1)}\varphi_o^2p^2+\lambda_k^{(4)}\varphi_o^2)}\right]+O(\tilde{\varphi})\,,
\end{align}
from which, by the usual expansion, we can extract the scaling for the potential:
\begin{align}\nonumber
\frac{d}{dt}\tilde{m}^2_k=(-2-&\eta)\tilde{m}^2_k-\frac{1}{4(4\pi)^{D/2}\,\Gamma(3+D/2)}\;\times\\
&\times \left[(4+D)(2+D-\eta)\frac{\lambda_k^{(4)}}{(1+\tilde{m}^2_k)^2}+D(4+D-2\eta)\frac{\lambda_k^{(4,1)}}{(1+\tilde{m}^2_k)^2}\right]\;,\\
\nonumber
\frac{d}{dt}\tilde{\lambda}^{(4)}_k=(D-&4-2\eta)\tilde{\lambda}^{(4)}_k+\frac{3}{4(4\pi)^{D/2}\,\Gamma(4+D/2)} \left[\frac{\scriptstyle(D+4)(D+6)(D+2-\eta)}{(1+\tilde{m}_k^2)^3}(\lambda_k^{(4)})^2+\right.\\
& \left.+\frac{\scriptstyle 2D(D+6)(D+4-\eta)}{(1+\tilde{m}_k^2)^3}\lambda_k^{(4)}\lambda_k^{(4,1)}+\frac{\scriptstyle D(D+2)(D+6-\eta)}{(1+\tilde{m}_k^2)^3}(\lambda_k^{(4,1)})^2\right]\;.
\end{align}

Finally we compute the two remaining $\beta$-functions, by expanding (\ref{floweta}) in terms of $\varphi(x)=\varphi_o+\tilde{\varphi}(x)$ and using (\ref{commrel}):
\begin{align}
\eta=&-\frac{2(D+2-\eta)}{(2\pi)^{D/2}\,\Gamma(2+D/2)}\frac{\tilde{\lambda}_k^{(4,1)}}{(1+\tilde{m}^2)^2}\;,\\
\nonumber
\frac{d}{dt}\tilde{\lambda}^{(4,1)}_k=&(2D-6-2\eta)\tilde{\lambda}_k^{(4,1)}+\frac{3}{8(4\pi)^{D/2}\,\Gamma(3+D/2)}\times\\
&\times\left[\frac{(4+D)(2+D-\eta)}{(1+\tilde{m}^2_k)^3}\tilde{\lambda}_k^{(4)}\tilde{\lambda}_k^{(4,1)}+\frac{D(4+D-\eta)}{(1+\tilde{m}^2_k)^3}(\tilde{\lambda}_k^{(4,1)})^2\right]\;,
\end{align}
which yields a nonzero flow for $\eta$.

 \begin{figure}[!t]
  \begin{center}
\includegraphics[width=3.4cm, angle=-90]{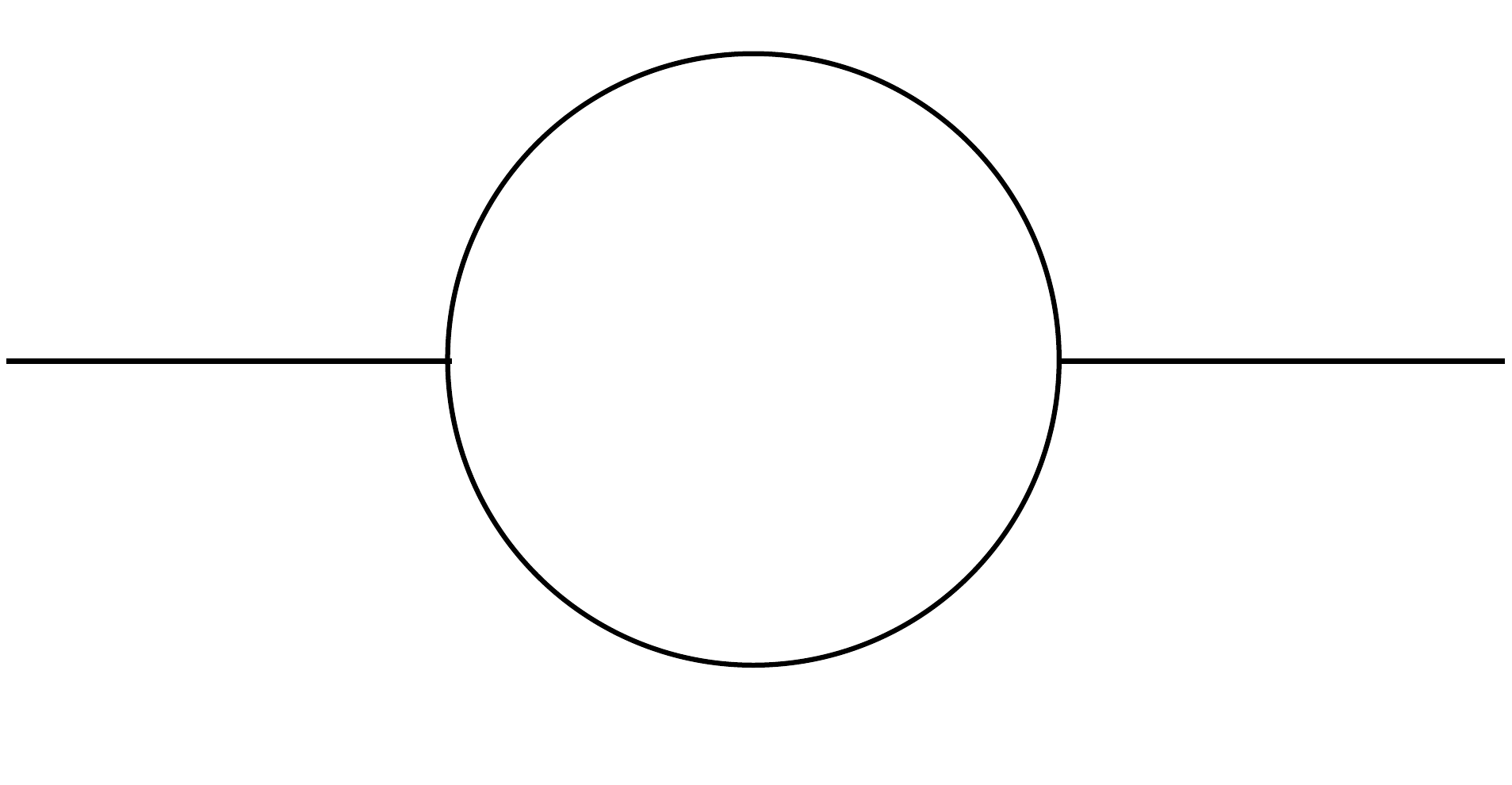}
  \caption{The one-loop contribution to the wavefunction renormalization in $\lambda \phi^3$.}
  \label{fig:phi3}
  \end{center}
\end{figure}

 \begin{exercise}[Regulator dependence]
Repeat the FRGE calculation to find the $\b$-functions for (\ref{eq:phi4ansatz}) using the more general regulator
 \begin{equation}
R_k(p;\;\alpha)=\alpha\;\mathcal{Z}_k \left(k^2-\frac{1}{\alpha}p^2 \right)\; H(k^2-\frac{1}{\alpha}p^2)\,,\quad\quad\a>0\,,
\end{equation}
 to show that the quadratic divergences are non-universal. You should find
 \begin{align}
\frac{d }{dt}\tilde{m}^2_k=&-2\tilde{m}^2_k-\frac{1}{32\pi^2} \frac{\alpha^3\tilde{\lambda}}{(\alpha+\tilde{m}^2)^2}\;\approx\;(-2+\frac{\tilde{\lambda}}{16\pi^2})\tilde{m}^2_k-\frac{\tilde{\lambda}_k}{32\alpha\pi^2}\,, \\
\frac{d }{dt}\tilde{\lambda}_k=&\,\frac{3}{16\pi^2} \frac{\alpha^3\tilde{\lambda}^2}{(\alpha+\tilde{m}^2)^3}\;\approx\;\frac{3\tilde{\lambda}^2}{16\pi^2}\,. 
\end{align}
 \end{exercise}
 
 \begin{exercise}[Wave-function renormalization]
 Consider $\lambda\phi^3$ theory in $D=6$. Perturbation theory calculations (\cite{collins}, \S 7.3) yield
\begin{align}
{\eta}=\frac{\tilde{\lambda}^2}{6(4\pi)^3}\;,&\\
\frac{d}{dt}\tilde{m}^2_k=-2\tilde{m}^2-\frac{5\,\tilde{\lambda}_k^2}{6(4\pi)^3}\,\tilde{m}_k^2\;,&\quad\quad
\frac{d}{dt}\tilde{\lambda}_k=-\frac{3\,\tilde{\lambda}_k^3}{4(4\pi)^3}\;,
\end{align}
where the running of $\eta$ comes from the graph in Figure \ref{fig:phi3}.\\
Using Litim's regulator, show that the $\b$-functions and $\eta$ predicted by the FRGE are
\begin{align}
\eta=0\;,&\\
\frac{d}{dt}\tilde{m}^2_k=-2\tilde{m}^2_k+\frac{1}{3(4\pi)^3}\frac{\tilde{\lambda}^2_k}{(1+\tilde{m}^2_k)^3}\;,&
\quad\quad
\frac{d}{dt}\tilde{\lambda}_k=-\frac{1}{(4\pi)^3}\frac{\tilde{\lambda}^3_k}{(1+\tilde{m}^2_k)^4}\;.
\end{align}
Explain the apparent discrepancy.
 \end{exercise}

\subsection{The Wilson-Fisher fixed point in $D=3$}\label{sec:wilsonfisher}
The techniques that we have developed up to now will allow us to consider a rather complicated statistical system.  It is an experimental fact that several three-dimensional magnetic systems exhibit a second order phase transition (PT) when their temperature $T$ approaches a critical value $T_\infty$. In that vicinity, for a class of them, it is observed that the correlation length $\xi$ diverges as
\be
\label{eq:correldiv}
\xi\approx (T-T_\infty)^{-\nu}\approx\vartheta^{-\nu}\,,\quad\quad \nu\approx0.63\,,
\ee
where we introduced the reduced temperature $\vartheta=(T-T_\infty)/T_\infty$. A typical model for this kind of magnets is the \emph{Ising model} that, in the approximation of continuous spins (and in zero magnetic field), can be described by Euclidean $\lambda\phi^4$ theory in $D=3$. However, the behavior (\ref{eq:correldiv}) is  \emph{universal}, i.e. common to many magnetic systems, which may include more general spin interactions.

Our experience suggests that this behavior should be explained in terms a property of the theory space $\mathcal{T}$ common to all these theories, namely the existence of a fixed point with one relevant eigenvalue $\zeta$ (related to $\nu$). Then there will be a codimension-one stable manifold, and the phase transition at $T=T_\infty$ will happen as the one-parameter curve in $\mathcal{T}$ described by the one-parameter family of QFTs under consideration intersects $\Ws$. This will be confirmed by the RG analysis.

In a statistical theory we are interested in the long-wavelength (IR) behavior, i.e. $k\to0$. When we integrate down from a scale $k_1$ to $k_2<k_1$, the free energy will transform inhomogeneously, offsetting by a constant term which represents the energy of the modes that have been integrated out. Schematically, recalling that the free energy is given by $\Gamma^{(0)}$,
\be
\Gamma^{(0)}(k_2;\,\vartheta_2)\approx \Delta \mathcal{F}_{k_1,k_2}+\left({k_2}/{k_1}\right)^{\rm some\ scaling}\Gamma^{(0)}(k_1;\,\vartheta_1)\,,
\ee
where the offset $\Delta \mathcal{F}_{k_1,k_2}$ is a regular function\footnote{This is similar to the offset $\Delta S_k$ that we subtracted from $\tilde{\Gamma}_k$ when deriving the FRGE.}. If we restrict to the singular contribution to the free energy, therefore, we get a homogeneous scaling equation. The scaling exponent is simply given by the dimension, since $\Gamma^{(0)}$ has no external lines, in agreement with the solution of Callan-Symanzik equation (\ref{eq:CSsolution}). As for the relation between $\vartheta_1$ and $\vartheta_2$, just like what happened with iterated map of the interval, since we are approaching the stable manifold $\Ws$ its scaling is given by the relevant eigenvalue at the fixed point, so that
\be
\Gamma^{(0)}_{\rm sing.}(s\,k;\,\vartheta)\approx s^3\ \Gamma^{(0)}_{\rm sing.}(k;\,s^{\zeta}\,\vartheta)\,,
\ee
where $s=k_2/k_1$, or equivalently
\be
\Gamma^{(0)}_{\rm sing.}((\vartheta_2/\vartheta_1)^{1/\zeta}k;\,\vartheta_2)\approx \left(\vartheta_1/\vartheta_2\right)^{3/\zeta}\Gamma^{(0)}_{\rm sing.}(k;\,\vartheta_1)\,.
\ee
This indicates that, close to the PT, the relative scaling of wavenumber and temperature is $k\approx \vartheta^{-1/\zeta}$, so that the correlation length diverges as
\be
\xi\approx \vartheta^{1/\zeta}\quad\quad\Longrightarrow\quad\quad \zeta=-1/\nu\,.
\ee

Now that we know how to relate $\zeta$ to the physics of the problem, it is time to investigate $\mathcal{T}$. As always, there is one fixed point in the theory space when the theory is free. Clearly it is not the one we are after (the Ising model at the PT is not a free theory!). There, both the $\phi^2$ and $\phi^4$ couplings are relevant, i.e. they sink into the GFP going to short wavelengths. This means that if we define an microscopic theory with given values of the corresponding couplings in the UV, these flow away from the perturbative region in the IR, while all $n$-points interactions are generated as effective vertices. 

To se what happens non-perturbatively we can use the FRGE \cite{wetterich93,adams}. The simplest way to proceed is the ansatz (\ref{eq:phi4ansatz}) which at $D=3$ gives
\begin{align}\label{wilsonflow}
\eta = 0,\ \ \ \ \frac{d }{dt}\tilde{m}^2_k=&-2\tilde{m}^2_k- \frac{\tilde{\lambda}_k}{6\pi^2\;(1+\tilde{m}_k^2)^2},\ \ \ \ 
\frac{d }{dt}\tilde{\lambda}_k=-\tilde{\lambda}_k+ \frac{\tilde{\lambda}_k^2}{\pi^2\,(1+\tilde{m}_k^2)^3}\;.
\end{align}
Besides the Gaussian FP, there is a nontrivial one at
\be
 \tilde{\lambda}^*\approx7.7627,\quad\quad\tilde{m}^{2*}\approx-0.0769\,.
\ee
and the critical exponents can be estimated from the Jacobian
\begin{equation}
\mathrm{Jac}_{NGFP}=\left(\begin{array}{cc}
-1.6667 & -0.0198\\
-25.229 & 1
\end{array}\right)\cong\left(\begin{array}{cc}
-1.8426 & 0\\
0 & 1.1759
\end{array}\right)\;.
\end{equation}
The estimate for the irrelevant exponent gives
\be
\zeta\approx-1.8426\quad\quad\Longrightarrow\quad\quad
\nu\approx0.5427\,,
\ee
which is not far from the measured value.

\begin{figure}[!t]
  \begin{center}
\includegraphics[width=0.6\textwidth]{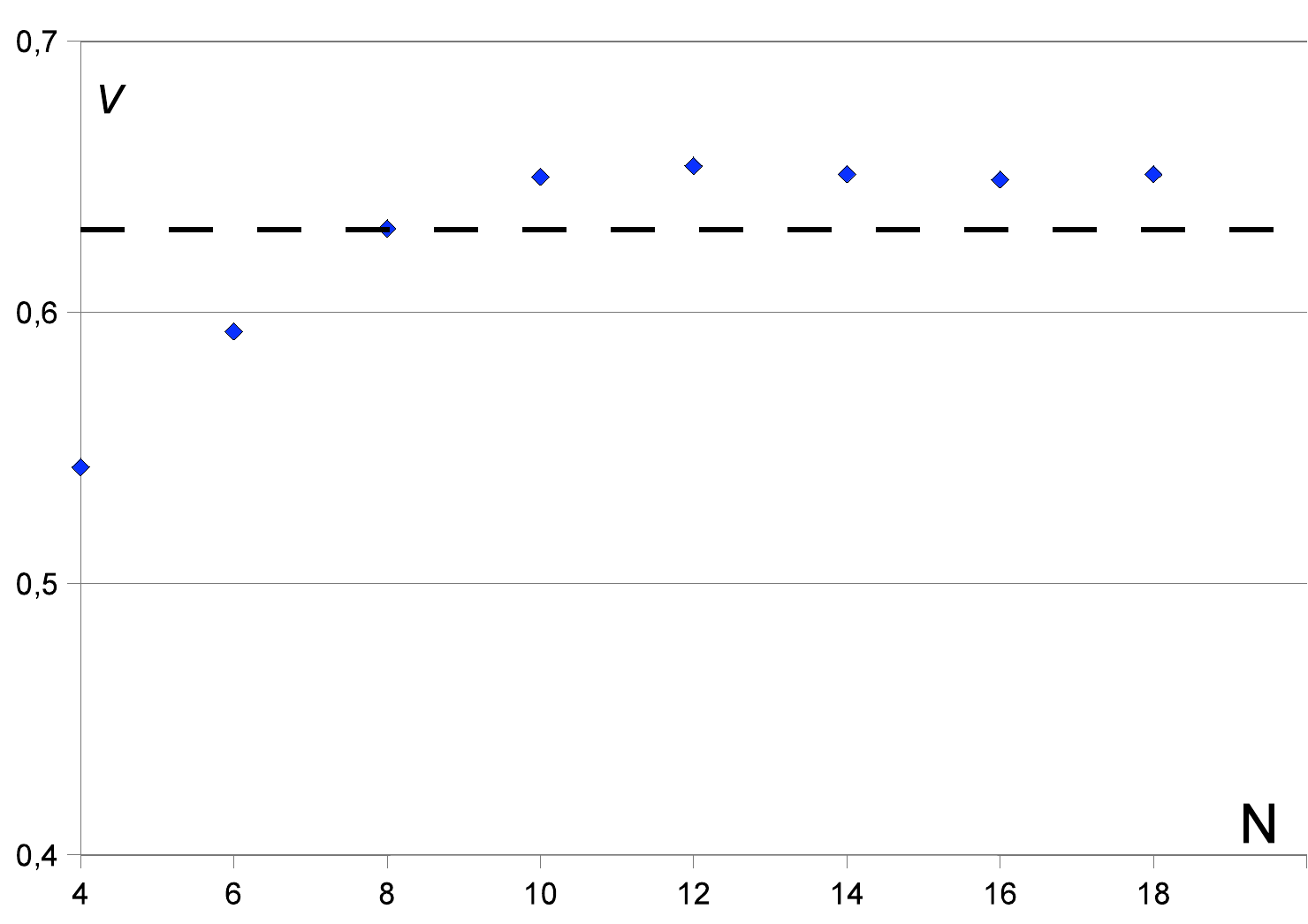}
  \caption{The critical exponent $\nu$ at the Wilson-Fisher FP, in various polynomial truncations of degree $N$. The dashed line is the prediction of seven-loops perturbation theory.}
  \label{fig:WF}
  \end{center}
\end{figure}

To be sure that we are dealing with a physical effect, we need to consider more general truncations, and ask ourselves:
\begin{enumerate}
\item Is the Wilson-Fisher fixed point present in all the truncations considered?
\item Is it the only non gaussian fixed point?
\item Does the dimension of the attractor at the fixed point change?
\item Are critical exponents such as $\nu$ well behaved when considering larger truncations?
\end{enumerate}
Considering more general ans\"atze one sees that, even if some other fixed points may appear, the Wilson-Fisher one is the only one to persist in all truncations. Furthermore, it appears that there is only ore relevant direction, and the estimate for $\nu$ is quite stable, as shown in Figure \ref{fig:WF}. There we compare the estimate with the result of resummed seven-loops perturbation theory $\nu=0.6304$, finding a discrepancy of about 3\% \cite{zinnjustin}. Remark that we have taken the local potential approximation where $\eta\equiv0$ whereas the the perturbative calculations gives $\eta^*=0.0335$.

We have therefore established a scenario similar to the one of Section \ref{sec:feigenbaum}. This was first done in this context by Wilson and Fisher  \cite{wilsonfisher} using a different RG based technique\footnote{The interested reader will find nice pedagogical expositions in\cite{cardy,weinbergQFT, zinnjustin}.} \cite{wilson72,wilsonkogut}. The implications of this fact are that when we consider a one-parameter family of QFTs (i.e. a magnetic system at different temperatures) and approach the PT, the critical exponents are determined by the RG properties. All the theories on the stable manifold have infinite correlation lengths, and have long-wavelength properties similar to the ones of the fixed point theory. 

\begin{figure}[!t]
  \begin{center}
\includegraphics[width=0.785\textwidth]{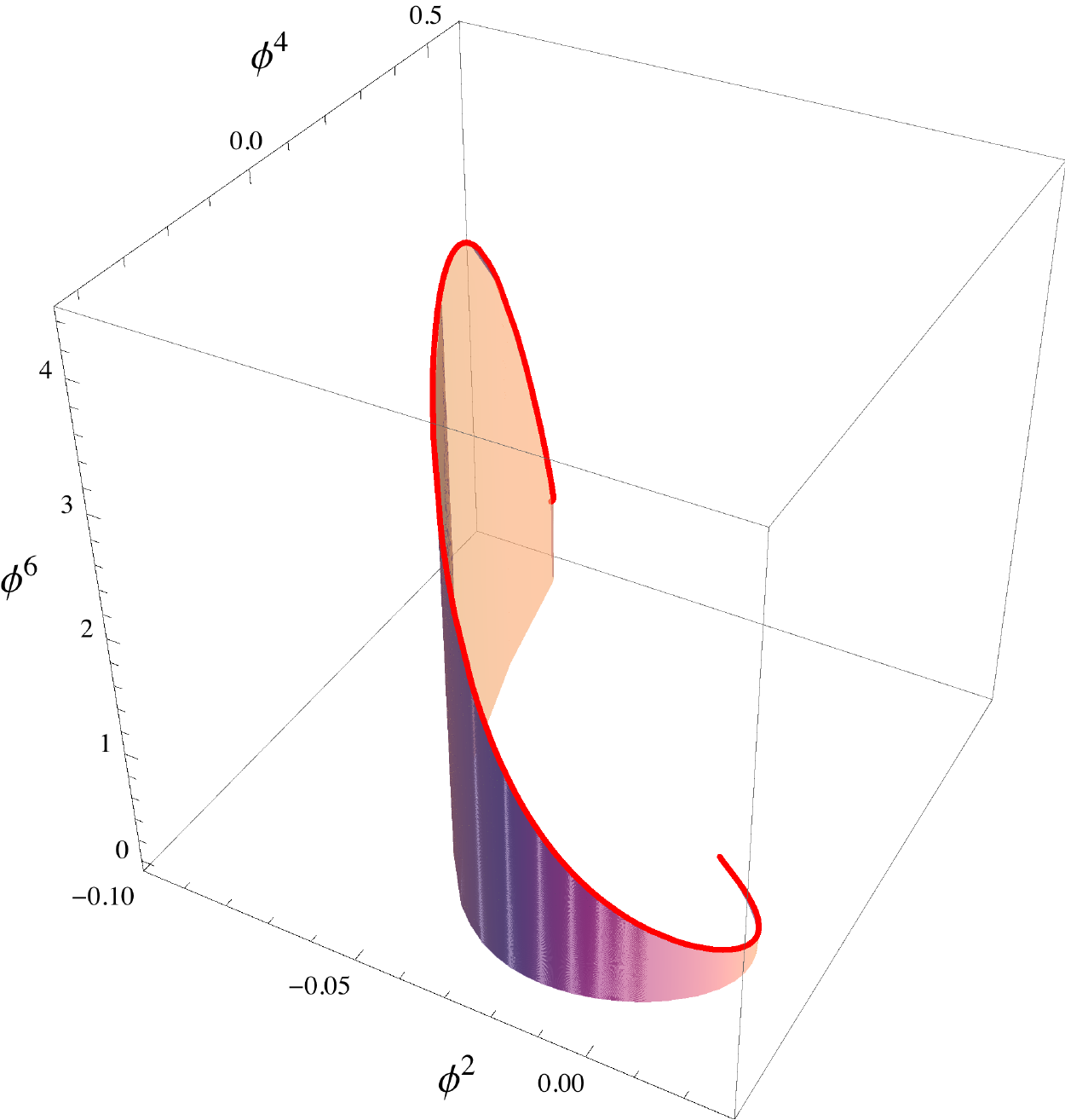}
  \caption{Flow from the GFP to the Wilson-Fisher fixed point. The shaded area represents the strength of the $\phi^6$ coupling.}
  \label{fig:WFflow}
  \end{center}
\end{figure}

Furthermore, the existence of two fixed points means that in this theory space there exists asymptotically safe theories. In the vicinity of the Gaussian fixed point there exists a two-dimensional manifold tangent to the plane $\{\tilde{m}^2,\tilde{\lambda}\}$ consisting of theories that in the UV sink into the GFP. We can consider theories that in the IR are attracted to the NGFP, so that for them the RG flow is bounded as we change the scale, justifying \textsl{a posteriori} our crude treatment of the cutoff $\Lambda$. Of course in this particular case this would not be an issue, since this theory is not supposed to be fundamental, and we could have a natural UV cutoff $\Lambda\approx1/a$ where $a$ is the lattice spacing (nanometers), as well as an IR one of order of the meters. Figure \ref{fig:WFflow} depicts the flow from the GFP into the Wilson fisher one in a truncation including up to $\phi^6$ interactions; notice how the initial condition has no six-points interaction, which is generated as an effective vertex. 

As mentioned, the FRGE technique can be extended to fermions, gauge theories and curved backgrounds, with some technical difficulties\footnote{To mention two: respecting gauge invariance, that can be done by the background field method \cite{abbott}, and evaluating traces in curved backgrounds, that requires heat-kernel techniques \cite{seeley,dewitt65}.}. A significant application of these ideas is to establish whether Gravity may be asymptotically safe as a QFT, due to a UV attractive NGFP, in a scenario similar to the one described here, involving however many subtleties \cite{christensenduff,weinberg76,weinberg79,smolin,reuter96,lauscherreuter02b,machado10} (see \cite{niedermaier} for a review). 

Let us conclude emphasizing that, to us, the most interesting feature of FRGE is the possibility of extracting predictions for a class of systems with relatively few computational difficulties and, what is most important, knowing only the general features of the systems under consideration, regardless of the details of its microscopic dynamics. In this sense, the RG approach explains our success in quantitatively describing Nature despite our ignorance of its most fundamental laws.

\begin{exercise}[The Ising universality class]
Show that
\be
H=-\frac{1}{2}\sum_{r,\,r'} J(r'-r)\,S(r)\,S(r')+\sum_r K\,(S(r)^2-1)^2\,,
\ee
that is an Ising-like Hamiltonian, can be written in the continuous limit as
\be
H=\int d^Dx\left(-\frac{1}{2}J\,a^2\,R^2\;S(x)\Delta S(x)-(2K+J)\, S(x)^2+K\,S(x)^4\right)\,a^{-D}\,,
\ee
 where $a$ is the lattice spacing and
$
J=\sum_r J(r)\,,$ $R^2=\sum_r r^2\,J(r)\,.
$\end{exercise}

\begin{exercise}[The local potential approximation]
An efficient way of finding $\nu$ is, rather than resorting to a truncation, to make an ansatz of the form
\begin{equation}
\Gamma_k[\varphi]=\int d^Dx \left(-\mathcal{Z}_k \varphi\Delta\varphi+\mathcal{Z}_kV_k(\varphi^2)\right)\;.
\end{equation}
Show that in this case the fixed point condition is the differential equation for the dimensionless potential $\tilde{V}(\varphi)$ (with $\eta=0$).
Interestingly, this approach can be extended beyond the local potential approximation  \cite{bonannozappala,bergestetradis}.
\end{exercise}

\end{document}